\documentclass[11pt]{article}
\usepackage{amssymb,amsmath,amsthm,graphicx,ulem}

\pdfoutput=1

\usepackage{graphicx,subfigure}
\usepackage{epsfig}
\usepackage{amsmath}
\usepackage{amsfonts}
\usepackage{amssymb}
\usepackage[usenames]{color}
\usepackage[a4paper,left=2.cm,right=2.cm,top=2.5cm,bottom=2.5cm]{geometry}

\newcommand{\beq}{\begin{equation}}
\newcommand{\eeq}{\end{equation}}
\newcommand{\be}{\begin{equation}}
\newcommand{\ee}{\end{equation}}
\newcommand{\beqa}{\begin{eqnarray}}
\newcommand{\eeqa}{\end{eqnarray}}
\newcommand{\beqar}{\begin{eqnarray*}}
\newcommand{\eeqar}{\end{eqnarray*}}
\newcommand{\bea}{\begin{eqnarray}}
\newcommand{\eea}{\end{eqnarray}}

\newcommand{\reef}[1]{(\ref{#1})}

\newcommand{\ie}{{\it i.e.}\ }

\def\CD{{ \cal D }}

\def\ii{\textrm{i}}
\def\Y{\mathbb{Y}}

\begin{document}

\allowdisplaybreaks

\normalem

\title{A scalar field condensation instability of rotating anti-de Sitter black holes}

\vskip1cm
\author{\'Oscar J. C. Dias${}^{\dagger a}$, Ricardo Monteiro${}^{\dagger,\ast b}$, Harvey S. Reall${}^{\dagger c}$,
Jorge E. Santos${}^{\dagger,\ast d}$\\ \\ ${}^{\dagger}$ DAMTP, Centre for Mathematical Sciences, University of Cambridge, \\
 Wilberforce Road, Cambridge CB3 0WA, UK \\ ${}^{\ast}$ CFP, Faculdade de Ci\^encias da Universidade do Porto, \\
Rua do Campo Alegre, 4169-007 Porto, Portugal \\ \\
 \small{ ${}^{a}$ O.Dias@damtp.cam.ac.uk, ${}^{b}$ R.J.F.Monteiro@damtp.cam.ac.uk,} \\
 \small{ ${}^{c}$ H.S.Reall@damtp.cam.ac.uk, ${}^{d}$ J.E.Santos@damtp.cam.ac.uk}}

\maketitle

\begin{abstract}
\noindent Near-extreme Reissner-Nordstr\"om-anti-de Sitter black
holes are unstable against the condensation of an  uncharged scalar
field with mass close to the Breitenl\"ohner-Freedman bound. It is
shown that a similar instability afflicts near-extreme large
rotating AdS black holes, and near-extreme hyperbolic
Schwarzschild-AdS black holes. The resulting nonlinear hairy black
hole solutions are determined numerically. Some stability results
for (possibly charged) scalar fields in black hole backgrounds are
proved. For most of the extreme black holes  we consider, these
demonstrate stability if the ``effective mass" respects the
near-horizon BF bound. Small spherical Reissner-Nordstr\"om-AdS black
holes are an interesting exception to this result.
\end{abstract}


\tableofcontents

\section{Introduction}
Charged black holes in anti-de Sitter space have played an important
role in recent  discussions of superconductivity based on the
gauge/gravity correspondence. The simplest model of this phenomenon
consists of gravity coupled to a Maxwell field and a charged scalar
field \cite{Gubser:2008px}. The non-superconducting phase is dual to
the Reissner-Nordstr\"om-AdS black hole solution, which is stable at
sufficiently high temperature. However, at temperature below a
certain critical value, this solution is unstable against
condensation of the scalar field. At these low temperatures there
exists a new charged black hole solution, with a non-trivial
scalar field present. This describes the superconducting phase.

Surprisingly, the condensation can occur even when the scalar field
is  uncharged \cite{Hartnoll:2008kx}. A nice way of understanding
this is to consider the special case of an extreme (planar)
Reissner-Nordstr\"om-AdS black hole. This has $AdS_2 \times R^2$
near-horizon geometry. The Breitenl\"ohner-Freedman bound
\cite{Breitenlohner:1982jf,Mezincescu:1984ev} required for stability
of the $AdS_2$ lies above that required for stability of the
asymptotic $AdS_4$. This suggests that an uncharged scalar field
which satisfies the $AdS_4$ BF bound but violates the $AdS_2$ BF
bound will be unstable in the extreme RN-AdS background.\footnote{Ref. \cite{Hartnoll:2008kx} attributed this argument to M. Roberts.}  This is
confirmed by solving numerically the scalar equation of motion in
the black hole background \cite{Hartnoll:2008kx,Denef:2009tp}.

The motivation for the present paper is the observation that
extreme,  {\it rotating}, AdS black holes also have near-horizon
geometries containing an $AdS_2$ factor. Hence one might expect that
a scalar condensation instability will afflict these black holes
too.\footnote{It was shown in Ref. \cite{Sonner:2009fk} that the
charged scalar condensation instability of static charged AdS black
holes extends also to rotating charged AdS black holes. In our work,
neither the scalar field nor the black hole is charged.} We shall
confirm that this is indeed the case for black holes sufficiently
close to extremality.

We shall consider the (topologically spherical) rotating $AdS_5$
black hole  solutions discovered by Hawking, Hunter and Taylor (HHT)
\cite{Hawking:1998kw} (sometimes called Myers-Perry-AdS). In general
these have two angular momenta but we shall set these equal since
the solution then depends non-trivially only on a single radial
coordinate $r$. These solutions can be parameterized by the horizon
radius $r_+$ and the angular velocity $\Omega_H$.  At extremality,
the $AdS_2$ BF bound is above the $AdS_5$ bound if $r_+/\ell > 1$
(where $\ell$ is the $AdS_5$ radius), i.e., for ``large" black
holes. In this case, we might expect scalar condensation to occur if
the scalar mass $\mu$ lies between the two BF bounds.

\begin{figure}[t]
\centerline{\includegraphics[width=.55\textwidth]{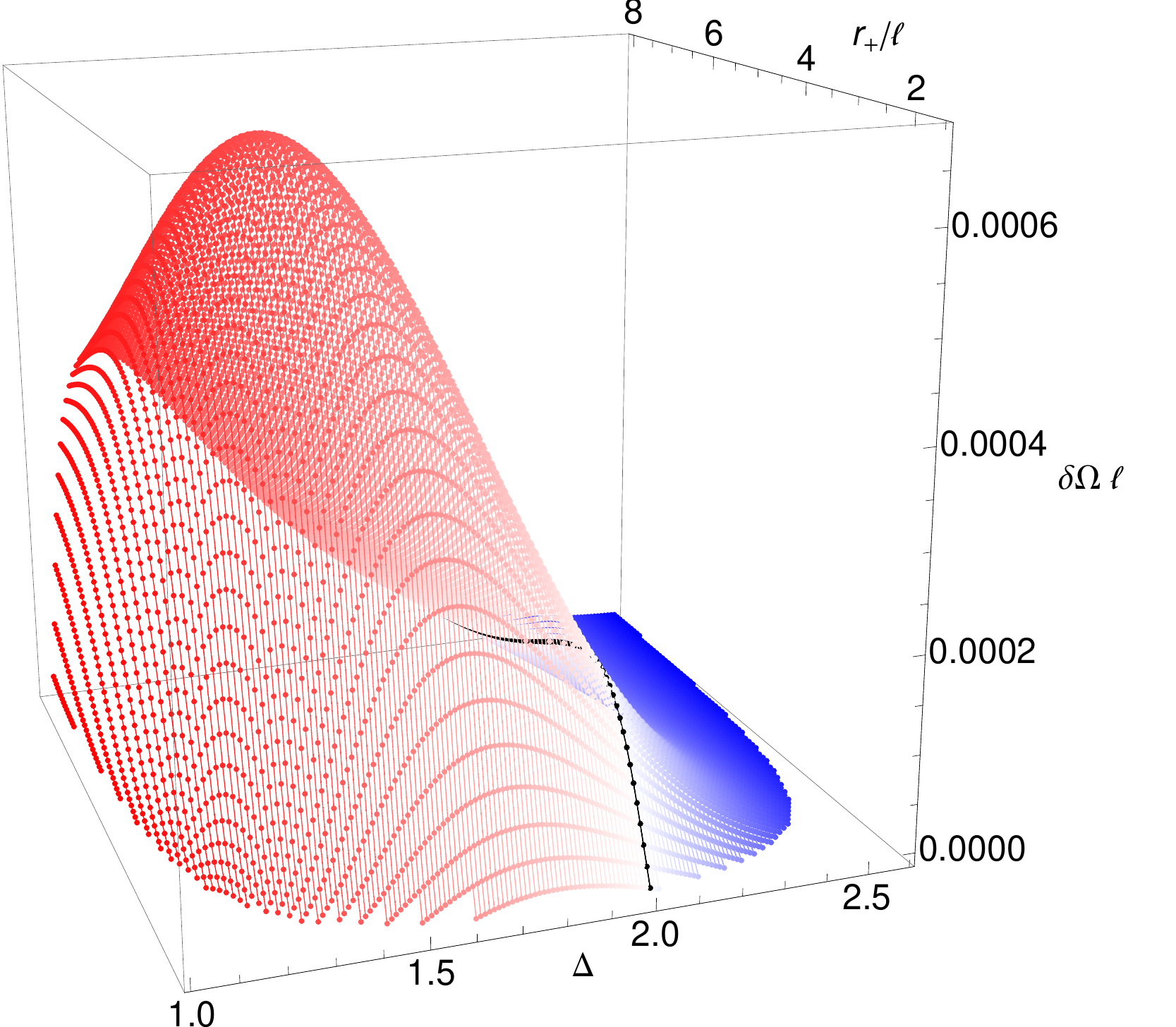}}
\caption{Scalar condensation in the HHT black hole with equal angular momenta. We consider a free scalar field
whose CFT dual is an operator of dimension $\Delta$. The vertical
axis is the difference of $\Omega_H$ from its extreme value
$\delta\Omega  \equiv \Omega_H^{\rm ext}(r_+)-\Omega_H$. The surface
corresponds to values of $\Omega_H$, $r_+$ and $\Delta$ for which
there exists a regular, time-independent solution of the scalar
equation of motion. The surface extends to arbitrarily large $r_+$.
The region enclosed by this surface corresponds
to values of $\Omega_H$, $r_+$ and $\Delta$ for which the black hole
is unstable against scalar condensation. The intersection of the
surface with the $\delta\Omega=0$ plane corresponds to saturation of
the near-horizon $AdS_2$ BF bound for extreme black holes. The black curve is for $\Delta=2$, i.e., saturation of the $AdS_5$ BF bound.}
\label{Fig:linearMP}
\end{figure}

We have determined numerically the values of $r_+$, $\Omega_H$ and
the  scalar mass $\mu$ for which there exists a time-independent
solution of the scalar field equation of motion which preserves the
symmetries of the background. Such a solution is expected to arise
at the threshold of the instability. Figure \ref{Fig:linearMP}
summarizes the result of this calculation. The condensate appears
only when the black hole is very close to extremality: typical temperatures are $T_H \ell \sim 10^{-3}$.

The analogy with the charged case suggests that the endpoint of the
instability should be a 2-parameter rotating black hole solution
with scalar hair. This new solution should have the same symmetries
as the HHT solution, so it will have equal angular momenta. We have
determined certain 1-parameter subfamilies of this solution
numerically (taking $\mu$ to saturate the $AdS_5$ BF bound). We find
that the hairy black holes have higher entropy than the HHT black
holes with the same mass and angular momenta. Hence it is consistent
with the second law for the HHT black hole to evolve to our new
hairy solutions. Our numerical results suggest that the entropy of
the hairy black holes will vanish as the temperature approaches
zero, just as in the charged case
\cite{Horowitz:2009ij,Basu:2010uz,Bhattacharyya:2010yg}.

We have extended the linear analysis to the four-dimensional
Kerr-AdS solution.  Again, we found that a scalar condensation
instability occurs in a region close to extremality. Since this
solution depends non-trivially on two coordinates ($r$ and
$\theta$), we did not attempt to construct numerically the
corresponding hairy black holes. Note that it has been proved that a scalar field that respects the $AdS_4$ BF bound is stable in a sufficiently slowly-rotating Kerr-AdS background \cite{Holzegel:2009ye}. Our results show that this conclusion does not extend to rapidly rotating Kerr-AdS black holes.

Our results have implications for (conjectured) no hair theorems.
Consider AdS gravity coupled to an uncharged scalar field. Ref.
\cite{Hertog:2006rr} conjectured that static spherically symmetric
hairy black hole solutions do not exist if the theory admits a
positive energy theorem with boundary conditions that preserve the
symmetries of AdS. Our results demonstrate that this conjecture
cannot be extended to non-static black holes.

For Reissner-Nordstr\"om-AdS, the instability against condensation
of an uncharged scalar can co-exist with an instability against
condensation of a charged scalar. In the rotating case, the analogue
of the scalar field charge is the angular momentum quantum number
$m$ of the scalar field, which we have set to zero. However, for $m
\ne 0$, black holes with $\Omega_H \ell > 1$ exhibit a
superradiant instability \cite{Kunduri:2006qa}. This is the analogue
of charged scalar condensation. Now, since extreme black holes
always have $\Omega_H \ell > 1$, and scalar condensation with $m=0$
occurs only very close to extremality, it follows that the scalar
condensation instability studied in this paper always coexists with
the superradiant instability. Moreover, the new hairy black holes
that we find have $\Omega_H \ell > 1$. Therefore they are likely to
suffer from a superradiant instability themselves. The endpoint of
this kind of instability is unknown.

Two other topics are studied in this paper. First, we note that a
nice toy model for the study of uncharged scalar field condensation
is the Schwarzschild-AdS black hole with hyperbolic spatial slices.
This solution has a regular extreme limit (in contrast to the
spherical or planar Schwarzschild-AdS black hole) with an $AdS_2$ in
the near-horizon geometry. We determine the range of values for
$r_+$ and $\mu$ for which scalar condensation occurs. The resulting
1-parameter nonlinear hairy black hole solution is determined numerically (for particular $\mu$).\footnote{We note that previous work has found a similar solution {\it analytically}  by including a suitably chosen scalar potential \cite{Martinez:2004nb}. We emphasize that we are working with a {\it free} (and minimally coupled) scalar. Another difference is that we use AdS invariant boundary conditions at infinity, which does not appear to be the case for the solution of Ref. \cite{Martinez:2004nb}.}

Second, we shall present some stability results concerning scalar
condensation. The idea is to construct an ``energy" functional for
the scalar field (which might not coincide with the usual energy)
which is manifestly non-increasing in time. If one can show that
this energy is positive definite then an initially small fluctuation
in the field must remain small, \ie condensation does not occur. Demonstrating positivity is non-trivial because we are mainly interested in the situation in which $\mu^2<0$. Fortunately, this problem has been addressed in Ref. \cite{Holzegel:2009ye} for spherical Schwarzschild-AdS black holes so we can generalize this method.
For
the extreme black holes considered in this paper, we show that the
energy is positive definite if, and only if, the $AdS_2$ BF bound is
satisfied. Our numerical results demonstrate that an instability
appears as soon as the bound is violated. For non-extreme black
holes, our method gives a lower bound on $\mu^2$ that guarantees
stability although this bound is not sharp.

The method can also be applied to a charged scalar field
in a Reissner-Nordstr\"om-AdS background. For planar, hyperbolic or sufficiently large spherical black holes,
our energy functional is positive definite at extremality if, and only if, the near-horizon
``effective mass" satisfies the $AdS_2$ BF bound. For the planar case, Ref.~\cite{Denef:2009tp} found
numerically that an instability occurs precisely when this bound is violated. Small spherical extreme RN-AdS black holes are an interesting exception: for these our stability bound is more restrictive than the near-horizon BF bound and indeed Ref.~\cite{Basu:2010uz} found an instability even when the near-horizon BF bound is respected.

This paper is organized as follows. We state the problem in
Section~\ref{sec:scalarcond}. We then discuss in
Section~\ref{sec:SchwAdS} the hyperbolic Schwarzschild-AdS black
hole as a toy model to illustrate the scalar condensation
phenomenon. Next we discuss rotating AdS black holes:
cohomogeneity-1 solutions in Section~\ref{sec:CondensationHHT} and
also the four-dimensional Kerr-AdS in
Section~\ref{sec:CondensationKerrAdS}. In
Section~\ref{sec:stability} we present our stability results for
scalar condensation.

\section{\label{sec:scalarcond}Scalar condensation}

Consider a scalar field with mass $\mu$ in an asymptotically $AdS_d$
background. It satisfies the Klein-Gordon equation,
 \be \label{KGeq}
 \nabla^2 \Phi - \mu^2 \Phi = 0\,, \ee
and always decays at infinity as \cite{Breitenlohner:1982jf,Mezincescu:1984ev}
 \be
\label{AdSdecay}
 \Phi(r)\simeq
 \frac{A^{(+)}}{r^{\Delta_+}}+\frac{A^{(-)}}{r^{\Delta_-}}\,, \qquad
 \hbox{where} \quad \Delta_\pm=\frac{d-1}{2}\pm
 \sqrt{\frac{(d-1)^2}{4}+\mu^2\ell^2}
\ee
and $\ell$ is the AdS radius.  Stability of the AdS background requires
reality of $\Delta_\pm$, \ie that the mass of the scalar field must
obey the Breitenl\"ohner-Freedman (BF) bound
\cite{Breitenlohner:1982jf,Mezincescu:1984ev}
 \be
\label{BFbound}
   \mu^2 \ge \mu^2{\bigr |}_{BF}\equiv -\frac{(d-1)^2}{4\ell^2}.
\ee
 For scalars with mass above this BF bound and below the unitarity bound, \ie
 \be
\label{Unitbound}
 \mu^2{\bigr |}_{BF} < \mu^2 < \mu^2{\bigr |}_{unit}\,, \qquad \hbox{with} \quad \mu^2{\bigr |}_{unit}\equiv
 -\frac{(d-1)^2}{4\ell^2}+\frac{1}{\ell^2}\,,
\ee
there is a choice of boundary conditions: one can impose either that $\Phi$ decays as $r^{-\Delta_+}$ or as $r^{-\Delta_-}$, since both give normalizable solutions. In the AdS/CFT correspondence, this choice dictates whether the operator dual to $\Phi$ has dimension $\Delta_+$ or $\Delta_-$ \cite{Klebanov:1999tb}. For masses above the unitarity bound only the mode with the faster fall-off, $r^{-\Delta_+}$, is normalizable.

Now consider an extreme, asymptotically $AdS_d$ black hole whose near-horizon geometry contains an $AdS_2$ factor with radius $R$. The BF bound associated to this $AdS_2$ is
\be
 \mu^2 \ge \mu^2{\bigr |}_{NH\,BF} \equiv -\frac{1}{4R^2}.
\ee
Hence if
\be
\label{UnstableRange}
 \mu^2{\bigr |}_{NH\,BF}  > \mu^2 \ge \mu^2{\bigr |}_{BF}
\ee
then the asymptotic $AdS_d$ will be stable but the near-horizon geometry is unstable. This suggests that the full black hole solution will be unstable against scalar condensation \cite{Hartnoll:2008kx}. This has been confirmed by numerical calculation in certain cases \cite{Denef:2009tp}.

\section{\label{sec:SchwAdS}Toy model: hyperbolic Schwarzschild-AdS}

\subsection{Introduction}

The simplest black hole that exhibits the scalar condensation phenomenon is the Schwarzschild-AdS solution with hyperbolic spatial sections. Consider the general Schwarzschild-AdS solution in $d$ dimensions:
\be \label{SchwAdS}
  ds^2 = -f(r) dt^2 + f(r)^{-1} dr^2 + r^2 d\Sigma_k^2,
 \ee where $d\Sigma_k^2$ is the
metric on a unit sphere ($k=1$), hyperboloid ($k=-1$), or flat space ($k=0$) that satisfies $R_{ab} = k(d-3)\hat{g}_{ab}$. The function $f(r)$ is given by
\be \label{SchwAdSf}
 f = k \frac{(r^{d-3}-r_+^{d-3})}{r^{d-3}} + \frac{r^{d-1} - r_+^{d-1}}{r^{d-3}
 \ell^2}\,,
 \ee
where the horizon is at $r=r_+$.

For $k=0,1$, the horizon is always non-degenerate. However, for $k=-1$ a black hole solution
exists for
 \be \label{Sch:rMext}
 r_+\geq r_+^{\rm ext}\,, \qquad \hbox{with} \quad r_+^{\rm ext} = \sqrt{\frac{d-3}{d-1}}
 \,\ell \,,
\ee
 and has Hawking temperature
  \be \label{Sch:T}
  T_H = \frac{(d-1)r_{+}^{2} - (d-3)\ell^{2}}{4\pi \ell^{2}r_{+}}.
  \ee
An extreme (zero temperature) regular solution is present when the bound in \eqref{Sch:rMext} is saturated. This extreme solution has near-horizon geometry $AdS_2 \times H^{d-2}$, where the $AdS_2$ has radius $R = \ell/\sqrt{d-1}$. Hence scalar condensation seems likely when the scalar mass lies between the $AdS_d$ and $AdS_2$ BF bounds:
\be
 -\frac{(d-1)^2}{4\ell^2} \le \mu^2 < - \frac{d-1}{4 \ell^2}.
 \ee
If $\mu$ lies strictly above the $AdS_d$ BF bound then the stability
argument that  we shall present in Section~\ref{sec:stability} below
demonstrates that the scalar field is stable for sufficiently large
$r_+$.  However, if $\mu$ also lies below the near-horizon BF bound for
the extreme black hole then we expect the scalar field to become
unstable as extremality is approached, \ie as $r_+$ decreases. At
the value of $r_+$ corresponding to the threshold of instability, we
expect there to exist a time-independent solution $\Phi(r)$ of the
scalar equation of motion\footnote{
See Refs. \cite{Chan:1999sc,Aros:2002te} for previous studies of scalar fields in the hyperbolic Schwarzschild-AdS spacetime.}
 \be \label{Schw:KGeq}
 \Phi''(r)+\left(\frac{d-2}{r}+\frac{f'(r)}{f(r)}\right) \Phi'(r)-\frac{\mu^2}{f(r)} \Phi(r)=0\,.
 \ee
We shall confirm the existence of this solution numerically.

We note that the entropy and energy of the $k=-1$ black hole are given by
\be
 S=\frac{r_+^{d-3} V_\Sigma}{4}, \qquad E=\frac{(d-2)r_+^{d-3} V_{\Sigma}}{16 \pi}\left(\frac{r_+^2}{\ell^2}-1\right), \ee where $V_\Sigma$ is the
volume of the hyperboloid (which we shall assume to be
compactified).

The solution with $r_+/\ell=1$ is special: this solution is locally isometric to $AdS_5$ and has zero energy. Solutions with $r_+/\ell<1$ have negative energy with the extreme solution having the lowest energy.

\subsection{\label{subsec:linearSchwAdS}Numerical results: linear}

For definiteness, we set $d=5$. For scalar mass below the unitarity
bound we shall  consider both choices of boundary condition. For
this reason, it is more convenient to work with the dimension
$\Delta$ of the operator dual to the scalar field than with $\mu$.
For given $\Delta$ we wish to determine the value of $r_+$ for which
there exists a solution $\Phi=\Phi(r)$ that is regular at the
horizon and decays as $r^{-\Delta}$ at infinity. We do this using a
standard shooting method where we construct power series solutions
near the horizon and at infinity up to seventh order and then evolve
the solution away from these points numerically. Fixing the
amplitude of the solution at spatial infinity, we impose matching
conditions, namely the continuity of $\Phi$ and $\Phi'$, at some
fixed value of $r>r_+$. We find that matching is possible only for a
particular value of $r_+/\ell$.

\begin{figure}[t]
\centerline{\includegraphics[width=.45\textwidth]{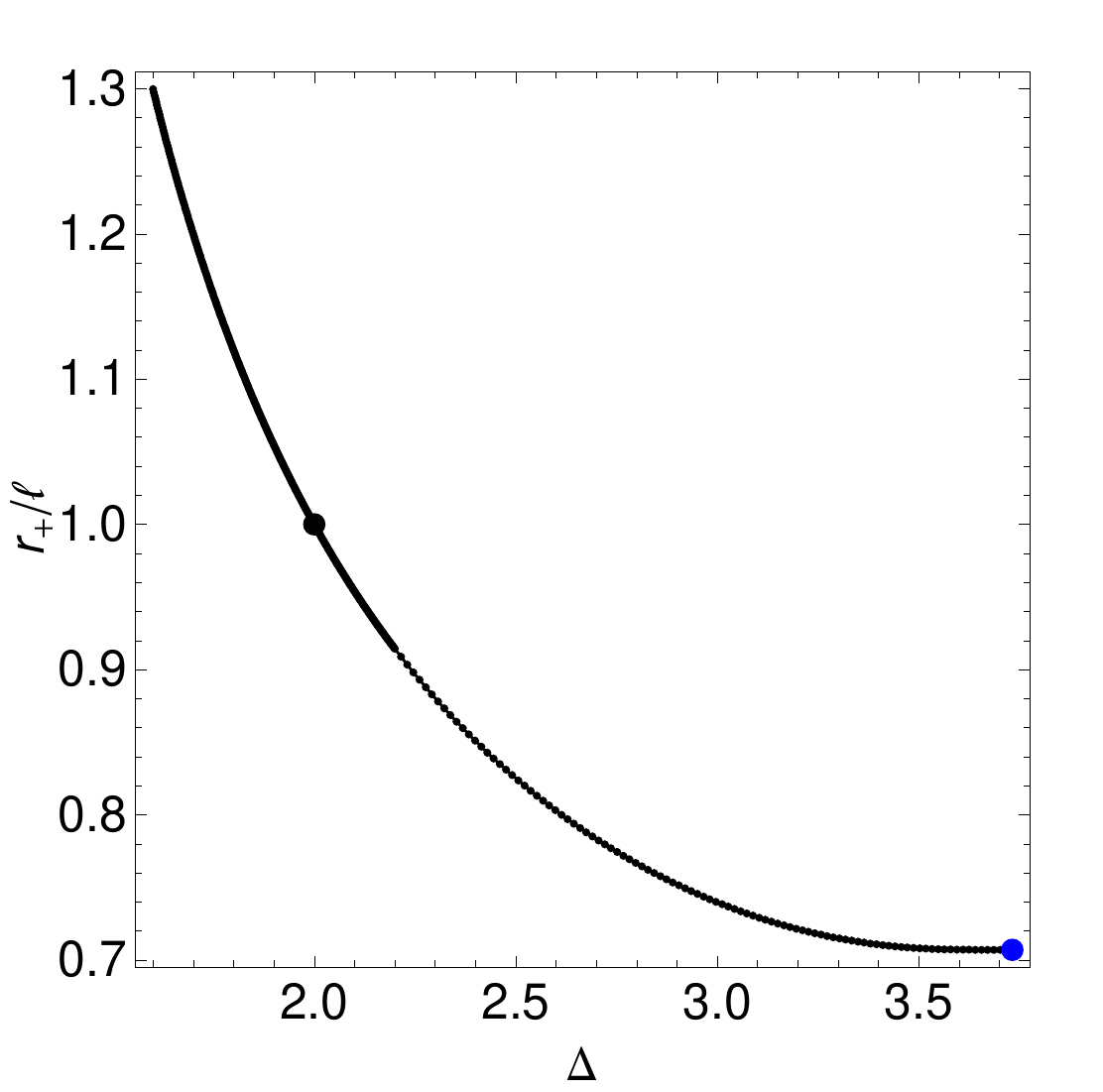}}
\caption{Threshold of the scalar condensation instability in the
$d=5$ hyperbolic Schwarzchild-AdS black hole. The plot shows the
critical dimensionless horizon radius $r_+/\ell$ where the
instability sets in as a function of the conformal dimension
$\Delta$ of the CFT operator dual to the scalar field. Points below the curve correspond to instability.
The blue dot corresponds to the extreme black hole. The critical value of $\Delta$ at this point is precisely the value predicted from the near-horizon BF bound. The black dot corresponds to the $AdS_5$ BF
bound, i.e., $\Delta=2$.} \label{Fig:SchwHyperb}
\end{figure}

The results are presented in Figure \ref{Fig:SchwHyperb}. The black
hole  is extreme for $r_+=r_+^{\rm ext}=2^{-1/2}\simeq 0.707$. The
near-horizon BF bound (which lies above the unitarity bound)
corresponds to $\Delta=2+\sqrt{3}\simeq 3.732$. This is indeed the
limiting value of the curve at the extremal blue point.\footnote{
This blue point is a limit of the curve, it does not belong to the curve. This is because, at extremality, there is no regular time-independent solution associated to the threshold of instability. The scalar field diverges at the horizon if one takes a limit of the time-independent solution associated to the threshold of instability of non-extreme solutions.}
Hence our
results agree with the expectation that scalar condensation occurs
for the extreme solution if, and only if, the near-horizon BF bound
is violated. For $\Delta > 2+\sqrt{3}$, we find no solution, so
there is no scalar condensation instability for any $r_+$. However,
as $\Delta$ decreases through $2+\sqrt{3}$, first the extreme black
hole becomes unstable and then the critical temperature below which
the black hole is unstable increases monotonically.

Saturation of the $AdS_5$ BF bound corresponds to $\Delta=2$. Our
numerical results suggest that the threshold of instability for
$\Delta=2$ occurs at $r_+  = \ell$. One can confirm analytically that the
black hole with $r_+ = \ell$ admits the regular solution
$\Phi(r)=\ell^2/r^2$ if $\Delta=2$. As mentioned above, the $r_+ = \ell$ solution is
special: it is locally isometric to $AdS_5$. Our results imply that
this solution is unstable against scalar perturbations with
$\Delta<2$. But $AdS_5$ is stable! The point is that our
solution is isometric only to {\it part} of $AdS_5$. If one
analytically extended an unstable mode of our solution to global
$AdS_5$ then either it would not be regular everywhere or it would
not satisfy the appropriate boundary conditions at infinity. Hence
it would not correspond to an instability of $AdS_5$ so there is no
contradiction.

\subsection{\label{subsec:nonlinearSchwAdS}Numerical results: nonlinear}

In this subsection, we
construct the family of hairy black holes that branches-off from the
original Schwarzchild-AdS black hole family in a phase diagram of
static and neutral solutions with an hyperbolic horizon topology.

We employ two different methods that
agree in the regime where they have to. First, we follow an
analytical perturbative approach. Starting with an appropriate ansatz for the hairy black hole solution, we expand out the
gravitational and scalar fields in a power series of a small
parameter $\epsilon$. This is a measure of the value of the scalar
field at infinity and indicates the deviation of the hairy solution
from the unstable Schwarzchild-AdS black hole. We do this
perturbative expansion up to $15^{\rm th}$ order. To obtain the
exact result, we then do a full non-linear numerical construction of
the hairy black hole solution. We present our results in subsection
\ref{sec:NonLinearSchwPD} where we describe our findings in a phase
diagram of static and neutral solutions with an hyperbolic horizon
topology.

\subsubsection{\label{sec:NonLinearSchwAnsatz}Hairy black hole ansatz. Equations of motion}

We want to look for hairy black hole solutions with an hyperbolic
horizon topology. As in the linear analysis, we consider $d=5$.
Therefore we take the following ansatz for the gravitational and
scalar fields,
 \begin{eqnarray} \label{HairySchw:ansatz}
 && ds^2 = -f(r) e^{\chi(r)}\,dt^2 + f(r)^{-1} dr^2 + r^2
 d\Sigma_{-1}^2\,, \nonumber\\
 && \Phi=\Phi(r)\,.
 \end{eqnarray}
When $\Phi(r)=0$, $\chi(r)=0$ and $f(r)$ is given by
\eqref{SchwAdSf} with $k=-1$, the system describes the hyperbolic
Schwarzchild-AdS black hole. This ansatz solves the equations
of motion derived from the Einstein-scalar action,
 \be \label{action5d} S=\frac{1}{16\pi}
\int_{\mathcal{M}}d^5 x\,\sqrt{-g}\left[R+\frac{12}{\ell^2}-\frac{1}{2}(\nabla\Phi)^2-\frac{1}{2}\mu^2\Phi^2\right]\,,\ee
if the following equations are satisfied,
\begin{subequations} \label{HairySchw:eom}
\begin{align}
 & \Phi''(r)+\frac{1}{3\ell^2 r f(r)} \left[\left(6 (2 r^2-\ell^2)+3\ell^2 f(r)-r^2 \mu^2\ell^2 \Phi(r)^2\right)\Phi'(r)
  -3 r \mu^2\ell^2 \Phi(r)\right]=0\,,  \label{HairySchw:eom1} \\
 & f'(r)+f(r)\left(\frac{2}{r}+\frac{r}{3}\Phi'(r)^2\right)+\frac{\mu^2}{3}\, r\Phi(r)^2+\frac{2}{r}-\frac{4 r}{\ell^2}=0\,,  \label{HairySchw:eom2} \\
 & \chi'(r)=\frac{r}{3} \Phi'(r)^2\,.  \label{HairySchw:eom3}
\end{align}
\end{subequations}
Given $\Phi(r)$, $\chi(r)$ can be easily found through the
integration of \eqref{HairySchw:eom3}. So the problem at hand
reduces to determining $\Phi(r)$ and $f(r)$.

The boundary conditions to be imposed are as follows. We use the
scaling symmetry of the equations of motion, $e^{\chi}\rightarrow
\lambda^2 e^{\chi}$ and $t\rightarrow\lambda t$ (with the other
variables/fields unchanged) to set $\chi=0$ at the asymptotic
boundary without loss of generality. At the horizon we demand that
$\chi$ is regular. The function $f(r)$ must vanish at the horizon,
and we take this condition as our definition for the location of the
black hole horizon. We require the hairy black hole to be
asymptotically AdS and thus $f(r)$ must behave as $r^2/\ell^2-1$ at
large distances. On the other hand, the scalar field has to be
regular at the horizon. Its asymptotic boundary condition is
determined by the requirement of normalisability at infinity. In our
linear analysis we found that a hairy black hole should exist only
for scalar masses above the BF bound and below the NH BF bound.
Therefore we restrict our attention to this range of masses. For
concreteness we consider the simplest case $\mu=\mu|_{BF}$ (and thus
$\Delta=2$) for which an analytical perturbative
construction is possible and where the numerics simplify considerably. Summarizing, the boundary conditions for the non-linear
problem are:
 \bea \label{HairySchw:BCs}
 && f{\bigl |}_{r=r_+}=0\,,\qquad  f{\bigl
 |}_{r\to\infty}\to\frac{r^2}{\ell^2}-1\,; \qquad
 \chi{\bigl |}_{r=r_+}=\mathcal{O}(1)\,,\qquad  \chi{\bigl |}_{r\to\infty}\to \mathcal{O}(r^{-4})\,; \nonumber\\
 && \Phi{\bigl |}_{r=r_+}=\mathcal{O}(1)\,,\qquad
 \Phi{\bigl |}_{r\to\infty}\to\phi_0 \,r^{-2}.
 \eea

For an explicit construction of the hairy black hole we should still
select a particular hyperbolic Schwarzchild-AdS black hole to which
the hairy black hole should reduce in the limit where the scalar
field vanishes, \ie a particular value of $r_+$ in \eqref{SchwAdSf}
(with $k=-1$). We are taking $\Delta=2$ so will find the hairy black hole that bifurcates
from the hyperbolic Schwarzchild-AdS black hole with $r_+=\ell$.
This choice is not arbitrary. Indeed in this case we can solve
analytically the equation \eqref{Schw:KGeq} that gives the
perturbative leading order scalar field. This can then be used to
generate the next-to-leading order expansion terms. Ultimately, we
will then use this perturbative result as the seed solution in the
relaxation method employed to construct numerically the exact hairy
black hole.  To sum up, our selection of parameters is
 \be \label{HairySchw:choiceParameter}
 \mu^2=\mu^2{\bigl|}_{BF}\,, \qquad \hbox{and}\qquad  f(r){\bigl|}_{\Phi \rightarrow
 0}= \frac{r^2}{\ell^2}-1\,.
 \ee

\subsubsection{\label{sec:NonLinearSchw}Perturbative and numerical construction of solution}

To construct perturbatively the hairy black hole, we expand all the
unknown functions of the system, namely $r_+$, $f(r)$ and $\Phi(r)$
in a power series of $\epsilon$,
 \be
\label{HairySchw:expansion}
  r_+(\epsilon)=\ell \sum _{j=0}^{n}\rho_{2j}\,\epsilon^{2j}\,,
  \qquad
  f(r,\epsilon)=\sum _{j=0}^n  f_{2j}(r)\,\epsilon^{2j}\,, \qquad
  \Phi(r,\epsilon)=\sum _{j=0}^n  \Phi_{2j+1}(r)\,\epsilon^{2j+1}\,,
 \ee
where $\epsilon$ is a measure of the scalar field at infinity. Using
a standard perturbation theory strategy, we plug these expansions in
the equations of motion \eqref{HairySchw:eom} to solve for the
coefficients $\rho_{2j}$, $f_{2j}(r)$ and $\Phi_{2j+1}(r)$. In
short, the procedure at each order is the following. Equation
\eqref{HairySchw:eom2} is used to solve for $f_{2j}(r)$, up to an
integration constant. The boundary condition $f(r_+)=0$ must be
satisfied at each order and is used to find $\rho_{2j}$, which
defines the horizon location where some of the boundary conditions
will be imposed. We can now use \eqref{HairySchw:eom1} to solve for
$\Phi_{2j+1}(r)$, again up to integration constants. The boundary
conditions \eqref{HairySchw:BCs} are now imposed to constraint the
integration constants generated in the process. Finally, $\chi(r)$
also has an expansion in $\epsilon$ that can at this point be easily
found by direct integration of \eqref{HairySchw:eom3}.

At leading order, $n=0$, the scalar field is given by
$\Phi(r)=\epsilon\,\ell^2/r^2 +\mathcal{O}(\epsilon^3)$ while
$f(r)=r^2/\ell^2-1 +\mathcal{O}(\epsilon^2)$ and $r_+=\ell
+\mathcal{O}(\epsilon^2)$. That is, we are in the linear regime
discussed in the previous section, where the back-reaction in the
gravitational field is neglected and $\Phi(r)$ solves
\eqref{Schw:KGeq} (note that the analytical solution is possible
only because we perturb the black hole with $r_+=\ell$). We will
find that for our purposes it will be enough to do the expansion up
to order $n=14$. The explicit expression for the expansion functions
$\{\rho_{2j}\,,f_{2j}(r)\,,\Phi_{2j+1}(r)\}$ up to $15^{\rm th}$
order would be far too cumbersome and not illuminating at all. For
illustration purposes we just give the result up to the $3^{\rm rd}$
order,
 \bea
  && \hspace{-0.8cm} r_+(\epsilon )=\ell
  -\frac{2\ell}{9}\,\epsilon^2-\frac{67\ell}{810}\,\epsilon^4 +\mathcal{O}(\epsilon^6)\,,\qquad
  \Phi(r,\epsilon)=\frac{\ell^2}{r^2}\,\epsilon -\frac{\ell^6}{9 r^6}\,\epsilon^3
  +\frac{\ell^6 \left(3 r^4+2 r^2\ell^2+54\ell^4\right)}{1620 r^{10}}\,\epsilon^5 +\mathcal{O}(\epsilon^7)\,, \nonumber\\
  && \hspace{-0.8cm}  f(r,\epsilon)=\left(\frac{r^2}{\ell^2}-1\right)+\frac{2\ell^2\left(5 r^2-3\ell^2\right)}{9 r^4}\,\epsilon^2
  -\frac{\ell^2 \left(r^4-30\ell^4\right)}{135 r^6}\,\epsilon^4
  +\mathcal{O}(\epsilon^6)\,.
 \eea
The result of the perturbative analysis up to $15^{\rm th}$ order
will be presented in the phase diagrams of subsection
\ref{sec:NonLinearSchwPD}.

The most efficient way to get the full information on hairy black
holes in the phase diagram of solutions is to resort to a full
non-linear numerical approach that solves non-perturbatively the
equations of motion \eqref{HairySchw:eom}. Our numerical strategy to
find the exact hairy black hole is to use a standard relaxation
method \cite{Nrecipes}, with a spectral discretization of the
integration grid \cite{Trefethen}. Again, we have to solve the
system of coupled ODEs \eqref{HairySchw:eom1} and
\eqref{HairySchw:eom2} subject to the boundary conditions
\eqref{HairySchw:BCs}, and for the selection of parameters
\eqref{HairySchw:choiceParameter}.

The implementation of the boundary conditions using the spectral
discretization is simpler if we implement the following redefinition
of the fields,
 \bea \label{HairySchw:fieldRedef}
 && q_f(r)=\left(\frac{r^2}{\ell^2}-1+\frac{\ell^4}{r^4}\right)^{-1}f(r)\,, \nonumber\\
 &&
 q_\Phi(r)=\frac{r^2}{r_+^2}\,\Phi(r)\,,
 \eea
in which case the  boundary conditions \eqref{HairySchw:BCs} reduce
to
 \bea
\label{HairySchw:BCsqs}
 && q_f{\bigl |}_{r=r_+}=0\,,\qquad q_f{\bigl |}_{r\to\infty}\to1\, ;
 \nonumber\\
&& \partial_r q_\Phi{\bigl |}_{r=r_+}=2 q_{\Phi}\left(\frac{1}{r_+}
+\frac{r_+}{\ell^2 -r_+^2 \left(q_{\Phi}^2 + 3\right)}
\right){\biggl |}_{r=r_+}\,,\qquad  \partial_r q_\Phi{\bigl
|}_{r\to\infty}\to0\,.
 \eea
 The factor $\ell^4/r^4$ in the relation for $q_f$ in
\eqref{HairySchw:fieldRedef} guarantees that the field redefinition
does not introduce a new critical (singular) point in the
differential equations to be solved (\ie besides the boundary
critical points). Our numerical results for the hairy black hole
will be presented in the next subsection.

\subsubsection{\label{sec:NonLinearSchwPD}Phase diagram}

In this subsection we present the properties of the hairy black
holes that we constructed using the perturbative and numerical
approaches described in the previous subsection. We present the
scalar condensate, the temperature and the entropy of the hairy
black hole as a function of its energy.

To compute the energy of the hairy black hole in AdS we use the
Astekhar-Das formalism \cite{Ashtekar:1999jx}. The temperature,
entropy and energy of the hairy black hole are given by
 \be T_H=\frac{|f'(r_+)|\,e^{\chi(r_+)}}{4\pi }\,, \qquad
  S=\frac{r_+^3 V_\Sigma}{4}\,, \qquad \hbox{and}
 \quad E=\frac{V_\Sigma \,\left(-3 r_0^2+2 \ell^2\phi_0^2\right)}{16 \pi }\,, \ee
where the energy is measured with respect to the AdS background.
$V_\Sigma$ is the volume of the (compactified) hyperboloid of unit
radius, $\phi_0\ell^2$ is the $\mathcal{O}(r^{-2})$ coefficient in
the large $r$  expansion of $\Phi(r)$ and $r_0^2$ is the
$\mathcal{O}(r^{-2})$ coefficient in the large $r$ expansion of
$f(r)$. We will compare these thermodynamic quantities with those
for the family of hyperbolic Schwarzchild-AdS black holes. Henceforth we shall set $V_\Sigma=1$, equivalently $E$ and $S$ can be regarded as energy and entropy per unit hyperboloid volume.

\begin{figure}[h]
\centerline{\includegraphics[width=.45\textwidth]{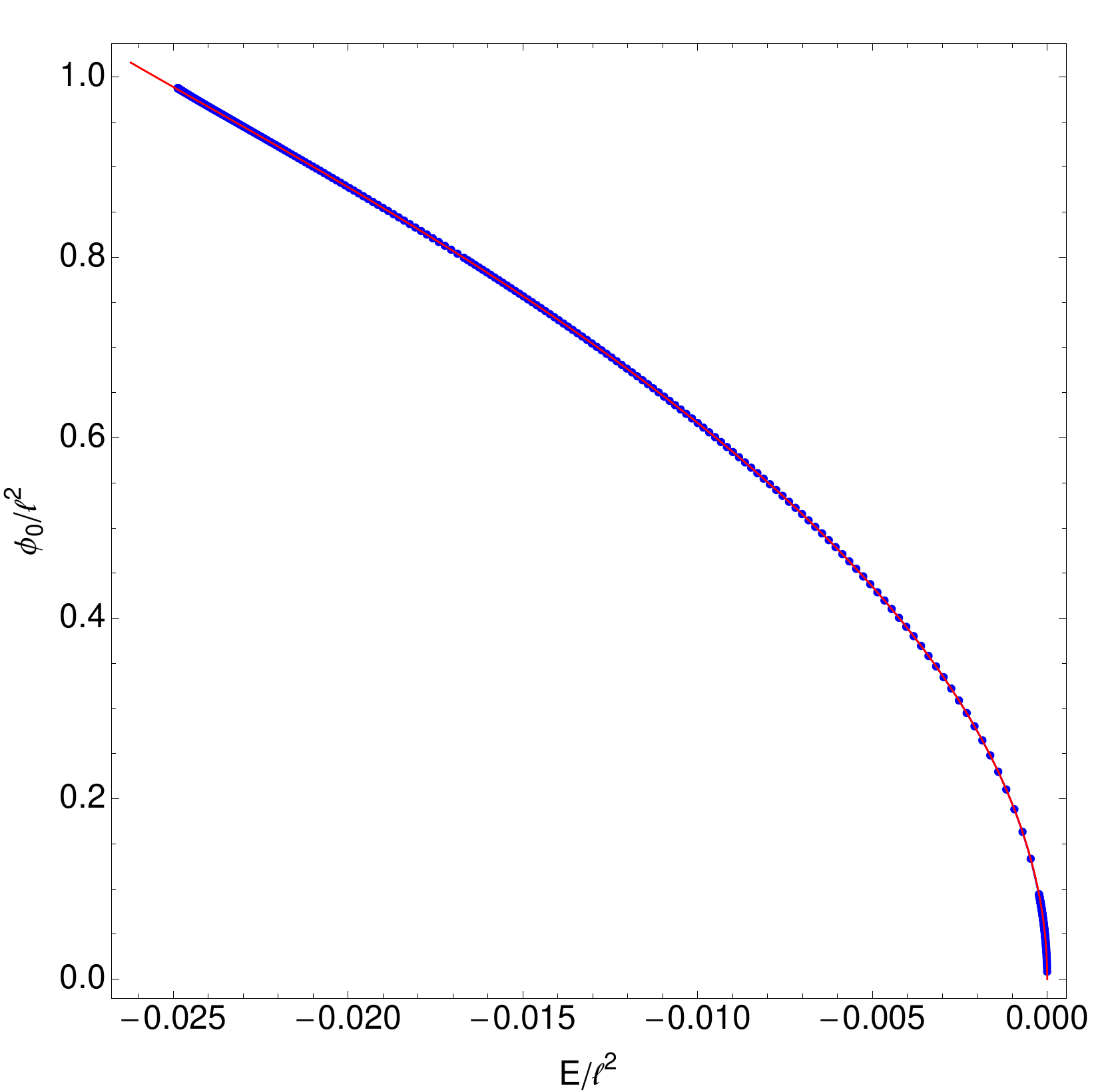}}
\caption{Asymptotic value of the condensate, $\phi_0/\ell^2$ as
defined in \eqref{HairySchw:BCs} (equivalently: vev of $\Delta=2$
operator dual to the scalar field),  as a function of the
dimensionless energy $E/\ell^2$ of the hyperbolic hairy black hole.
The red curve is the perturbative result, the blue curve is
determined numerically.} \label{Fig:HairySchw1}
\end{figure}

In Fig. \ref{Fig:HairySchw1}, we plot the value of the condensate at
infinity, $\phi_0/\ell^2$ as defined in \eqref{HairySchw:BCs}, as a
function of the dimensionless energy $E/\ell^2$ of the hairy black
hole (equivalently: we are plotting the vev of the operator dual to
our scalar field). The condensate vanishes in the limit where the
hairy black hole reduces to the Schwarzchild-AdS black hole with
$r_+=\ell$, which has zero energy. The condensate then increases as
the energy grows negative. The red curve describes the results
obtained using the analytical perturbative construction of section
\ref{sec:NonLinearSchw}, while the blue curve describes the exact
numerical results using the relaxation method of section
\ref{sec:NonLinearSchw}. The matching of these two curves is very
good for absolute values of the energy that are not too large, as it
should be.

\begin{figure}[h]
\centerline{\includegraphics[width=.45\textwidth]{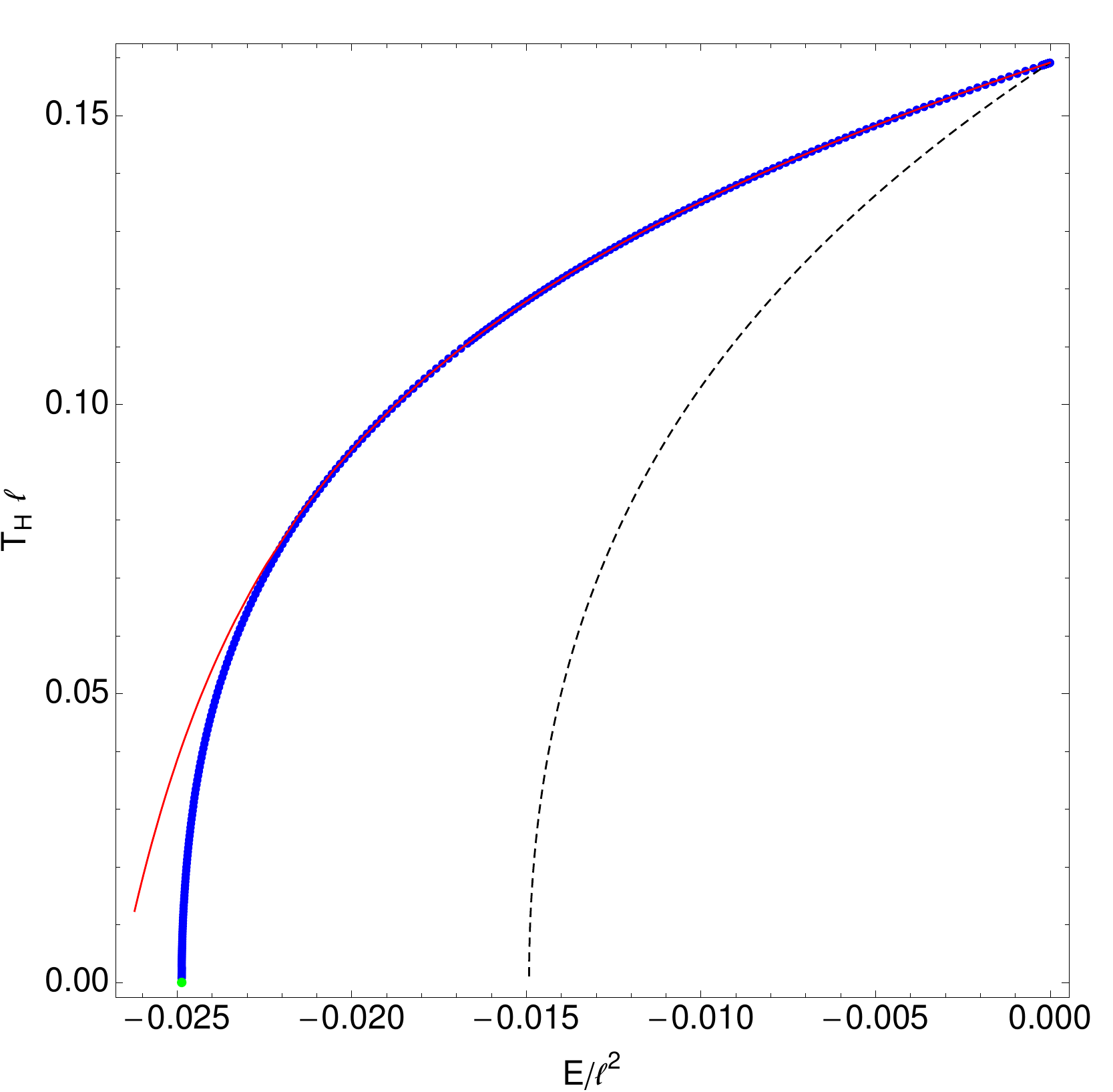}
\hspace{0.7cm}\includegraphics[width=.45\textwidth]{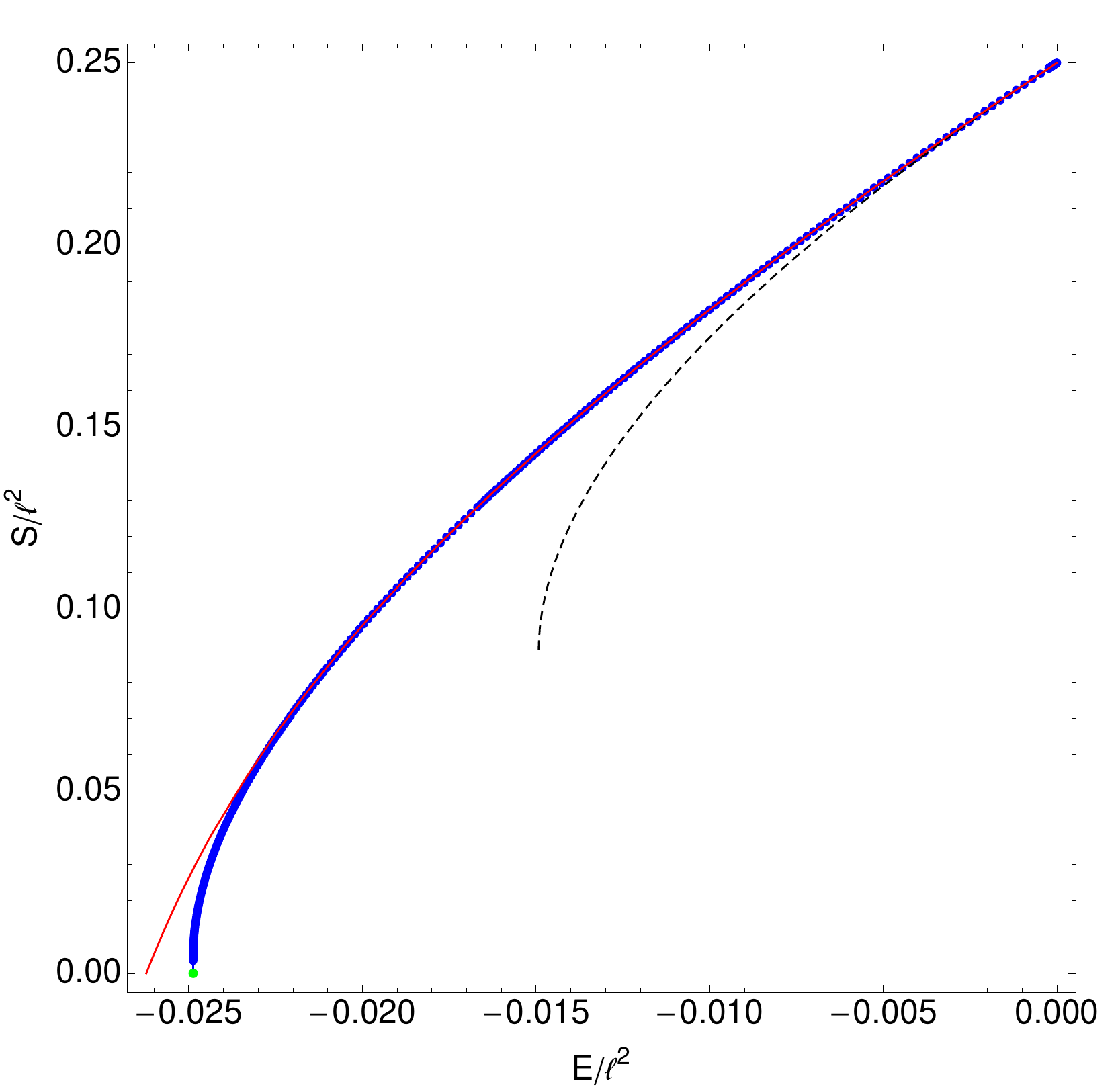}}
\caption{Dimensionless temperature, $T_H \ell$ (left), and
dimensionless entropy, $S/\ell^3$ (right), of the hyperbolic hairy
black hole as a function of its dimensionless energy $E/\ell^2$. The
scalar is at the BF bound: $\mu^2 \ell^2 = -4$ ($\Delta=2$). The red curve is the
perturbative result, and the blue curve is the numerical result (it
ends in a zero temperature and entropy configuration indicated by a
green dot). The dashed black curve corresponds to the hyperbolic
Schwarzchild-AdS black hole. This terminates at an extreme solution
with non-zero entropy. Note that the hairy black hole always has
greater entropy than the Scharzschild-AdS solution with the same
mass.} \label{Fig:HairySchw2}
\end{figure}

In Fig.~\ref{Fig:HairySchw2}, we plot the dimensionless temperature,
$T_H \ell$ (left), and dimensionless entropy, $S/\ell^3$ (right), of
the hairy black hole as a function of its dimensionless energy
$E/\ell^2$. Again, the red (blue) curve describes the results
obtained using the perturbative (numerical) construction of section
\ref{sec:NonLinearSchw}. Once more these two curves coincide for
absolute values of the energy that are small. The perturbative
construction up to the $15^{\rm th}$ order in the expansion starts
deviating from the numerical curve only for energies smaller than
$E/\ell^2~\sim -0.02$ (say). In these plots, we also display in a
dashed black line the temperature and entropy of the hyperbolic
Schwarzchild-AdS black hole family. We display this family only for
$r_+^{\rm ext} \leq r_+\leq \ell$ which corresponds to a range of
(non-positive) energies where it co-exists with the hairy black
hole. At $r_+=\ell$, we have the Schwarzchild-AdS black hole with
zero energy, from which the hairy black hole solution bifurcates. As we decrease the energy of the Schwarzchild-AdS family, the
temperature and entropy starts decreasing until the extreme black
hole with $r_+=r_+^{\rm ext}$, $T_H=0$ and non-vanishing entropy is
reached.

Our results above show that hyperbolic Schwarzschild-AdS black holes
with $r_+/\ell<1$ ($E<0$) are unstable against condensation of a
free scalar field dual to a $\Delta=2$ CFT operator. The threshold
mode of this instability is identified as a black point in
Fig.~\ref{Fig:SchwHyperb}. In the phase diagrams of
Fig.~\ref{Fig:HairySchw2}, this zero-mode signals a bifurcation
point at $E=0$ in the Schwarzchild-AdS family of solutions into a
new branch of hairy black holes. For fixed energy where the two
families co-exist, the hairy black holes always have higher
temperature and entropy than the Schwarzschild-AdS black holes. So
we have a second order phase transition. As the energy decreases,
the temperature and entropy of the hairy black hole decrease. This
family of solutions ends in a zero temperature state where both the
temperature {\it and} the entropy of the hairy solution vanish. This
configuration is represented by a green point in the phase diagrams.
Using the information from both phase diagrams we find this state to
have an energy of $E/\ell^2\simeq -0.02487$ with an estimated error
based on both plots of $10^{-3}\,\%$.

It is worth commenting on the nature of the zero temperature limit
of these hairy black holes.  The fact that the entropy appears to
vanish in this limit is good evidence that the limiting geometry is
not an extreme black hole. Further evidence for this comes from the
fact that if a hairy extreme black hole did exist then the scalar
field would be constant in its near-horizon geometry. But then the
scalar equation of motion would imply that the scalar field must be
zero in the near-horizon geometry (this argument is due to Ref.
\cite{FernandezGracia:2009em}). This would imply that the scalar
must vanish at the horizon of the full extreme black hole. However,
our numerical results indicate that the value of the scalar field at
the horizon {\it increases} as the temperature decreases. Hence the
zero temperature limit cannot be an extreme black hole. A similar
conclusion has been obtained in the case of charged scalar
condensation in Reissner-Nordstr\"om-AdS \cite{Horowitz:2009ij}.

\section{\label{sec:CondensationHHT}Rotating black holes with equal angular momenta}

\subsection{\label{sec:HHTintro}Introduction}

The HHT black hole \cite{Hawking:1998kw} with equal angular momenta is the $d=5$ case of a family of ``Myers-Perry-AdS'' black holes \cite{Gibbons:2004uw,Gibbons:2004js} with cohomogeneity-1. These solutions exist in odd dimensions $d=2 N+3$, and the metric can be written as:\footnote{The radial coordinate used here can be related to the
standard Boyer-Lindquist radial coordinate of \cite{Gibbons:2004uw,Gibbons:2004js}
through $r^2 \to (r^2+a^2)\left( 1-\frac{a^2}{\ell^2}\right)^{-1}$.
The functions $(f,h)$ used here are related with the functions
$(F,G,H)$ used in \cite{Kunduri:2006qa} through $f(r)/h(r)=F(r)^2$,
$f(r)^{-1}=G(r)^2$ and $r^2 h(r) =H(r)^2$.}
 \be \label{background} ds^2 = -\frac{f(r)}{h(r)}dt^2
+\frac{dr^2}{f(r)} + r^2 h(r)[d\psi +A_a dx^a - \Omega(r)dt]^2 + r^2
\hat{g}_{ab} dx^a dx^b\,, \ee
 where
 \bea \label{MP:metricAUX}f(r)=1+\frac{r^2}{\ell^2} -\frac{r_M^{2N}}{r^{2 N}}\left( 1-\frac{a^2}{\ell^2}\right)+\frac{r_M^{2 N}a^2}{r^{2(N+1)}},
 \qquad h(r)=1+\frac{r_M^{2 N} a^2}{r^{2 (N+1)}},
 \qquad\Omega(r)=\frac{r_M^{2 N}a}{r^{2 (N+1)}h(r)}\,, \nonumber \eea
and $\hat{g}_{a b}$ is the Fubini-Study metric on $CP^{N}$ with
Ricci tensor  $\hat{R}_{ab} =2(N+1) \hat{g}_{ab}\,$, and $A = A_a
dx^a\,$ is related to the K\"ahler form $J$ by $dA=2J$. Surfaces of
constant $t$ and $r$ have the geometry of a homogeneously squashed
$S^{2N+1}$, written as an $S^1$ fibred over $CP^{N}$. The fibre is
parameterized by the coordinate $\psi$, which has period $2\pi$.

The spacetime metric satisfies $R_{\mu\nu}
=-\ell^{-2}(d-1)g_{\mu\nu}$. Asymptotically, the solution approaches
anti-de Sitter (AdS) space with radius of curvature $\ell$.  The
event horizon is located at $r=r_+$ (the largest real root of
$f(r)$) and it is a Killing horizon of $\xi =
\partial_t+\Omega_H \partial_\psi\,$, where the angular velocity of
the horizon is given by:
 \be \label{angvel}
  \Omega_H = \frac{r_M^{2N} a}{r_+^{2N+2}+r_M^{2N}a^2}
 \leq \Omega_H^{\rm ext}\,\qquad \hbox{where}\quad \Omega_H^{\rm ext}=\frac{1}{\ell}\sqrt{1+\frac{N \ell^2}{(1+N)r_+^2}} \,. \ee
The solution saturating the bound in the angular velocity
corresponds to an extreme black hole with a regular, but degenerate,
horizon. Note that this upper bound in $\Omega_H\ell$ is always
greater than $1$ and tends to the unit value in the limit of large
$r_+/\ell$.

We shall find it convenient to parameterize the solution in terms of $(r_+,\Omega_H)$
 instead of $(r_M,a)$ through the relations,
 \be
 r_M^{2N}=\frac{r_+^{2(N+1)} \left(r_+^2+\ell^2\right)}{r_+^2 \ell^2-a^2 \left(r_+^2+ \ell^2\right)}\,,
 \qquad a=\frac{r_+^2 \ell ^2 \Omega_H}{r_+^2+\ell^2}\,.
 \ee

The temperature of the black hole,
 \be
 T_H = \sqrt{\frac{r_+^2+\ell^2}{\ell^2-r_+^2 \left( \Omega_H^2\ell^2-1\right)}}
 \frac{N \ell^2-(N+1) r_+^2 \left(\Omega_H^2\ell^2-1\right)}{2 \pi  r_+
 \ell^2}\,,
 \ee
vanishes for the extreme configuration. The mass $E$ and angular
momentum $J$, defined with respect to $\partial_{\psi}$,
are~\cite{Gibbons:2004ai} (taking $G=1$)
 \be \label{EJmpeq} E =
\frac{A_{2N+1}}{8\pi}r_M^{2N}\left (N+\frac{1}{2} + \frac{r_+^4
\ell^2 \Omega_H^2}{2(r_+^2+\ell^2)^2} \right)\,, \qquad J  =
\frac{A_{2N+1}}{8\pi} \frac{(N+1) r_M^{2N} r_+^2 \ell ^2
\Omega_H}{r_+^2+\ell^2}\,, \ee where $A_{2N+1}$ is the area of a
unit ($2N+1$)-sphere.

Once again, we shall look for a time-independent solution of the scalar wave equation that we expect to signal the onset of instability. We shall assume the scalar field to be axisymmetric (with respect to
$\partial_\psi$), i.e., it preserves the symmetry associated to the rotation of the background. However, we shall allow for the possibility that the scalar breaks the symmetry of $CP^N$. We take the field to have the separable form $\Phi_\kappa =\Phi(r)
\Y_\kappa(x)$. Here, $\Y_\kappa$ is a scalar harmonic of $CP^N$,
with Laplacian eigenvalue given by $-4\kappa(\kappa+N)$,
$\kappa=0,1,2,\ldots$. The scalar wave equation reduces to \be
\label{MP:radialEq}
 \Phi''(r)+\left(\frac{2N+1}{r}+\frac{f'(r)}{f(r)}\right) \Phi'(r)-\frac{r^2\mu^2+4\kappa(\kappa+N)}{r^2f(r)} \Phi(r)=0\,.
 \ee
In the extreme case, the near-horizon geometry is homogeneous and
takes the form of a fibration over $AdS_2$. From the above equation
we can read off an ``effective mass" for the scalar field near the
horizon: $\mu_{eff}^2=\mu^2+4\kappa(\kappa+N)/r_+^2$. The
near-horizon BF bound (determined by the radius of the $AdS_2$) is
\be
 \label{MP:nhBF}
 \mu_{eff}^2 \ge \mu^2{\bigr |}_{NH\,BF}^{(MP)} \equiv -\frac{N+1}{2\ell^2}\left(N+2 +N\, \frac{\ell ^2}{r_+^2}\right)
 \ee
We are interested in whether it is possible to violate this bound whilst respecting the $AdS_d$ BF bound (\ref{BFbound}). A short calculation reveals that this is possible if, and only if,
\be
\label{MP:rMminNHBF}
 \frac{r_+}{\ell} > \sqrt{ 1 + \frac{8 \kappa(\kappa+N)}{N(N+1)} }.
\ee
So we expect an instability only for black holes larger than the AdS length $\ell$. Furthermore, modes with $\kappa=0$ should be the most unstable since they have the lowest $\mu_{eff}^2$. Written out explicitly, we expect an instability of the extreme black hole if the scalar mass $\mu$ lies in the range
\be
 \label{MP:instcondition}
 \mu^2{\bigr |}_{BF} \le \mu^2 < \mu^2{\bigr |}_{NH\,BF}^{(MP)} - \frac{4 \kappa (\kappa+N)}{r_+^2}.
\ee
Another interesting 5d black hole solution is the supersymmetric $AdS_5$ black hole of Ref. \cite{Gutowski:2004ez}. In this case, we find that the near-horizon BF bound coincides with the $AdS_5$ bound so there is no reason to expect a scalar condensation instability. We shall not consider this solution further.

\subsection{Numerical results: linear}

\subsubsection{Numerical methods}

We will solve Eq.~\eqref{MP:radialEq} numerically in search of the
onset of the scalar condensation.  This is a boundary value problem.
The appropriate boundary conditions can be found through a standard
Fr\"obenius analysis. At the horizon, regularity of the solution
requires that the scalar field must be finite, while at the
asymptotic region the scalar field decays as
\begin{eqnarray} \label{MP:BCoo}
 && \hspace{-0.8cm}\Phi(r) \simeq \Phi^{(+)}\left(\frac{r_+}{r}\right)^{\Delta_+}\left[1-
 \frac{\ell^2 \Delta_+ (\Delta_+ -2N)}{4 r_+^2 (\Delta_+ -N)}\left(\frac{r_+}{r}\right)^2\right]
 + \Phi^{(-)}\left(\frac{r_+}{r}\right)^{\Delta_-}\left[1-
 \frac{\ell^2 \Delta_- (\Delta_- -2N)}{4 r_+^2 (\Delta_-
 -N)}\left(\frac{r_+}{r}\right)^2\right].\nonumber\\
 &&
\end{eqnarray}
 Stability of the AdS background requires that the scalar field
must obey the Breitenl\"ohner-Freedman bound \eqref{BFbound}. Both
fall-offs in \eqref{MP:BCoo} can be normalizable so we can impose
one of the two boundary conditions: $\Phi^{(+)}=0$ or
$\Phi^{(-)}=0$.

At this point we make the important observation that the radial
equation \eqref{MP:radialEq} can be written as a generalized
eigenvalue problem in $\Omega_H^2$. That is, we can cast
\eqref{MP:radialEq} in the form \be \label{MP:eigenvalue}
 L(r) \,\Phi(r) = \Omega_H^2\ell^2 \, \Lambda(r)\, \Phi(r),
\ee
 where $L(r)$ and $\Lambda(r)$ are both second order differential
operators that do not depend on $\Omega_H$. For fixed $r_+$ and
$\Delta$, our strategy for finding a zero-mode for the scalar
condensation will be to determine the eigenvalue $\Omega_H(r_+)^2$
for which there exists a solution of \eqref{MP:eigenvalue} that is
regular at $r=r_+$ and satisfies the appropriate boundary condition
(either $\Phi^{(+)}=0$ or $\Phi^{(-)}=0$) at $r=\infty$. This
strategy is motivated by the availability of numerical techniques
for solving eigenvalue equations of the form \eqref{MP:eigenvalue}.
More concretely, we solve this equation using a spectral numerical
method. This method already proved to be suitable to study
perturbations in black hole backgrounds
\cite{Monteiro:2009ke,Dias:2009iu,Dias:2010eu}. The application of
the method is simpler for Dirichlet and/or Neumann boundary
conditions. Therefore we introduce the following perturbation
functions,
 \be \label{MP:spectralFunction}
 q^{(\pm)}(r) = \Phi(r) \left( 1-\frac{r_+^2}{r^2}\right)\left[1+
 \frac{\ell^2 \Delta_\pm (\Delta_\pm -2N)}{4 r_+^2 (\Delta_\pm -N)}\left(\frac{r_+}{r}\right)^2\right]
 \left(\frac{r_+}{r}\right)^{\Delta_\pm},
\ee and we impose the Dirichlet boundary condition at the horizon
and the Neumann boundary condition at infinity,
 \be \label{MP:spectralFunctionBC}
 q^{(\pm)}{\bigl |}_{r=r+} =0\,, \qquad  \partial_r q^{(\pm)}{\bigl
 |}_{r\to\infty} \to0\,, \ee
 that guarantee that the boundary conditions discussed above for
 $\Phi(r)$ are satisfied.
In these equations the choice of $q^{(+)}$ applies if we want to
study the case where $\Phi^{(-)}=0$ in \eqref{MP:BCoo}, while
choosing the function $q^{(-)}$ allows us to address the case
$\Phi^{(+)}=0$. For the numerical implementation, it is convenient
to use the variable
 \be \label{yCoord} y= 1-\frac{r_+^2}{r^2}, \ee
  instead of the radial coordinate $r$, since $y$ is dimensionless and bounded, $0
\leq y \leq 1$. We will also use a scaling symmetry of the radial
equation to normalize all our quantities in units of the
cosmological length. Then equation \eqref{MP:radialEq} or
\eqref{MP:eigenvalue} depends only on four
dimensionless parameters, namely, $\kappa$, $r_+/\ell$,
$\Omega_H\ell$ and $\Delta_\pm$ (or, equivalently, $\mu\ell$). The
strategy is now to fix $\kappa$ and run the spectral numerical code
for several values of $r_+/\ell$ and $\Delta_\pm$ and find the
eigenvalues $\Omega_H^2\ell^2$ of \eqref{MP:eigenvalue}. These will be
presented in Section~\ref{MPresults}.

As a double check, we will also use a shooting method to reproduce
some of our results. A further motivation to use this second
numerical method is that it allows us to find the important tail on the
lower $r_+$ region of Figure \ref{Fig:MP3dk0} with better accuracy.

In the shooting method, the numerical strategy is the following.
Again we want to fix $\kappa$, $r_+/\ell$ and $\Delta_\pm$ and find
$\Omega_H\ell$ at the onset of the instability. Equation
\eqref{MP:radialEq} has two critical points: at the horizon and at
infinity. We first focus our attention at the horizon. Using a
Taylor series expansion, we construct the solution in the
near-horizon region up to the eighth order in the radial distance to
the horizon. Fixing $\kappa$, this solution depends on
$r_+/\ell,\Delta_\pm,\Omega_H\ell$ but also on an arbitrary
amplitude $A_H$.  We then integrate
numerically the radial second order ODE, using a standard fourth
order Runge-Kutta method, up to a large radial distance. We repeat
the procedure, this time at the asymptotic critical point where we
start by obtaining the asymptotic solution up to eighth order. Again
this solution is a function of $r_+/\ell,\Delta_\pm,\Omega_H\ell$
and of an arbitrary amplitude $A_\infty$ that we set equal to $1$. We integrate this solution down to very small
values of the radial distance. In the overlapping region of the two
solutions we then do their matching. The requirement that both the
scalar field and its radial derivative must be continuous fixes the
values of the two unknowns, namely the desired $\Omega_H$ and the
amplitude $A_H$. The whole process is now repeated for a grid of
values $\{r_+/\ell\,,\Delta_\pm\}$ at fixed $\kappa$ to generate the
results in Section~\ref{MPresults}.

A subtlety intrinsic to the system at hand and not so standard in
shooting applications is that the onset of our instability occurs
close to the extreme solution. The Taylor expansion of the near-horizon solution is in terms of powers of
$y/\left[\Omega_H^2-(\Omega_H^{\rm ext})^2\right]$, where $y$ is the
compact radial coordinate defined in \eqref{yCoord}. Therefore,
since the instability we search for sets in very close to
extremality, typically the near-horizon expansion breaks down. To avoid
this we work with a new compact radial coordinate defined as
 \be \label{yCoordNew} \tilde{y}= \left[\Omega_H^2-(\Omega_H^{\rm ext})^2\right]^{-1} \left(1-\frac{r_+^2}{r^2}\right). \ee

\subsubsection{\label{MPresults}Results}

We expect an instability if we satisfy \eqref{MP:instcondition}.
The simplest equal spin MP system that captures all the features of
the scalar condensation instability is the $d=5$ case with
$\kappa=0$. Therefore we will present most of our results for this
case, and in the end we will discuss the $d\geq 7$ and/or
$\kappa\neq 0$ cases.

\begin{figure}[t]
\centerline{\includegraphics[width=.45\textwidth]{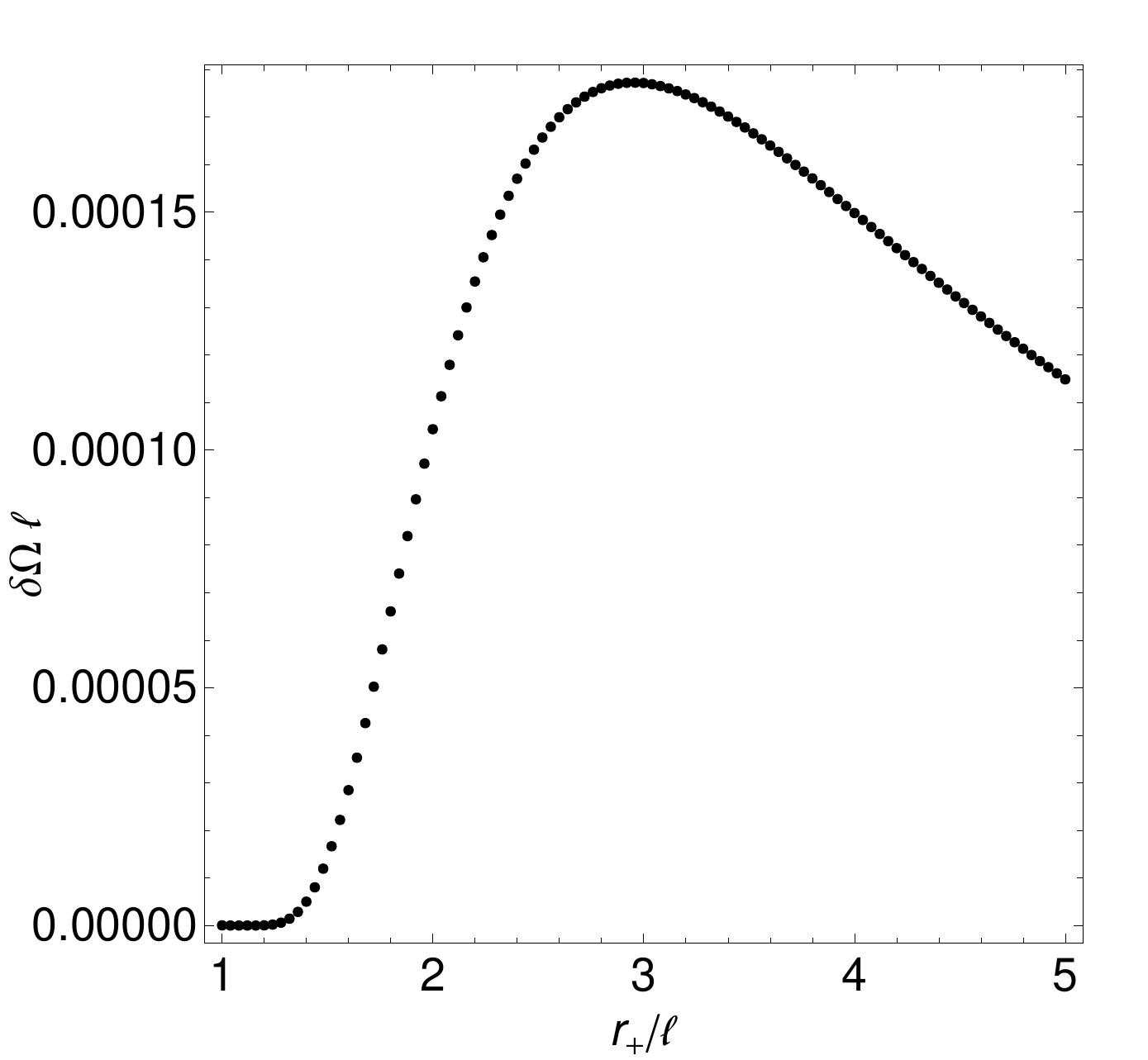}}
\caption{Detail of the black curve in Figure \ref{Fig:linearMP}
where we fix the scalar mass to be at the BF bound
  \ie when $\Delta_+=\Delta_-$. The plot shows the dimensionless angular velocity wrt the extreme value,
$\delta\Omega \ell=(\Omega_H^{\rm ext}-\Omega_H) \ell$,  as a
function of the horizon radius, $r_+/\ell$.} \label{Fig:MP3dk0}
\end{figure}

In $d=5$, and $\kappa=0$ one has $\mu^2|_{BF}\,\ell^2=-4$,
$\mu^2|_{NH\,BF}\,\ell^2=-(3+\ell^2/r_+^2)$ and $\mu^2|_{unit}\,\ell^2=-3$,
 with $\mu^2|_{NH\,BF}< \mu^2|_{unit}$.  Therefore both normalizable
modes with fall-off $r^{-\Delta_\pm}$ should be unstable for masses
obeying the condition \eqref{MP:instcondition} (for $d \ge 7$ this is not true for large enough $r_+$). This is precisely
what we find numerically, following the strategy outlined in the
previous section, as shown in Figure \ref{Fig:linearMP}. In this
figure, we plot the dimensionless angular velocity $\Omega_H \ell$
(measured wrt the extreme angular velocity $\Omega_H^{\rm
ext}\ell$) at which the HHT black hole has the zero-mode for the
scalar condensation instability, as a function of the dimensionless
horizon radius $r_+/\ell$ and of the conformal dimension(s)
$\Delta=\Delta_{\pm}$ defined in \eqref{AdSdecay}. The red curves describe
the modes with faster fall-off $r^{-\Delta_+}$, the blue curves
describe the modes with fall-off $r^{-\Delta_-}$, and the single
black curve is precisely at the BF bound where $\Delta_+=\Delta_-=2$.

The expectation is that for fixed $\Delta$, all black holes that are in the parameter space below this
surface are unstable and decay into hairy black holes with a scalar condensate
surrounding the horizon. That is, the HHT black hole is stable for sufficiently high temperatures, but, for given $r_+$ and $\Delta$, there is a critical temperature where it becomes marginally unstable
against scalar condensation, and for lower values of the temperature the black hole is unstable. Note that the surface extends to arbitrarily large $r_+$.

For clarity, we also plot the curve $(\Omega_H^{\rm
ext}-\Omega_H)\ell$ as a function of $r_+/\ell$ for a fixed value of
the scalar mass in Figure \ref{Fig:MP3dk0}; for concreteness we
picked the black curve in Figure \ref{Fig:linearMP} where
$\Delta=2$. Note that the ``tail" on this curve extends
all the way down to an extremal solution with $r_+/\ell=1$
(something that is not apparent from Figure \ref{Fig:linearMP}).
But, for $\Delta=2$, this is precisely the threshold of instability
predicted by the near-horizon BF bound. In more detail, for
$\Delta=2$ the near-horizon BF bound is violated if, and only if,
$r_+/\ell>1$. Hence we have numerical evidence that this bound is
sharp: an instability appears as soon as it is violated.\footnote{
Just as for the hyperbolic black hole, the extreme solution does {\it not} admit a regular time-independent solution associated to the threshold of instability. If one takes the limit of the time-independent solution associated to non-extreme solutions then the value of the scalar field at the horizon diverges at extremality.}


The instability against the scalar condensation is not the only
instability present. Indeed, it competes with the well-known
superradiant instability that afflicts rotating AdS black holes whenever
$\Omega_H\ell>1$ \cite{Kunduri:2006qa}\footnote{The superradiant instability is associated to perturbations that break the symmetry generated by $\partial/\partial \psi$. This instability is not confined to the scalar field, e.g., it is also present for gravitational perturbations.}.
Since there is competition between the two instabilities it is relevant to ask whether the scalar
condensation instability occurs when $\Omega_H \ell \le 1$.
We find a negative answer. Recall that extreme black holes always have $\Omega_H \ell>1$. Figure \ref{Fig:linearMP} shows that condensation occurs only when $\Omega_H$ is very close to $\Omega_H^{ext}$, and we have checked that this implies that $\Omega_H \ell>1$. So the scalar condensation instability studied here always co-exists with the superradiant instability.


\begin{figure}[t]
\centerline{\includegraphics[width=.45\textwidth]{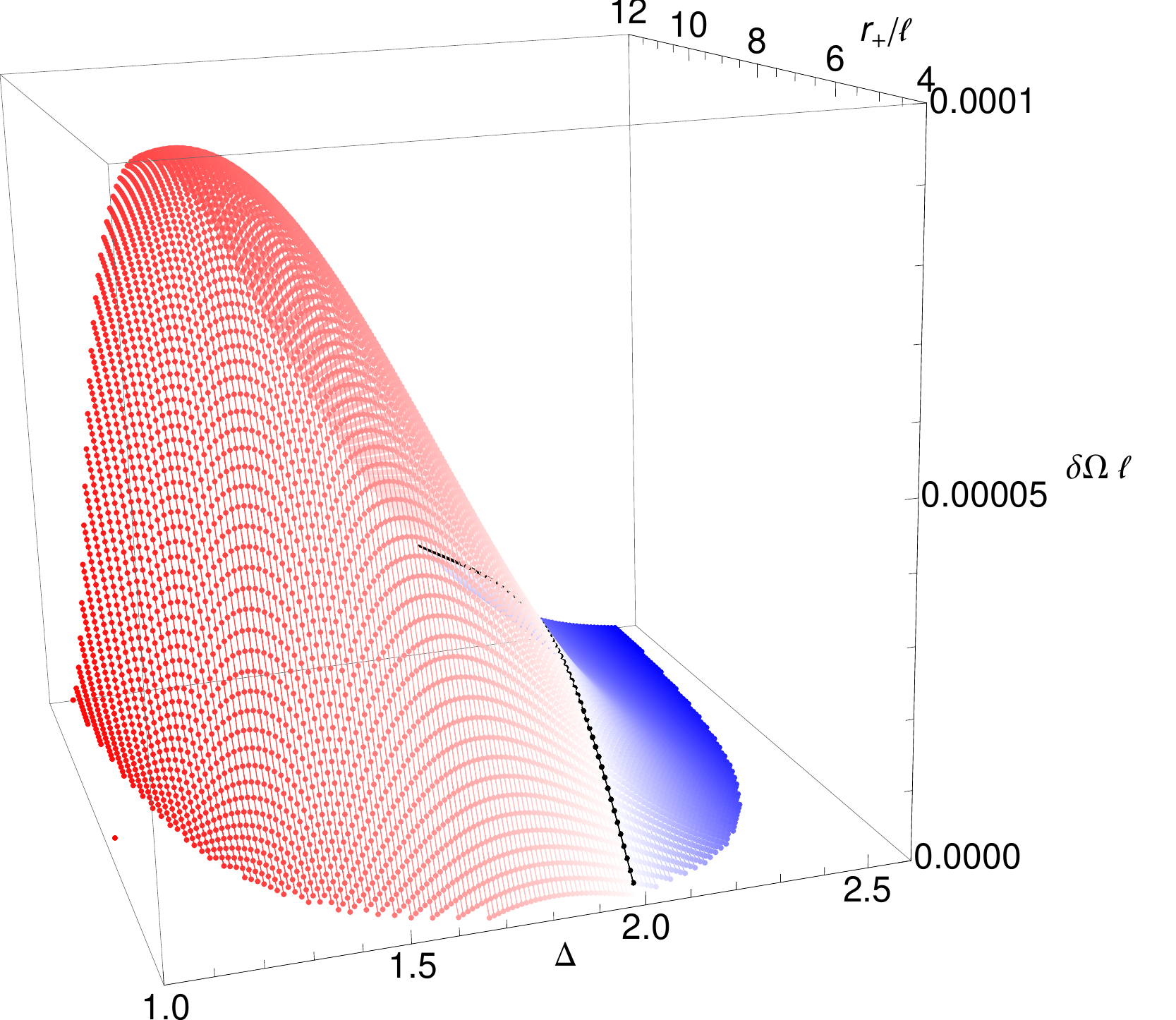}
\hspace{0.7cm}\includegraphics[width=.45\textwidth]{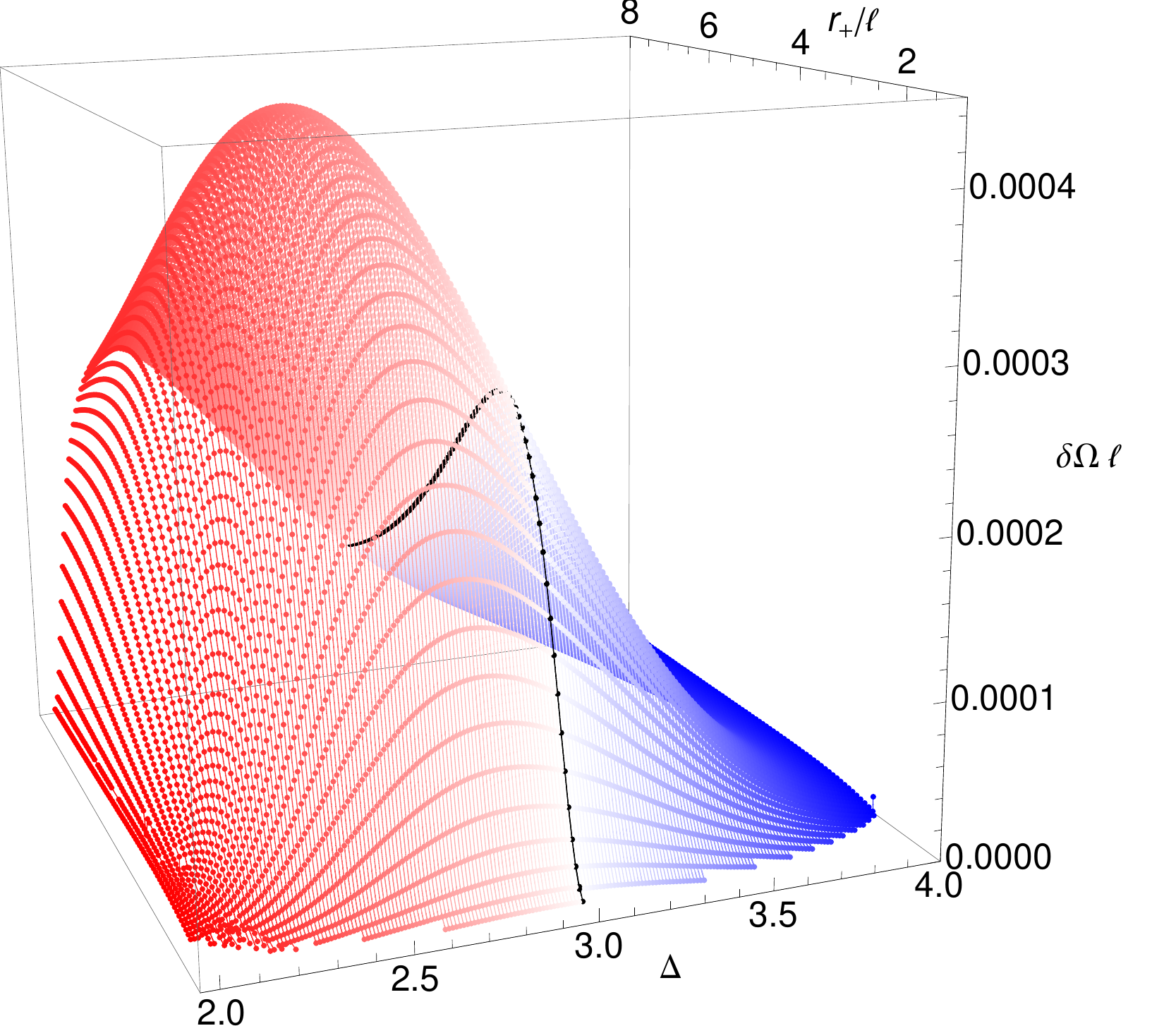}}
\caption{{\bf a)} Similar to Figure \ref{Fig:linearMP} but this time
for modes with angular eigenvalue $\kappa=1$ (in the HHT
background). {\bf b)} Similar to Figure \ref{Fig:linearMP} but this
time for the $d=7$ MP-AdS background (and $\kappa=0$ modes).}
\label{Fig:MP3dk1}
\end{figure}

We can now discuss what happens when we turn on the $CP^N$
dependence defined in terms of the non-negative integer $\kappa$. We
find zero-modes of the instability for $\kappa>0$, at least up to
$\kappa=3$. As the simplest example, in the left plot of Figure
\ref{Fig:MP3dk1}, we show the properties of the instability for
$d=5$ and $\kappa=1$. The stationary axisymmetric mode is indeed
still present but the instability sets in for lower temperatures.
Indeed the maximum difference between the extreme value
$\Omega_H^{\rm ext}$ of the HHT angular velocity and the angular
velocity $\Omega_H$ where the instability switches on is smaller in
the left plot of Figure \ref{Fig:MP3dk1} than in the $\kappa=0$ case
shown in Figure \ref{Fig:linearMP}. For $\kappa\neq
0$ it is still true that the instability persists all the way up to
the NH BF bound. To check this we take again as a reference point
the curve $\Delta_+=\Delta_-$. In this case, according to the
condition \eqref{MP:instcondition}, the zero-mode of the instability
should kick in at extremality at the critical value of $r_+$ given
by (\ref{MP:rMminNHBF}).
This value increases as $\kappa\neq 0$ grows. For $\kappa=1$, it is
$r_+/\ell=3$. Our numerical results indeed confirm that the
zero-mode of the instability is present for values of $r_+$ at and
above this value. This can be inferred by the range of $r_+/\ell$ in
the left panel of Figure \ref{Fig:MP3dk1}, although a precise check
requires analyzing the tail that develops close to extremality like
we did for $\kappa=0$ in Figure \ref{Fig:MP3dk0}.

Finally, let us discuss what happens as the dimension parameter $d=2N+3$
grows. We have explicitly checked the cases $d=7,9$ and the
instability is still present. The numerical results for $d=7$ and
$\kappa=0$ are presented in the right panel of Figure
\ref{Fig:MP3dk1}. As $d$ grows, the instability sets in for black
holes that are closer to extremality, \ie that have lower values of
$(\Omega_H^{\rm ext}-\Omega_H)/\Omega_H^{\rm ext}$. For $\kappa=0$,
the instability persists all the way down to $r_+/\ell=1$ when
$\Delta_+=\Delta_-$, in agreement with the conditions
\eqref{MP:rMminNHBF} and \eqref{MP:instcondition} for the existence of
the instability.

\subsection{\label{sec:NonLinearMP}Numerical results: nonlinear}

In the previous subsection, we found the properties of the onset of the scalar condensation instability in the codimension-1 Myers-Perry-AdS black hole. In this subsection, we include the backreaction in the gravitational field and construct the hairy codimension-1 rotating black holes that are associated with this instability.

\subsubsection{\label{sec:NonLinearMPAnsatz}Hairy black hole ansatz. Equations of motion}

We want to look for hairy black hole solutions that reduce to the codimension-1 Myers-Perry black hole \eqref{background} when the scalar condensate vanishes, and that preserve the isometries of the latter solution. Therefore we take the following ansatz for the gravitational and scalar fields,
 \begin{eqnarray}
 &&  ds^2 = -\frac{f(\tilde{r})}{h(\tilde{r})}dt^2+\frac{g(\tilde{r})dr^2}{4\tilde{r}f(\tilde{r})} + \tilde{r}h(\tilde{r})[d\psi +A_a dx^a - \Omega(\tilde{r})dt]^2 + \tilde{r}\hat{g}_{ab} dx^a dx^b\,, \nonumber\\
 && \Phi=\Phi(r)\,,
 \label{HairyMP:ansatz}
 \end{eqnarray}
where, as before, $\hat{g}_{a b}$ is the Fubini-Study metric on
$CP^{N}$, and $A = A_a dx^a\,$ is the K\"ahler potential of $CP^N$.
When the condensate vanishes, $\Phi(\tilde{r})=0$, and
$g(\tilde{r})=1$, \eqref{HairyMP:ansatz} describes the MP solution
\eqref{background} after performing the radial coordinate
transformation $\tilde{r}\rightarrow r^2$. This ansatz solves the
equations of motion derived from the Einstein-scalar action
\eqref{action5d} when the following equations are satisfied
(taking $N=1$, i.e., $d=5$ henceforth)
\begin{subequations}
\begin{align}
& \{ \hbox{System  of  three coupled ODEs of $2^{\rm nd}$ order for $( f,h,\Phi)$}  \} \,,  \label{HairyMP:eom123}  \\
\nonumber \\
& g=\frac{2 \tilde{r}^3 \ell^2 \left[f \left(h \tilde{r} h'+\tilde{r}^2 h'^2+h^2 \left(2 \tilde{r}^2 \Phi'^2-3\right)\right)
 -h \tilde{r} f' \left(3 h+\tilde{r} h'\right)\right]}{h^2 \tilde{r}^3 \left[2 (h-4) \ell^2+\tilde{r} \left(\mu^2\ell^2
 \Phi^2-12\right)\right]+2 \ell^2 C_{\psi}^2} \,,\label{HairyMP:eom5}  \\
\nonumber \\
& \Omega' =-C_\psi \frac{\sqrt{g(\tilde{r})}}{\tilde{r}^3h(\tilde{r})^2} \,,  \label{HairyMP:eom4}
\end{align}
\label{HairyMP:eom}
\end{subequations}
where the system of three ODEs described in \eqref{HairyMP:eom123}
is cumbersome and thus we leave its  explicit expressions for
equation \eqref{HairyMP:eom3} of Appendix \ref{sec:eomMPnonlinear}.

The construction of the hairy black hole amounts to determining
$f(\tilde{r})$, $h(\tilde{r})$ and $\Phi(\tilde{r})$  that solve the
system \eqref{HairyMP:eom}, \ie \eqref{HairyMP:eom3}, and
\eqref{HairyMP:eom4}. Once these are known, $g(\tilde{r})$ is
straightforwardly obtained from \eqref{HairyMP:eom5}. The constant
$C_\psi$ in \eqref{HairyMP:eom4} is linearly proportional to the
angular momentum of the solution, as we shall confirm later.

The boundary conditions are the following.  The function $f(\tilde{r})$ must vanish at the horizon, and we take this condition as our definition for the location of the black hole horizon. We require the hairy black hole to be asymptotically AdS and thus $f(\tilde{r})$ must approach $1+\tilde{r}/\ell^2$ asymptotically. For the same reason, $h(\tilde{r})$ must go to the unit value at infinity, while $\Omega(\tilde{r})$ vanishes there. At the horizon we require $g,\Omega,\,h,\Phi$ to be regular. The scalar field asymptotic boundary condition is determined by the requirement of normalisability at infinity. Hairy black holes exist for scalar masses above the BF bound and below the NH BF bound. For concreteness, we work with the scalar mass $\mu=\mu|_{BF}$ for which
the numerics considerably simplifies. Summarizing, the boundary conditions for the non-linear problem are:
 \bea
 && f{\bigl |}_{\tilde{r}=r_+^2}=0\,,\qquad  f{\bigl |}_{\tilde{r}\sim\infty}\to\frac{\tilde{r}}{\ell^2}+1+\mathcal{O}(\tilde{r}^{-1})\,; \qquad  h{\bigl |}_{\tilde{r}=r_+^2}=\mathcal{O}(1)\,,\qquad  h{\bigl |}_{\tilde{r}\to\infty}\sim 1+\mathcal{O}(\tilde{r}^{-2})\,; \\
 && \Omega{\bigl |}_{\tilde{r}=r_+^2}=\mathcal{O}(1)\,,\qquad  \Omega{\bigl |}_{\tilde{r}\to\infty}\sim \frac{C_{\psi}\ell^3}{2\tilde{r}^2}+\mathcal{O}(\tilde{r}^{-3})\,;\qquad \Phi{\bigl |}_{\tilde{r}=r_+^2}=\mathcal{O}(1)\,,\qquad  \Phi{\bigl |}_{\tilde{r}\to\infty}\sim\frac{\phi_0}{\tilde{r}}+\mathcal{O}(\tilde{r}^{-3}).\nonumber
 \label{HairyMP:BCs}
 \eea

\subsubsection{\label{sec:NonLinearMPresult}Results}
In this subsection, we present the scalar condensate, the temperature, the angular velocity and the entropy of the rotating hairy black hole as a function of either its energy or angular momentum.

To construct the exact hairy black hole, we cannot use spectral relaxation methods, in contrast to what we did in Section~\ref{subsec:nonlinearSchwAdS}. Relaxation methods crucially hinge on the positivity of the discretisation matrix representing the differential system at hand. When rotation is included, this is no longer possible, and the method at its most basic form does not converge.

However, because \eqref{HairyMP:eom123} is a one-dimensional system of non-linear differential equations, we can resort to a shooting method. Here we regard as the boundaries of our integration domain the horizon and the AdS spatial infinity. This method attempts to solve a boundary value problem, by reducing it to two initial value problems starting at each boundary. To determine the initial data at each boundary, we Taylor expand the equations in the neighbourhood of both the horizon and the AdS spatial infinity, determining the asymptotic solutions in a series expansion at each of the boundaries. We then integrate from both boundaries to an interior point in the integration domain, using a standard fourth order Runge-Kutta method, and demand that the two solutions match.

A couple of comments, regarding the specific solution we are looking
at, are now in order. Unlike previous non-linear hairy black hole
solutions, such as the ones found in \cite{Hartnoll:2008kx} and the ones studied in section~\ref{subsec:nonlinearSchwAdS}, the
rotating non-linear hairy black holes that we are going to study
here only exist for very small temperatures ($T_H \ell \sim
10^{-3}$) - they are \emph{ultra-cold} black holes. Furthermore,
small variations of the temperature often correspond to large
gradients of the entropy, as we shall see later in the phase
diagram. Physically, the almost ``infinite'' throat characteristic
of near-extreme black holes increases the proper length between the
horizon and the cut-off scale where we choose to truncate the AdS
space. This means that, in order to stabilise any numerical
approach, we need a large resolution in both our discretisation
scheme and moduli space of solutions. Finally, because the number of
integration constants of the above system of equations is seven
(recall that $g$ is obtained algebraically once $f,h$ and $\Phi$ are
known, and that $\Omega$ obeys to a first order differential
equation), the total number of parameters to be determined in the
shooting process is seven. This makes the possible phase space of
solutions very large, and thus very difficult to explore. Of course,
these problems would be solved by using relaxation methods, but, as
we have explained above, these are not available when rotation is
included.

To compute the energy of the hairy black hole in AdS we use the
Astekhar-Das formalism \cite{Ashtekar:1999jx}.  The temperature,
entropy, angular momentum and energy of these solutions are given by
 \be
T_H=\frac{|f'(r_+^2)|r_+}{2 \pi \sqrt{h(r_+^2) g(r_+^2)}}\,, \qquad  S=\pi^2\frac{\sqrt{h(r_+^2)}}{2}r_+^3\,,\quad J = \pi \ell^3\frac{C_{\psi}}{4} \qquad \hbox{and}  \quad E=\left(\frac{4\gamma_1-3\gamma_2}{8}\right)\pi\ell^2\,,
 \ee
where $\gamma_1\ell^4$ is the $\mathcal{O}(\tilde{r}^{-2})$
coefficient in the large $\tilde{r}$  expansion of $h(\tilde{r})$
and $\gamma_2\ell^2$ is the $\mathcal{O}(\tilde{r}^{-1})$
coefficient in the large $\tilde{r}$ expansion of $f(\tilde{r})$.

Ideally, we would like to present three-dimensional plots for the variation of the physical parameters, such as the entropy and temperature, as functions of the energy and angular momentum. However, due to the difficulties alluded above, this does not seem feasible with the numerical methods we implemented. As such, we decided to determine the solution phase space at either lines of constant energy or lines of constant angular momentum. At the end of this section, we will explore the phase diagram along lines of constant horizon size $r_+$.

For the HHT black hole, the energy along lines of constant
angular momentum is bounded below by extreme solutions.
Alternatively, the angular momentum along lines of constant energy
is bounded above by extremality. We expect to see this zero
temperature bound when analysing the phase diagram. However, we do
not expect to find regular extreme \emph{hairy} black hole solutions
in the zero temperature limit. The argument is the same as we used for the hairy hyperbolic black hole (following Ref. \cite{FernandezGracia:2009em}). An extreme black hole has a near-horizon limit. The scalar field must be constant in the near-horizon geometry, but then its equation of motion (for non-zero mass) implies that it vanishes there. Hence the scalar must vanish at the horizon of the full black hole solution. But we expect the condensate to grow, not decrease, as we lower the temperature. Hence the zero temperature solution cannot be a black hole.

\begin{figure}
\centering
\subfigure[Condensate as a function of the energy.]{
\includegraphics[width = 6.5 cm]{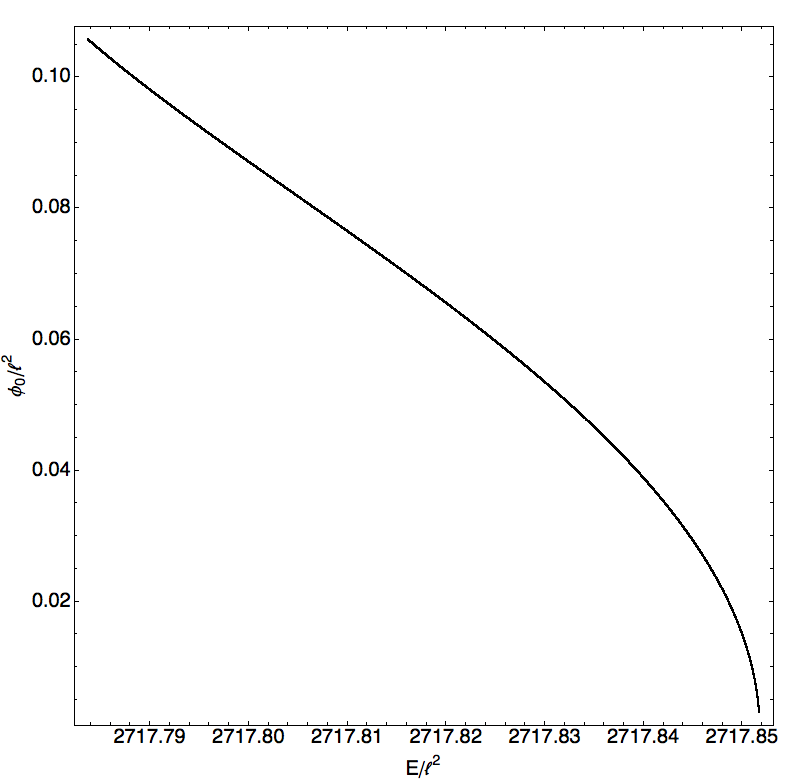}
\label{fig:HairyMP1a}
}
\subfigure[Temperature as a function of the energy.]{
\includegraphics[width = 6.5 cm]{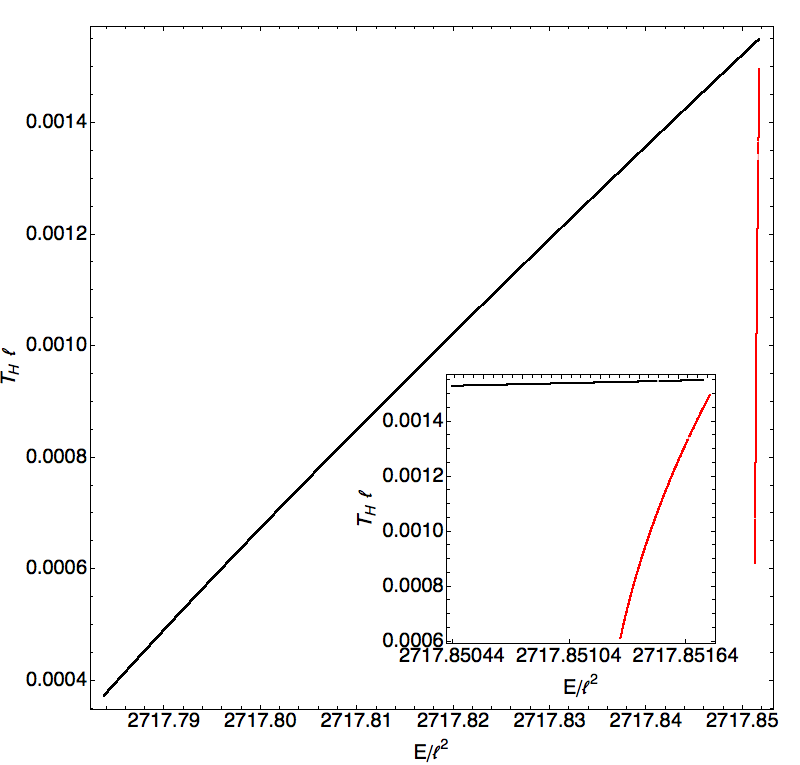}
\label{fig:HairyMP1b}
}
\\
\subfigure[Angular velocity as a function of the energy.]{
\includegraphics[width = 6.5 cm]{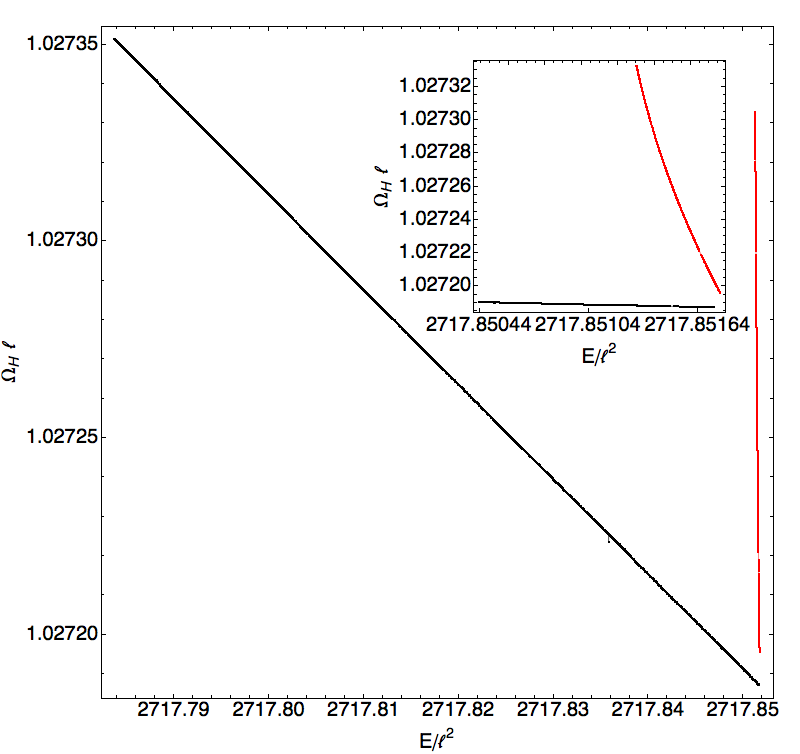}
\label{fig:HairyMP1c}
}
\subfigure[Entropy as a function of the energy.]{
\includegraphics[width = 6.5 cm]{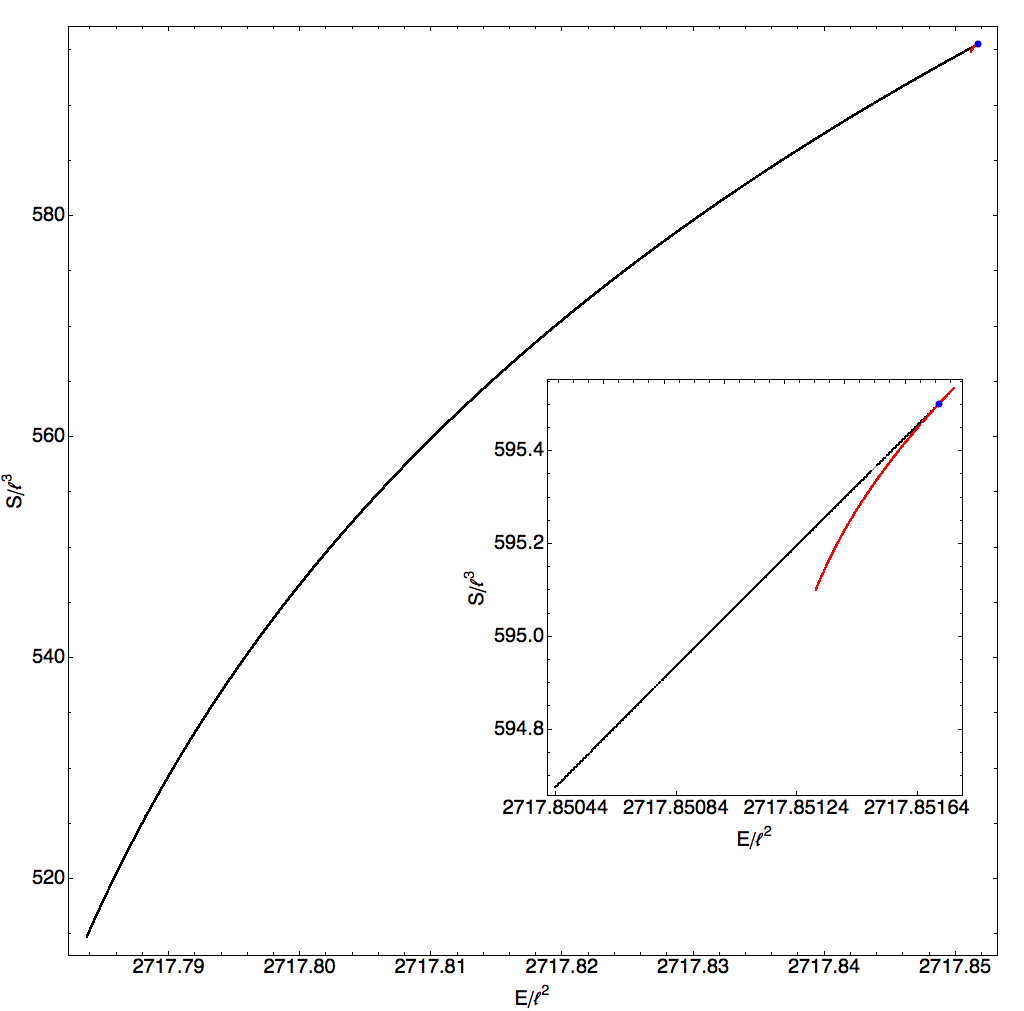}
\label{fig:HairyMP1d} } \caption{\label{figs:HairyMP1}Different
physical quantities for 5d rotating black holes as a function of the energy for $J = 830
\pi \ell^3$. The black curve corresponds to our hairy black hole, the red curve to the HHT solution. The inset figures provide an expanded view of the region near the bifurcation point where the two solutions merge.}
\end{figure}

In Figs.~\ref{figs:HairyMP1}, we plot several physical quantities as
a function of the dimensionless energy $E/\ell^2$ of the 5d rotating
hairy black hole, for $C_{\psi} = 3320$ (i.e. $J=830\pi\ell^3$). In each of
these plots there are two coexisting curves for some range of the
energy. The range of energy considered is very small. This is because the temperature at which the bifurcation to the hairy solution occurs is so small that (from the first law) a tiny change in energy corresponds to a large change in entropy.
The black curves are the numerical solutions corresponding
to the rotating hairy black hole and the red curves correspond to
the HHT black hole with the same angular momentum. There
is excellent agreement between the linear results above and the
non-linear solution when the condensate is sufficiently small, and
in particular the point where the condensate vanishes coincides with
the linear result with a $0.05\%$ precision. To control the
numerics, we have explicitly checked that the first law of
thermodynamics is readily satisfied to a precision of $0.09 \%$.

The
results are very similar to the hairy hyperbolic black hole:
the condensate (or vev of dual CFT operator) becomes larger as the energy decreases, and both the
temperature and entropy decrease with decreasing energy (in
accordance with the first law of thermodynamics). When the hairy black hole has the same mass and angular momentum as a HHT black hole, it is the former that has the larger entropy, which suggests that it should be more stable. However, from
Fig.~\ref{fig:HairyMP1c}, we see that the hairy black hole angular
velocity is always above $\ell^{-1}$, indicating that it probably will suffer a superradiant instability  \cite{Kunduri:2006qa}.
Our code is numerically unstable for temperatures below
$0.0004\,\ell^{-1}$, and that is the reason why we see an artificial
lower bound on the energy before we reach the zero temperature
solution. In analogy with the hairy hyperbolic black hole of the
previous section, we strongly believe that the zero temperature and
zero entropy solutions will coincide.

\begin{figure}
\centering
\subfigure[Condensate as a function of the angular momentum.]{
\includegraphics[width = 6.5 cm]{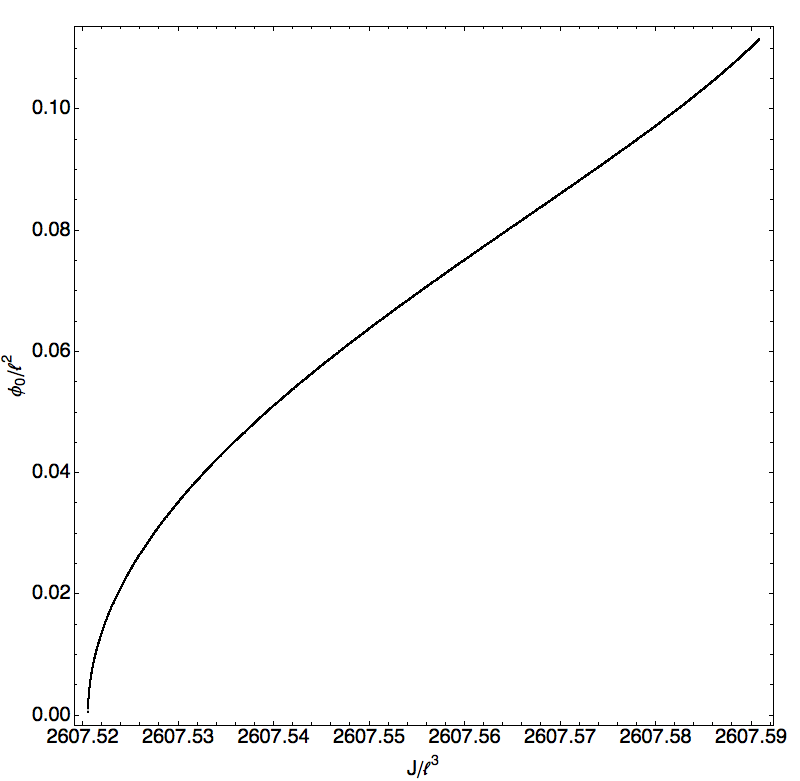}
\label{fig:HairyMP2a}
}
\subfigure[Temperature as a function of the angular momentum.]{
\includegraphics[width = 6.5 cm]{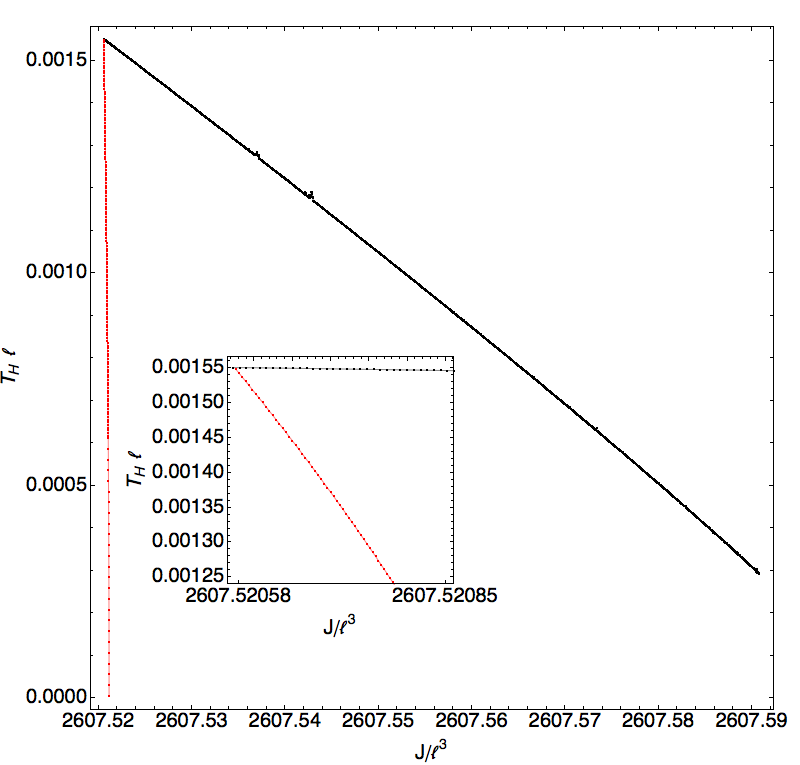}
\label{fig:HairyMP2b}
}
\\
\subfigure[Angular velocity as a function of the angular momentum.]{
\includegraphics[width = 6.5 cm]{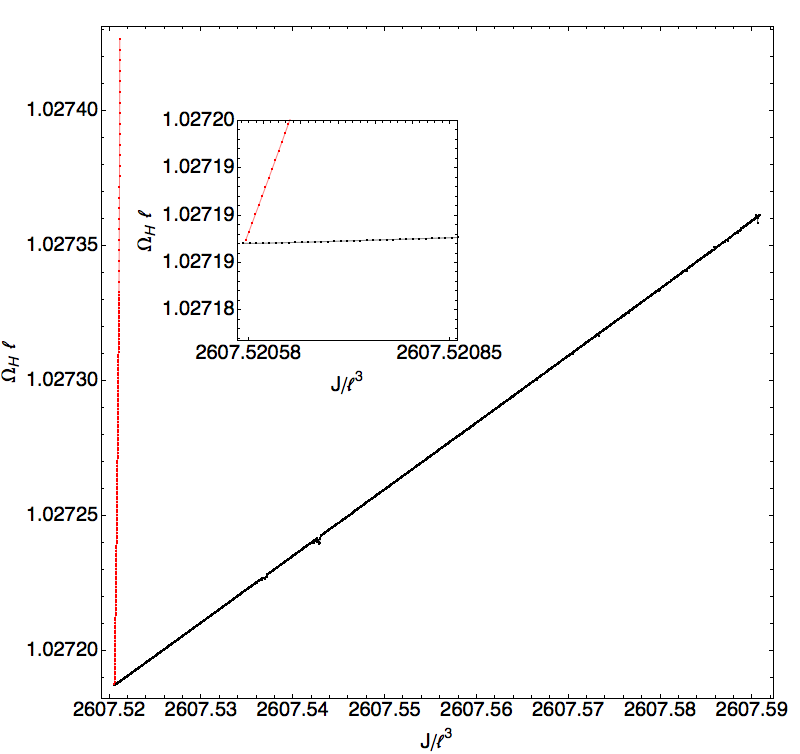}
\label{fig:HairyMP2c}
}
\subfigure[Entropy as a function of the angular momentum.]{
\includegraphics[width = 6.5 cm]{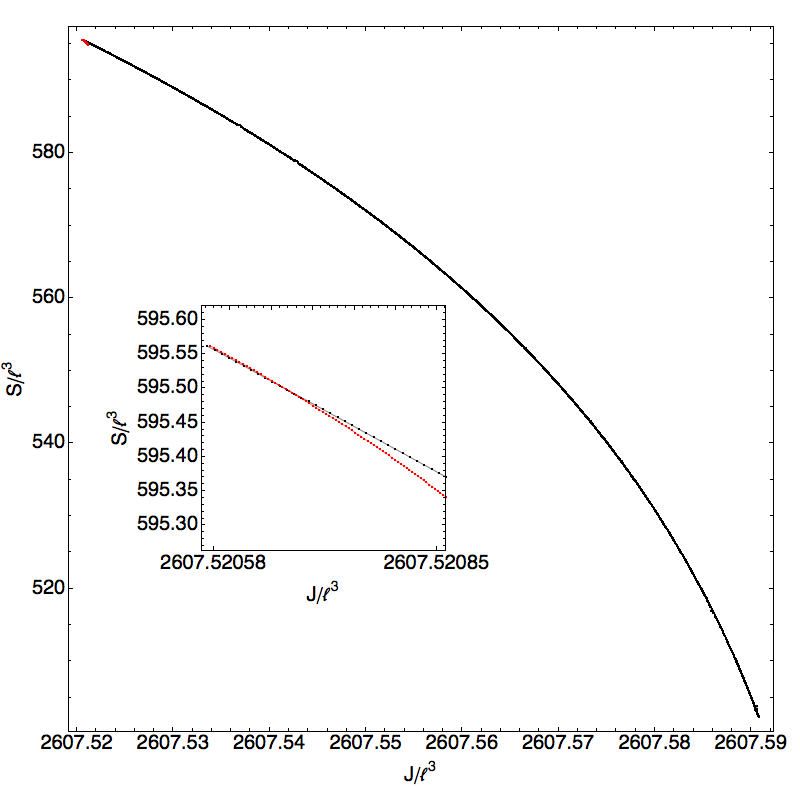}
\label{fig:HairyMP2d}
}
\caption{\label{figs:HairyMP2}Different physical quantities or 5d rotating black holes as a function of the angular momentum for $E = 2717.85044\,\ell^2$. The black curve corresponds to our hairy black hole, the red curve to the HHT solution. The inset figures provide an expanded view of the region near the bifurcation point where the two solutions merge.}
\end{figure}

In Figs. \ref{figs:HairyMP2} we plot several physical quantities as
a function of the dimensionless angular momentum $J/\ell^3$ of the
rotating hairy black hole, for $E = 2717.85044\,\ell^2$. (As mentioned above, the temperature is so small that a tiny change in the energy results in a large change in the solution, which is why we have to specify $E$ so precisely.)
Again, the smallness of the temperature implies that a small change in angular momentum results in a large change in the entropy, so the range of variation of $J$ is small.
Along the lines of constant energy, as the angular momentum increases, the condensate and angular velocity increase but the temperature and entropy decrease. The code is numerically
unstable whenever we reach a temperature of the order of
$0.0004\,\ell^{-1}$, and this is why we never reach zero temperature.
Again we observe that the angular velocity never decreases below
$\ell^{-1}$, and as such, we expect these black holes to be
superradiant-unstable.

We have also probed other directions of the phase space, namely
along lines of constant horizon size $r_+/\ell$.  This direction explores a larger region of the
phase diagram, \ie both the energy and angular
momentum vary more than in the previous directions, before reaching
either the HHT black hole or the numerical cut-off temperature
$0.0004\,\ell^{-1}$. In this way, we hoped to find black holes with
angular velocity below $\ell^{-1}$. However, that was never the
case. In all directions that we probed, the
angular velocity was always bigger than $\ell^{-1}$.

\section{\label{sec:CondensationKerrAdS}Rotating black holes: $d=4$ Kerr-AdS}

At this point, it will not surprise the reader to learn that a massive scalar field can also
condensate in the vicinity of a 4d Kerr-AdS black hole that is sufficiently close to extremality. In this section,
we will determine the threshold mode associated to the condensation in
this background.

\subsection{Introduction}

 The Kerr-AdS geometry is described by the line
element \cite{Carter:1968ks} \be ds^2
=-\frac{\Delta_r}{\Sigma^2}\left ( dt-\frac{a}{\Xi}\sin^2\theta
\,d\phi\right )^2 +\frac{\Sigma^2}{\Delta_r}\,dr^2
+\frac{\Sigma^2}{\Delta_{\theta}}\,d\theta^2
   +\frac{\Delta_\theta}{\Sigma^2} \sin^2\theta\left (
a\,dt-\frac{r^2+a^2}{\Xi} \,d\phi\right )^2 \,,
 \label{Kerr:metric}
\ee
 where
 \be \Delta_r=\left (r^2+a^2\right )\left (1+\frac{r^2}{\ell^2}
\right )-2Mr\,, \qquad  \Xi=1-\frac{a^2}{\ell^2}\,, \qquad
\Delta_{\theta}= 1-\frac{a^2}{\ell^2}\cos^2\theta\,, \qquad
\Sigma^2=r^2+a^2 \cos^2\theta  \,.
 \label{Kerr:metricAUX}
\ee This solution satisfies $R_{\mu\nu} =-3\ell^{-2}g_{\mu\nu}$, and
asymptotically approaches AdS space with radius of curvature $\ell$.
The ADM mass and angular momentum of the black hole are $M/\Xi^2$
and $J=M a/\Xi^2$, respectively \cite{Gibbons:2004ai}. The event horizon is located at $r=r_+$ (the
largest real root of $\Delta_r$). The
angular velocity measured with respect to a non-rotating frame at infinity is
 \be \label{kerr:OmegaH}
\Omega_H=\frac{a}{r_+^2+a^2}\left ( 1+\frac{r_+^2}{\ell^2} \right
)\,. \ee
The rotation parameter is bounded by
 \be
a<\ell \,.
  \label{kerr:constrainta}
\ee
Solutions saturating this bound do not describe black holes. In the limit $a \rightarrow \ell$ at fixed $r_+$, the mass and angular momentum of the black hole diverge. The circumference of the black hole as measured at the equator becomes infinitely large in this limit.

The temperature is given by
 \be \label{kerr:T}
 T=\frac{r_+}{2 \pi
}\left(1+\frac{r_+^2}{\ell ^2}\right)\frac{1}{r_+^2+a^2}-\frac{1}{4
\pi r_+}\left(1-\frac{r_+^2}{\ell ^2}\right)\,.
 \ee
 The extreme solution is the configuration with $a=a_{\rm ext}$, where
 \be
\label{kerr:extreme} a_{\rm ext}= r_+ \sqrt{\frac{3 r_+^2+\ell
^2}{\ell^2-r_+^2}} \quad \Rightarrow \quad \Omega_H^{\rm
ext}=\frac{\sqrt{\ell^4+2 r_+^2 \ell^2-3 r_+^4}}{2 r_+ \ell^2}\,,
\qquad \hbox{and} \quad \frac{r_+}{\ell}<\frac{1}{\sqrt{3}}\,.
 \ee
 Note that only ``small" black holes with $r_+/\ell<3^{-1/2}$ can reach zero temperature in virtue of
 \eqref{kerr:constrainta}.

The strategy is now similar to the one carried out in the HHT case. We
study the Klein-Gordon equation for a massive scalar field in
\eqref{Kerr:metric}. We seek a stationary axisymmetric solution, which we expect to arise at the threshold of the instability. The explicit details of this study are given in
Appendix \ref{sec:KerrParameterSpace}. In the next subsection we
present the results.

\subsection{Numerical results: linear}

\begin{figure}[t]
\centerline{\includegraphics[width=.45\textwidth]{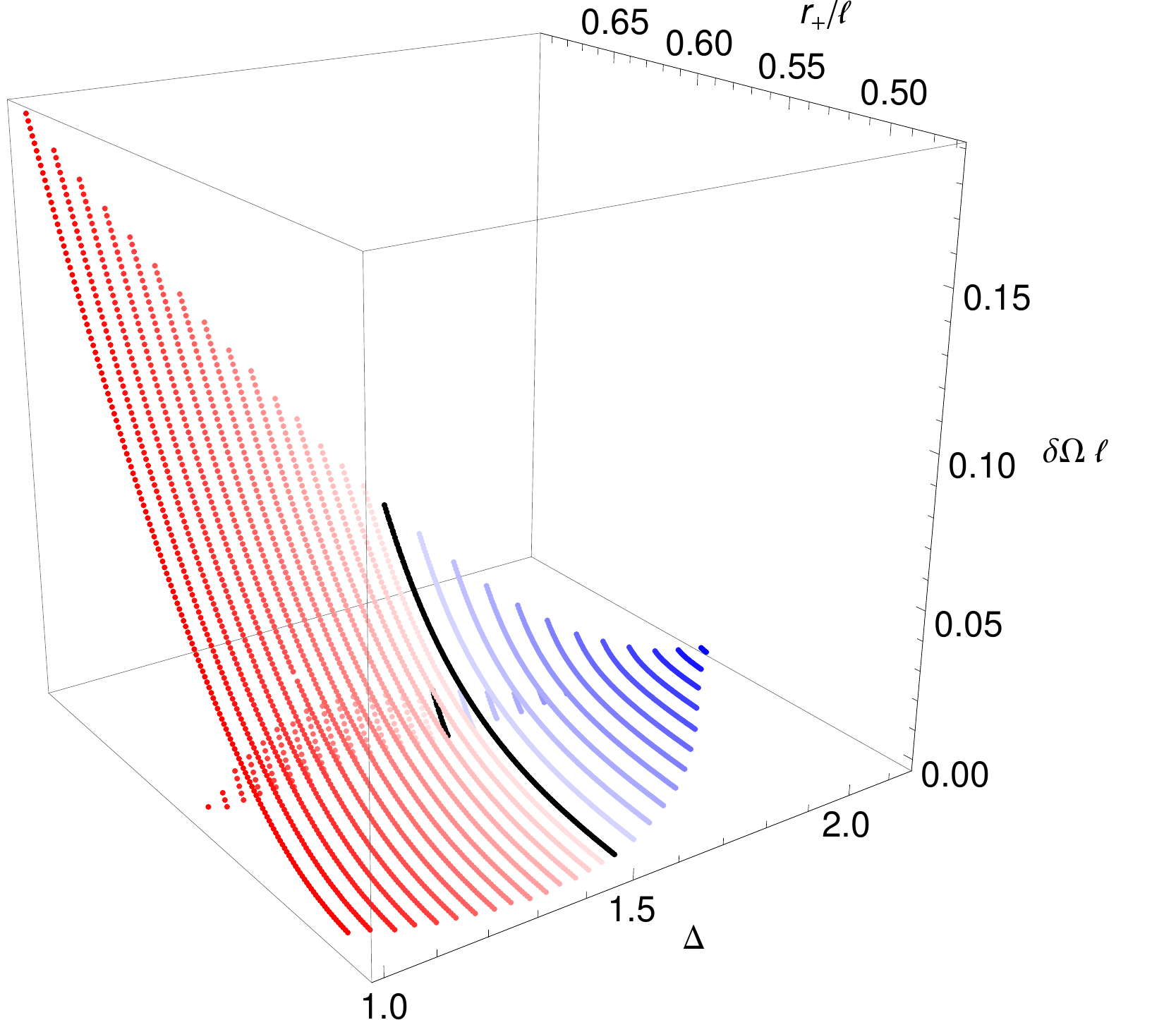}
\hspace{0cm}\includegraphics[width=.4\textwidth]{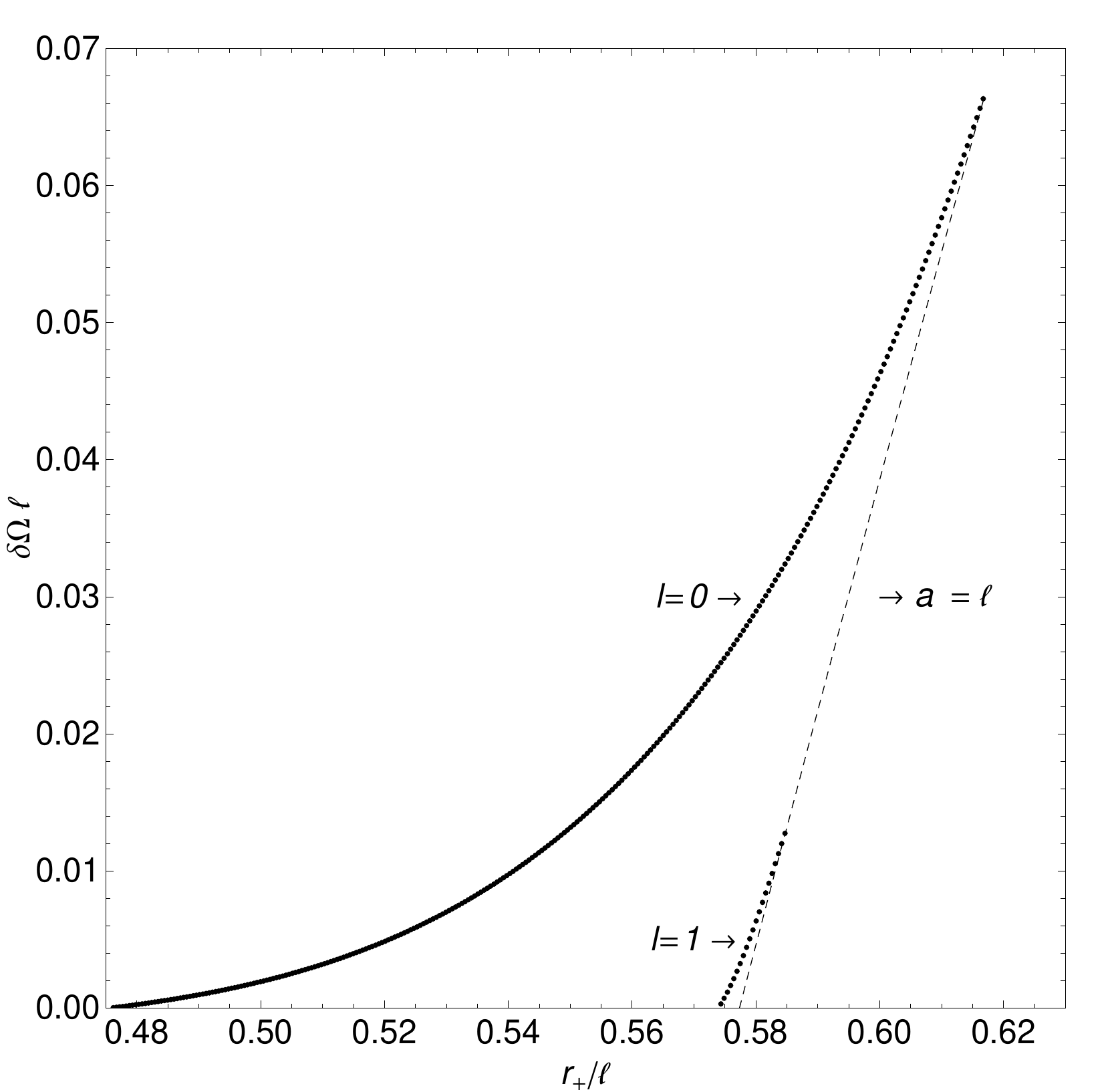}}
\caption{{\bf a)} Threshold mode of the scalar condensation in the
Kerr-AdS black hole. This plot shows the location of the threshold
mode as a function of $\delta\Omega \ell=(\Omega_H-\Omega_H^{\rm
ext}) \ell$, the parameter $r_+/\ell$, and the conformal dimension
$\Delta$.
 {\bf b)} Detail of the black curves in Figure a) with $\Delta=3/2$.
The upper (lower) curve describes the $l=0$ ($l=1$) harmonic. The
dashed line corresponds to $a=\ell$, i.e, infinitely large black
holes. An instability with given $l$ is present for points lying
below the surface/curve with that value of $l$.} \label{Fig:Kerr3d}
\end{figure}

\begin{figure}[t]
\centerline{\includegraphics[width=.45\textwidth]{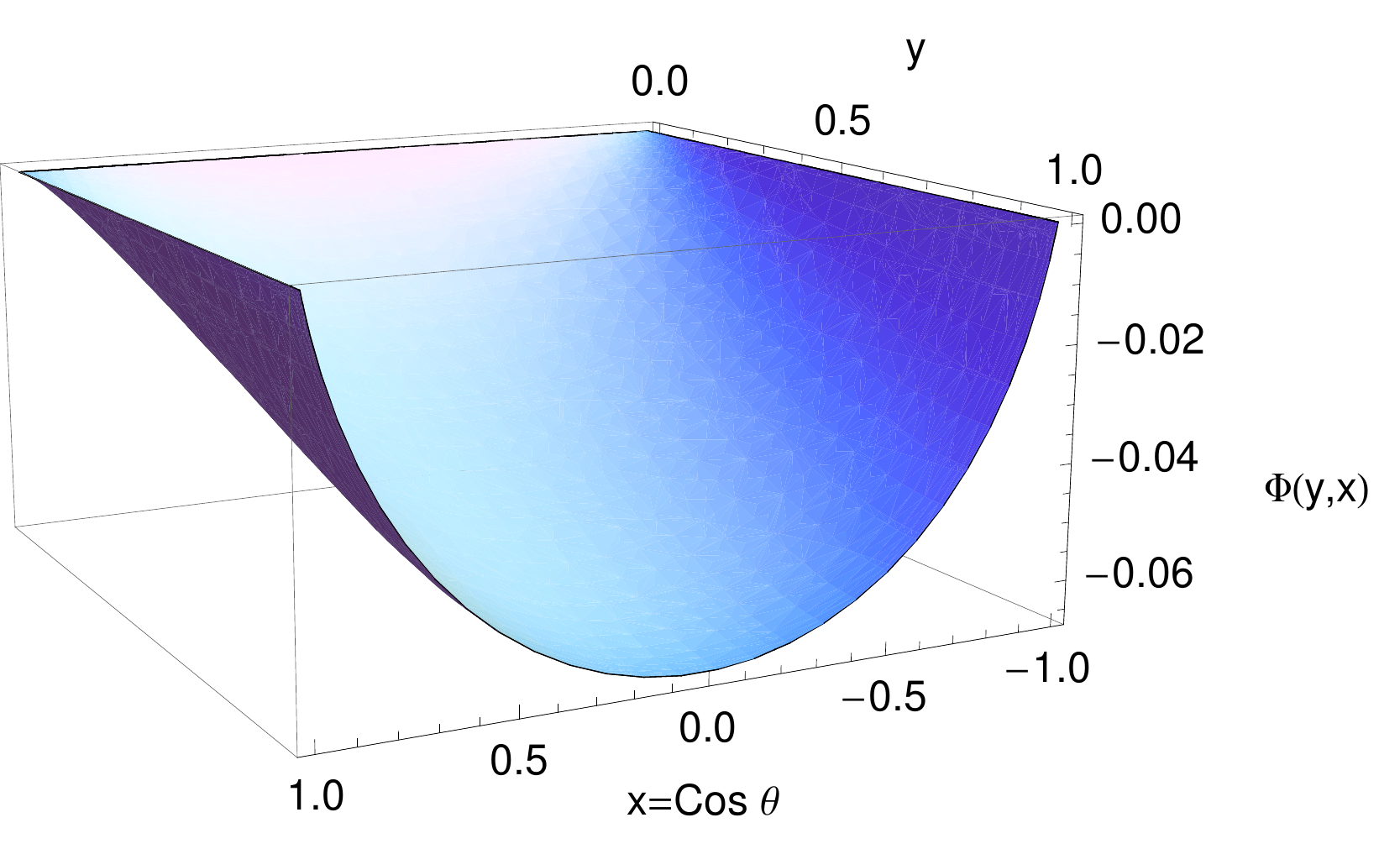}
\hspace{0.7cm}\includegraphics[width=.45\textwidth]{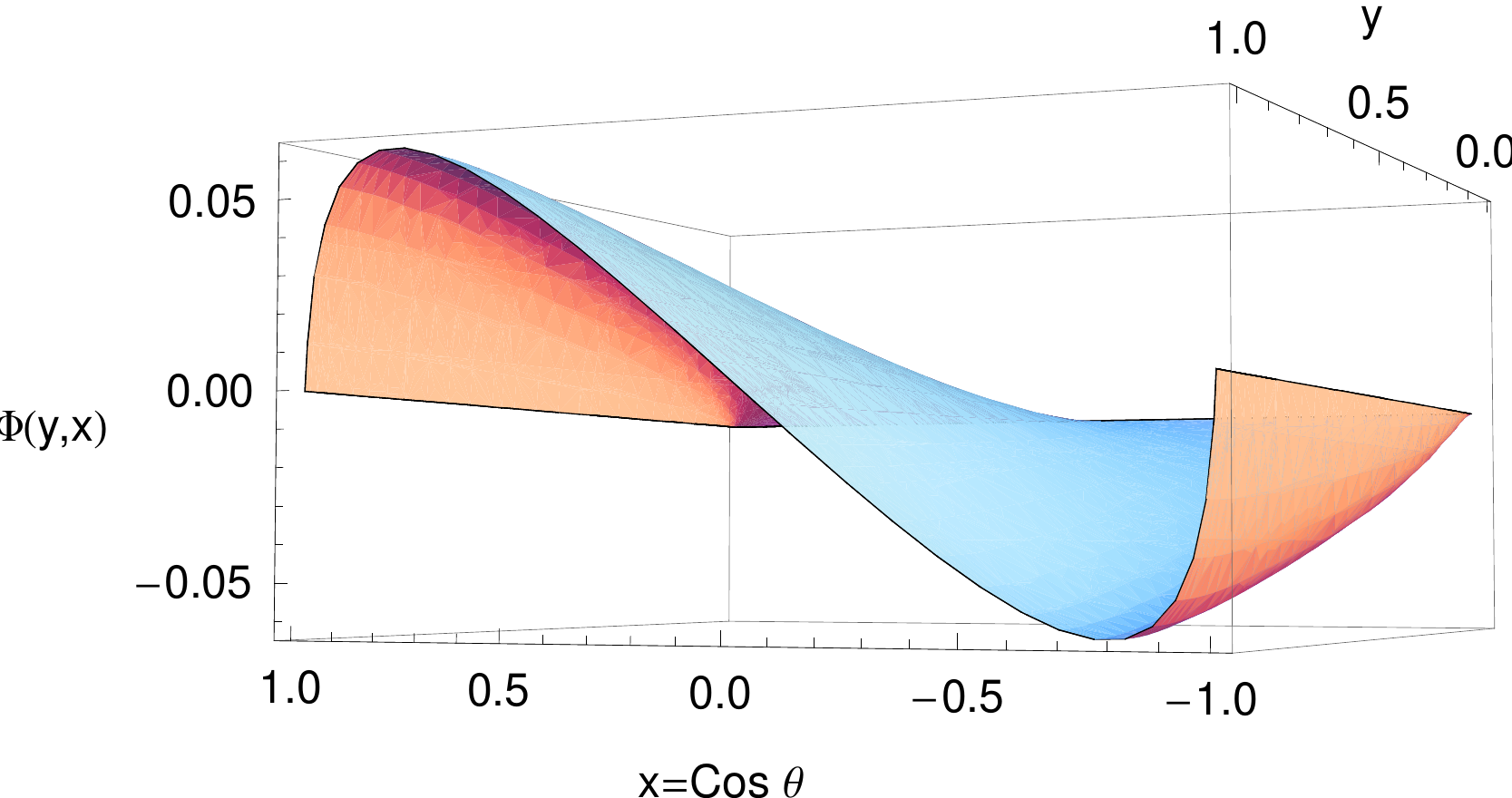}}
\caption{The harmonic structure of the zero-modes $l=0$ (left) and
$l=1$ (left) discussed in Figure \ref{Fig:Kerr3d}}
\label{Fig:kerrHarmonic}
\end{figure}

The properties of the zero-mode of the scalar condensation
instability are summarized in Figures \ref{Fig:Kerr3d} and
\ref{Fig:kerrHarmonic}. In the left panel of Figure
\ref{Fig:Kerr3d}, we plot again the dimensionless angular velocity
$\Omega_H \ell$ where a time-independent threshold mode appears (measured wrt
$\Omega_H^{\rm ext}\ell$), as a function of the dimensionless
horizon radius $r_+/\ell$ and of the conformal dimension
$\Delta$ defined in \eqref{AdSdecay}. The red curves
(which combine to form a surface) describe the modes with faster fall-off
$r^{-\Delta_+}$, the blue curves (surface) describe the modes with
fall-off $r^{-\Delta_-}$, and the black curves are precisely at the
BF bound, which corresponds to $\Delta=3/2$. Note that, for low temperatures, $\delta\Omega=(\Omega_H-\Omega_H^{\rm ext})$ and $r_+$ uniquely specify the solution (see Appendix~\ref{sec:KerrParameterSpace}).

In this figure, there are two distinct surfaces present. These correspond to threshold modes with different angular dependence, analagous to the modes with different $\kappa$ that we discussed above. The scalar wave equation is separable in Kerr-AdS and one can label the solution by an integer $l \ge 0$  equal to the number of zeros of the angular part of the solution. The upper surface in the figure is for $l=0$, the lower surface for $l=1$. We find no numerical evidence for zero-modes with higher harmonics, $l\geq 2$. Scalar condensation with given $l$ occurs for points below the appropriate surface.

In the right panel of Figure \ref{Fig:Kerr3d}, we plot in isolation the two black curves (i.e., with $\Delta=3/2$) of the left figure to see the details more clearly. This figure reveals that the instability occurs only above a minimum value of $r_+$ corresponding to an extreme black hole. The curves end on a curve with $a=\ell$, corresponding to infinitely large black holes.

In Figure \ref{Fig:kerrHarmonic}, we plot the amplitude of the
$l=0$ scalar mode (left panel) and of the $l=1$ scalar mode (right
panel) as a function of angular and radial coordinates,
$x=\cos\theta$ and $y=1-r_+/r$. We clearly see that the $l=0$
harmonic has indeed no zero in the interval $-1<x<1$ while the $l=1$
harmonic has precisely a single zero in the same interval.

In the examples discussed above, we have seen that, at extremality, the near-horizon geometry has an $AdS_2$ factor whose associated BF bound provides a sharp criterion for whether or not an instability is present. For Kerr-AdS, this is less clear because the near-horizon geometry is inhomogeneous. The radius of the $AdS_2$ varies according to position on the $S^2$ horizon. Hence there is no unique BF bound associated to the near-horizon geometry.

Finally we note that $\Omega_H\ell>1$ whenever scalar condensation occurs so this instability coexists with
the superradiant instability discussed in Refs. \cite{Hawking:1999dp}-\cite{Uchikata:2009zz}.

\section{\label{sec:stability}Stability results}

\subsection{Introduction}

We have discussed several examples in which a free scalar field with negative $\mu^2$ is unstable in a near-extreme AdS black hole background. The aim of this section is to obtain sufficient conditions for such a scalar field to be {\it stable}. More precisely, given a black hole background we wish to determine a minimum value of $\mu^2<0$ which guarantees that the scalar field is stable. Our arguments will be based on those of Holzegel \cite{Holzegel:2009ye}, who proved stability of a free scalar field in the $d=4$ spherical Schwarzschild-AdS background provided that the BF bound is respected, and (if below the unitarity bound) using boundary conditions defined by the fall-off $r^{-\Delta_+}$. The main idea is to construct an energy functional for the scalar field that is non-negative and non-increasing. Hence if initially small then it must remain small. This can be used to prove decay of the field. We shall generalize only the first step of this argument (construction of a suitable energy functional) to various different black holes, and also to charged scalar fields. We expect that the second part of the argument (proof of decay) also can be generalized.

\subsection{Static black holes, uncharged scalar}

We shall follow the same approach as Holzegel but allowing for planar or hyperbolic spatial sections and general $d$. Consider a static, asymptotically AdS solution in $d$-dimensions, with metric of the form (\ref{SchwAdS}) for some $f(r)$. This encompasses both the Schwarzschild-AdS and Reissner-Nordstr\"om-AdS solutions.

We define a new time coordinate $t_\ast$ by
\cite{Holzegel:2009ye}
\be \label{coodSch}
dt_\ast = dt + \left( \frac{1}{f} - \frac{1}{1+r^2/\ell^2}
\right)dr\,. \ee
The line element \eqref{SchwAdS} then reads \be
\label{SchwAdS2} ds^2 = -f \,dt_\ast^2 +
2\,\frac{1+r^2/\ell^2-f}{1+r^2/\ell^2} \,dt_\ast dr + \frac{2+2
r^2/\ell^2-f}{(1+r^2/\ell^2)^2} \,dr^2 + r^2 \,d\Sigma_k^2\,. \ee
These coordinates are regular on, and outside the future horizon. They have the nice property that surfaces of constant $t_\ast$ are spacelike and intersect the horizon, with $t_\ast$ increasing along generators of the horizon. The Killing vector field timelike outside the horizon is
\be
 \xi = \frac{\partial}{\partial t} = \frac{\partial}{\partial t_\ast}
\ee Consider the energy of an uncharged scalar field on a constant time slice
$\Sigma_{t_\ast}$, with normal $n^\mu$ and an inner boundary at the
horizon. The energy is given by\footnote{ In this section, we shall
consider only boundary conditions corresponding to the decay
$r^{-\Delta_+}$. If below the unitarity bound then one could also
consider decay as $r^{-\Delta_-}$. This requires modifying the
definition of $E$.} \be
 E(t_\ast)= \int_{\Sigma_{t_\ast}}
T_{\mu\nu} \xi^\mu n^\nu\,dS
\ee
where $dS$ is the volume element on $\Sigma_{t_\ast}$, and the energy-momentum tensor is
\be
T_{\mu \nu} = (\partial_\mu \Phi)(\partial_\nu \Phi) - \frac{1}{2}\,g_{\mu \nu} \big( (\partial \Phi)^2+\mu^2 \Phi^2 \big)\,.
\ee
On the background \eqref{SchwAdS2}, the energy is explicitly given by
  \begin{eqnarray} \label{ModifiedE:Schw}
 E(t_\ast) =\frac{1}{2} \int d\Sigma_k \int_{r_+}^\infty \left( \frac{2+2 r^2/\ell^2-f}{(1+r^2/\ell^2)^2}\, \dot{\Phi}^2 + f {\Phi'}^2 +
 \frac{1}{r^2} (\hat{\nabla} \Phi)^2 + \mu^2 \Phi^2 \right) r^{d-2}
 dr  \,,
 \end{eqnarray}
where a dot denotes a derivative with respect to $t_\ast$, a dash
denotes an $r$-derivative and $\hat{\nabla}$ is the connection on
$\Sigma_k$.

Consider two constant time slices at times $t_{\ast 2}>t_{\ast 1}$. The boundary conditions on the scalar field imply that the energy flux at infinity vanishes, so conservation of the energy current implies that any change in the energy must result from a flux across the horizon:
\be
 E(t_{\ast 2}) - E(t_{\ast 1}) = -\int_{t_{\ast 1}}^{t_{\ast 2}} \left( T_{\mu\nu} \xi^\mu \xi^\nu \,r^{d-2} \right)_{r=r_+} dt_\ast d\Sigma_k
= -\int_{t_{\ast 1}}^{t_{\ast 2}} \left( \dot{\Phi}^2 \,r^{d-2} \right)_{r=r_+} dt_\ast d\Sigma_k\,,
\ee
where we used the fact that $\xi$ is both tangent and normal to the horizon. The RHS is non-positive hence the scalar field energy outside the black hole is non-increasing.

If $\mu^2 \ge 0$ then $E(t_\ast)$ is manifestly non-negative. If it is initially small then, since it is non-increasing, it must remain small.\footnote{This is a standard argument which can be applied whenever the dominant energy condition is satisfied.} Since $E(t_\ast)$ is a sum of squares of $\Phi$ and its derivatives, it follows that $\Phi$ must remain small, hence there cannot be any scalar condensation instability.

The interesting case, however, is $\mu^2<0$ with $\mu^2 \ge \mu^2{\bigr |}_{BF}$. In this case, $E(t_\ast)$ is {\it not} manifestly positive (the dominant energy condition is violated). However, we can exploit an argument of Ref. \cite{Holzegel:2009ye} to demonstrate positivity.
We integrate by parts the last
term:
 \be \label{HolzegelPI}
 \int_{r_+}^\infty \Phi^2 r^{d-2} dr = \frac{1}{d-1} \left[ \Phi^2 (r^{d-1} - r_+^{d-1}) \right]_{r_+}^\infty - \frac{2}{d-1}
 \int_{r_+}^\infty \Phi \Phi' (r^{d-1} - r_+^{d-1} ) dr\,.
\ee The surface term vanishes if we assume $\Phi$ decays
sufficiently fast at infinity, namely as $r^{-\Delta_+}$. The integral Hardy inequality\footnote{This can be derived using the Schwarz inequality on the RHS of \eqref{HolzegelPI}.} implies that
\cite{Holzegel:2009ye}
\be
\label{PhidPhi}
 \int_{r_+}^\infty \Phi^2 r^{d-2} dr \le \frac{4}{(d-1)^2} \int_{r_+}^\infty {\Phi'}^2 r^{-(d-2)}
 \left( r^{d-1} - r_+^{d-1} \right)^2 dr\,.
 \ee
Hence, if $\mu^2 < 0$ we have
 \be \label{Ebound}
 E(t_\ast) \ge  \frac{1}{2} \int_{r_+}^\infty  F(r) {\Phi'}^2 dr d\Sigma_k,
 \ee
where
\be \label{defF}
 F(r) =  f r^{d-2} + \frac{4\mu^2 }{(d-1)^2} r^{-(d-2)}
  \left( r^{d-1} - r_+^{d-1} \right)^2
 \ee
So if we can prove that $F(r)$ is non-negative then the energy is non-negative. We can then argue as before: $E(t_\ast)$ is non-negative and non-increasing and so must remain small if initially small. We can then deduce from (\ref{Ebound}) that $\Phi'$ must remain small and then (\ref{PhidPhi}) implies that $\Phi$ must remain small so there cannot be a scalar condensation instability.\footnote{Ref. \cite{Holzegel:2009ye} gives more rigorous arguments to show boundedness of $\Phi$.}

We now examine the form of $F(r)$ for Schwarzschild-AdS. Using $r \ge r_+$ and $\mu^2<0$ gives
\be
 F(r)/r \ge \left( 1 - \frac{ \mu^2}{ \mu^2{\bigr |}_{BF}} \right) \left( r^{d-1} - r_+^{d-1} \right) + k \left( r^{d-3} - r_+^{d-3} \right).
\ee
Hence if $\mu^2 \ge \mu^2{\bigr |}_{BF}$ then $F(r)$ is manifestly non-negative for $k=0,1$. (This generalizes the argument of Ref. \cite{Holzegel:2009ye} to general $d$ and to $k=0$.) Hence a scalar field in the spherical or planar Schwarzschild-AdS background is stable if it obeys the BF bound.

For the $k=-1$ case, a positive integrand in \eqref{Ebound} requires
 \be \label{boundmax}
- \frac{ \mu^2}{ \mu^2{\bigr |}_{BF}} \ge {\rm max} \left\{ - \frac{\ell^2 f r^{2(d-2)}}{(r^{d-1} - r_+^{d-1})^2}
 \right\}.
\ee
 Sketching the function in curly brackets on the RHS of this relation for a non-extreme solution, it starts at $-\infty$ at $r=r_+$ then increases to a negative maximum value greater than $-1$, then decreases monotonically to $-1$ as $r \rightarrow \infty$. Hence $F(r) \ge 0$ implies that $\mu^2$ must lie strictly above the BF bound. (For large black holes $r_+ > 2\ell$ (say), the value at the maximum is only slightly greater than $-1$ so the new bound is very close to the BF bound.) However, as $r_+ \rightarrow r_+^{\rm ext}$, the location of the maximum tends towards $r=r_+$ and the above inequality reduces to
 \be
  \mu^2 \ge - \frac{d-1}{4 \ell^2}  = \mu^2{\bigr |}_{NH\,BF}^{(Schw)}.
 \ee
 In other words, $F(r)$ is non-negative if, and only if, the scalar field respects the BF bound associated to the {\it near-horizon} $AdS_2$. Hence if this bound is respected then there can be no scalar condensation.

This argument demonstrates that the extreme black hole is stable if
the near-horizon BF bound is respected. But is this bound sharp?
Maybe, although it is not manifest, the energy is still positive for
even lower $\mu^2$. This is easy to exclude: we shall exhibit a
trial function $\Phi$ for which the energy is negative if the
near-horizon BF bound is violated. For a general black hole with
metric of the form (\ref{SchwAdS}) extremality implies that \be
\label{fmuNH} f(r) = (r-r_+)^2/R^2 + {\mathcal O} ((r-r_+)^3)\,. \ee
where $R$ is the $AdS_2$ radius. Consider a trial function
$\Phi_\epsilon (r)$ defined by \be \label{trialfunc} \Phi_\epsilon
(r) = C_\epsilon \, (r-r_+ + \epsilon \,\ell)^{-1/2} r^{-d}\,, \ee
where $C_\epsilon$ is a normalization constant fixed by demanding
that $V_\Sigma \,\int dr \,\Phi_\epsilon^2\, r^{d-2}=1$ ($V_\Sigma$ is the volume of the unit radius compactified hyperboloid). This trial function is
regular at the horizon and satisfies the decay conditions at
infinity. For small enough $\epsilon$, the associated energy is
dominated by the near-horizon contribution, \be
\label{trialfuncresult} E_\epsilon = \frac{1}{2} (\mu^2 -
\mu^2|_{NH\,BF}) + {\mathcal{O}\left(
\frac{1}{\log{(\epsilon^{-1})}} \right)} \,. \ee Therefore, for any
extreme black hole of the form  (\ref{SchwAdS}), if $\mu^2 <
\mu^2|_{NH\,BF}$, then there are regular initial data for which the
energy functional is negative. For the hyperbolic Schwarzschild-AdS
black hole, our numerical results reveal that an instability is present in
this case.

\subsection{Rotating black holes}

We focus now on the rotating black hole solutions \eqref{background}. Let us introduce a coordinate system analogous to \reef{coodSch} which is well behaved on the horizon. Consider new coordinates $t_\ast$ and $\psi_\ast$, such that
\be
dt_\ast = dt + \left( \frac{1}{f} - \frac{1}{1+r^2/\ell^2} \right) \sqrt{h}\, dr\,, \qquad
d\psi_\ast = d\psi + \left( \frac{1}{f} - \frac{1}{1+r^2/\ell^2} \right) \sqrt{h}\, \Omega\, dr\,.
\ee
The line element \eqref{background} then reads
\begin{align}
\label{MPAdS2}
ds^2 = & -\frac{f}{h}\, dt_\ast^2 + \frac{2}{\sqrt{h}}\,\frac{1+r^2/\ell^2-f}{1+r^2/\ell^2} \,dt_\ast dr + \frac{2+2 r^2/\ell^2-f}{(1+r^2/\ell^2)^2} \,dr^2 \nonumber \\
& + r^2 h [d\psi_\ast +A_a dx^a - \Omega dt_\ast]^2 + r^2 \hat{g}_{ab} dx^a dx^b\,.
\end{align}
The Killing vector $\xi= \partial_{t_\ast} + \Omega_H
\partial_{\psi_\ast}$ is everywhere timelike outside the horizon for
$|\Omega_H|\ell<1$. Let us define the energy on a constant time
slice $\Sigma_{t_\ast}$ with normal $n^\mu$ as $E=
\int_{\Sigma_{t_\ast}} T_{\mu\nu} \xi^\mu n^\nu$, \ie we follow Ref.
\cite{Hawking:1999dp} and define the ``energy" with respect to the
Killing vector $\xi$, as opposed to $\partial_{t_\ast}$.\footnote{
This means that our ``energy" is actually ${\cal E} - \Omega_H {\cal J}$,
where ${\cal E}$ is the energy defined with respect to
$\partial/\partial t$ and ${\cal J}$ is the total scalar field angular momentum.}
  On the background \eqref{MPAdS2}, the energy is then explicitly given by
  \begin{align} \label{ModifiedE:MP}
 E =& \pi  \int d\Sigma_N \int_{r_+}^\infty \Bigg[ \frac{2+2 r^2/\ell^2-f}{(1+r^2/\ell^2)^2}\,h\, (\xi \Phi)^2 + \left( \frac{1}{r^2h}-\frac{h}{f}(\Omega_H-\Omega)^2 \right)\,(\partial_{\psi_\ast} \Phi)^2  \nonumber \\
 & + f \left(\partial_r \Phi + \frac{2+2 r^2/\ell^2-f}{(1+r^2/\ell^2)^2}\,\frac{h}{f}(\Omega_H-\Omega)\, \partial_{\psi_\ast} \Phi \right)^2 +
 \frac{1}{r^2} (\CD \Phi)^2 + \mu^2 \Phi^2 \Bigg] r^{d-2}
 dr  \,,
 \end{align}
where $\CD_a = \hat{\nabla}_a - A_a \partial_{\psi_\ast} $, while $\hat{\nabla}$ is the connection on $CP^N$, whose integration measure we denote as $\Sigma_N$. As before, it is straightforward to demonstrate that $E$ is a non-increasing function of $t_\ast$ (this uses the fact that $\xi$ is tangent to the generators of the horizon).

If $\mu^2 \geq 0$, every term in the integrand of Eq.~\eqref{ModifiedE:MP} is non-negative when $|\Omega_H|\ell \le 1$. Hence there is no scalar field instability \cite{Hawking:1999dp}. When $|\Omega_H|\ell>1$, the coefficient in front of $(\partial_{\psi_\ast} \Phi)^2$ becomes negative in a region near infinity, where $\xi$ is spacelike. This is the signal for the superradiance instability \cite{Kunduri:2006qa}, which affects perturbations breaking the $\partial_{\psi_\ast} = \partial_{\psi}$ rotational symmetry.

In the present paper we are interested in axisymmetric perturbations, i.e., $\partial_{\psi_\ast} \Phi=0$. In this case, the relations \eqref{Ebound}, \eqref{defF} and \eqref{boundmax} still hold (with $d\Sigma_k$ substituted by $2\pi d\Sigma_N$). Hence we have a lower bound \eqref{boundmax} on $\mu^2$ which guarantees non-negativity of $E$ and hence stability against axisymmetric scalar perturbations. In the
extreme limit, the maximum on the RHS of \eqref{boundmax} is located
at $r=r_+$ for large black  holes, and at spatial infinity for small
black holes. The energy is non-negative for \be \mu^2 \geq
-\frac{N+1}{2\ell^2}\left(N+2 +N\, \frac{\ell ^2}{r_+^2}\right)\,
\equiv \mu^2|_{NH\,BF}^{(MP)} \qquad \text{if}\;\; r_+ > \ell\,.
\ee
The trial function argument \eqref{fmuNH}-\eqref{trialfuncresult} shows that this bound is sharp in the sense that there exist negative energy initial data if it is violated. Our numerical results confirm that an instability appears if this bound is violated. For small black holes, $r_+\leq\ell$, the inequality is simply $\mu^2 \geq \mu^2|_{BF}$, \ie these solutions are stable against axisymmetric scalar field perturbations as long as the asymptotic BF bound is respected.

\subsection{\label{sec:RNcondition}Charged scalar field}

Since the condensation of a charged scalar field in Reissner-Nordstr\"om-AdS has attracted so much attention recently, we shall consider the extension of the above arguments to this case. The extension is not
completely straightforward because the charged scalar field current couples to the background gauge potential. The equation of motion
for the scalar field is \be \label{RNbox} D^2 \Phi = \mu^2 \Phi\,,
\ee with $D_\mu = \nabla_\mu - \ii\, q A_\mu$, where $q$ is the
charge of the scalar field and $A_\mu$ is the gauge potential. We
consider here Reissner-Nordstr\"om black holes with line element
\eqref{SchwAdS}, where \be f(r) = \left( k - \frac{Q^2}{(r\,
r_+)^{d-3}} \right) \frac{(r^{d-3}-r_+^{d-3})}{r^{d-3}} +
\frac{r^{d-1} - r_+^{d-1}}{r^{d-3}
 \ell^2}\,,
\ee and the gauge potential is given by \be \label{RNgaugeA} A = \frac{Q}{\gamma}
\,\left( \frac{1}{r_+^{d-3}} - \frac{1}{r^{d-3}} \right) dt = \left(
\phi - \frac{Q}{\gamma\,r^{d-3}} \right) dt\,, \ee where $\gamma=
\sqrt{2(d-3)/(d-2)}$. We are using the gauge in which the potential vanishes at the horizon.

The linear instability for scalar condensation can be found by
solving Eq.~\eqref{RNbox} leaving the gauge potential unperturbed
\cite{Gubser:2008px}, since this latter perturbation would have a
higher order contribution. However, energy considerations require a
more careful treatment. As noticed by \cite{Maeda:2010hf}, the
charged current \be j_\mu = \ii\, q[(D_\mu \Phi)^\dagger \Phi -
(D_\mu \Phi) \Phi^\dagger] \ee sources a gauge potential
perturbation through \be \nabla^\mu F_{\mu \nu} = - j_\mu\,. \ee The
consequence is that the energy-momentum tensor for the scalar field,
\be T_{\mu \nu} = (D_\mu \Phi)(D_\nu \Phi)^\dagger + (D_\mu
\Phi)^\dagger(D_\nu \Phi) -g_{\mu \nu} (|D\Phi|^2+\mu^2|\Phi|^2)\,,
\ee where $|\cdot|^2\equiv (\cdot)^\dagger(\cdot)$, is not
separately conserved, $\nabla^\mu T_{\mu \nu}=F_{\nu\mu} j^\mu$.
Only the total energy-momentum tensor, which includes the
contribution from the perturbed gauge potential, is conserved.
Therefore, the energy current $J_\mu=T_{\mu \nu} \xi^\nu$, where
$\xi$ is the timelike Killing vector, is not conserved either,
$\nabla^\mu J_\mu = F_{\mu \nu} \xi^\mu j^\nu $.

This can be easily fixed by considering instead the energy current
(in the Lorentz gauge $\nabla^\mu A_\mu =0$) \be J_\mu=T_{\mu \nu}
\xi^\nu - j_\mu (A_\nu \xi^\nu)\,, \ee which accounts for what would
be the perturbed gauge potential contribution. Note that this is not gauge invariant. The conservation of
this energy current is guaranteed by the condition $A^{[ a} \xi^{b
]}=0$. Using the coordinate system in \eqref{SchwAdS2}, the energy
$E= \int_{\Sigma_{t_\ast}} J_\mu n^\mu$ of the charged scalar field
is given by
\begin{eqnarray}
E = \int d\Sigma_k \int_{r_+}^\infty \left( \frac{2+2 r^2/\ell^2-f}{(1+r^2/\ell^2)^2}\, |D_{t_\ast} \Phi|^2 + f |D_{r}\Phi|^2 +
\frac{1}{r^2} |\hat{\nabla} \Phi|^2 + \mu^2 |\Phi|^2 - j^{t_\ast} A_{t_\ast} \right) r^{d-2} dr  \,,
\end{eqnarray}
whose only difference with respect to \eqref{ModifiedE:Schw}, apart from the fact that we have a complex scalar field and gauge covariant derivatives, is the presence of the last term.

The energy flux through the horizon is determined by $J \cdot \xi =
T_{\mu\nu} \xi^\mu \xi^\nu$ which is easily seen to be positive at
the horizon. Hence the energy is non-increasing in time.\footnote{
There is a close analogy with the rotating case here. The
``standard" choice of gauge would be $A_\mu=0$ at infinity. This
leads to the ``usual" notion of energy ${\cal E}$. This is analogous
to the energy in the rotating case defined with respect to the usual
generator of time translations $\partial/\partial t$. However, to
prove stability we make the alternative gauge choice, for which
$A_\mu=0$ at the horizon and we call the energy $E$. This
corresponds to the energy in the rotating case defined with respect
to the co-rotating Killing field $\xi$. In the charged case we have
$E = {\cal E} - {\cal Q} \phi$, where ${\cal Q}$ is the total charge of the scalar
field, which is analogous to the result $E = {\cal E} - \Omega_H {\cal J}$
in the rotating case.}

The squares in the integrand can be rearranged to give
\begin{align} \label{RNenergy}
E &= \int d\Sigma_k \int_{r_+}^\infty \left( \frac{2+2 r^2/\ell^2-f}{(1+r^2/\ell^2)^2}\, |\partial_{t_\ast} \Phi|^2 + f |\partial_{r}\Phi|^2 +
\frac{1}{r^2} |\hat{\nabla} \Phi|^2 + (\mu^2-f^{-1}q^2A_{t_\ast}^2) |\Phi|^2  \right) r^{d-2} dr  \nonumber \,\\
&\geq \int d\Sigma_k \int_{r_+}^\infty \left( \frac{2+2 r^2/\ell^2-f}{(1+r^2/\ell^2)^2}\, |\partial_{t_\ast} \Phi|^2 + f |\partial_{r}\Phi|^2 +
\frac{1}{r^2} |\hat{\nabla} \Phi|^2 + \mu_q^2 |\Phi|^2  \right) r^{d-2} dr\,,
\end{align}
where $\mu_q^2 \equiv \mu^2-\text{max}\{f^{-1}q^2A_{t_\ast}^2\}$.

If $\mu_q^2 \geq 0$ then the energy is manifestly non-negative. If $\mu_q^2 < 0$, we can proceed as in the uncharged case. Notice first that $|\partial_{r}\Phi|^2 \geq (\partial_{r}|\Phi|)^2$. Hence, if $\mu_q^2 < 0$,
 \be \label{Eboundq}
 E(t_\ast) \ge  \int_{r_+}^\infty  F_q(r) (\partial_{r}|\Phi|)^2 dr d\Sigma_k,
 \ee
where
\be \label{defFq}
 F_q(r) =  f r^{d-2} + \frac{4\mu_q^2 }{(d-1)^2} r^{-(d-2)}
  \left( r^{d-1} - r_+^{d-1} \right)^2.
 \ee
 A positive integrand in \eqref{Eboundq} requires
 \be \label{boundmaxq}
 -\frac{\mu_q^2}{ \mu^2|_{BF}} \ge {\rm max} \left\{ - \frac{\ell^2 f r^{2(d-2)}}{(r^{d-1} - r_+^{d-1})^2}
 \right\}.
\ee
This gives a sufficient condition for stability against charged scalar condensation.

Consider now extreme black holes. These solutions satisfy \be
\label{RNrMext}
\frac{r_+}{\ell}=\sqrt{\frac{d-3}{d-1}(\gamma^2\phi^2-k)} \,. \ee
The maximum of $f^{-1}q^2A_{t_\ast}^2$ is located at $r=r_+$, and
thus $\mu_q$ is the near-horizon ``effective mass'', \be \mu_q^2 =
\mu^2 - (f^{-1}q^2A_{t_\ast}^2)_{r=r_+} = \mu^2 - \frac{(d-3) q^2
\phi^2}{2(d-3) \phi^2 -k}\,. \ee The maximum on the RHS of
\eqref{boundmaxq} is also located at $r=r_+$ if $k=0$ or $k=-1$. On
the other hand, if $k=1$, that maximum is located at spatial
infinity for small $r_+/\ell$, otherwise it is located at $r=r_+$.
We get from \eqref{boundmaxq} the inequality \be \mu_q^2 \geq
-\frac{d-3}{4 \,r_+^2} \left(2(d-3) \phi^2 -k \right)  \equiv
\mu^2|_{NH\,BF}^{(RN)}  \qquad \textrm{if} \;\; k=0,-1, \;\;
\textrm{or}\;\textrm{if}\;\; k=1 \;\; \textrm{with}\;\;
\frac{r_+}{\ell} \geq \frac{d-3}{\sqrt{d-1}} \,, \ee which is
analogous to the uncharged cases (with $\mu \to \mu_q$). This is
shown to be a sharp bound by the trial function argument
\eqref{fmuNH}-\eqref{trialfuncresult}, i.e., there exist negative energy initial data if it is violated.

Small spherical black holes are an exceptional case. In this case,  \eqref{boundmaxq} reduces to
\be
\label{smallbh}
\mu_q^2 \geq \mu^2|_{BF} \qquad \textrm{if}\;\; k=1 \;\;
\textrm{with}\;\; \frac{r_+}{\ell} < \frac{d-3}{\sqrt{d-1}} \,, \ee
since in this case $\mu^2|_{BF}>\mu^2|_{NH\,BF}^{(RN)}$. The
difference to the uncharged cases is that this bound can be
violated, without violating the BF bound for $AdS_d$, because $\mu_q^2\leq\mu^2$.
Small extreme charged black holes can be unstable to the
condensation of a charged scalar field even if the near-horizon
``effective mass'' $\mu_q^2$ is above the near-horizon BF bound.
Such an instability was found in Ref.~\cite{Basu:2010uz}. The result \eqref{smallbh} is unlikely to be sharp: it is sufficient but probably not necessary for stability.

Let us make two final comments. First, notice that the charged scalar condensation on a charged black hole corresponds to the (charged) superradiance instability in AdS. The threshold of scalar condensation is a time-independent mode in the gauge \eqref{RNgaugeA}, where $A(r_+)=0$ and $A(+\infty)=\phi$. To relate to the known phenomenon of superradiance, we consider instead a gauge where the chemical potential vanishes at spatial infinity, $\tilde{A}=A-\phi dt$. The scalar field transforms as $\Phi \sim e^{-i\omega t} \to \tilde{\Phi}\sim e^{-i\tilde{\omega} t}$, where $\tilde{\omega}=\omega+q\phi$. The threshold of superradiance is $\tilde{\omega}=q\phi$, which corresponds to a time-independent mode ($\omega=0$) in the gauge \eqref{RNgaugeA}.

The second comment is on the asymptotically flat limit $\ell \to \infty$. In this case, only $k=1$ solutions exist and $\mu^2$ must be non-negative. A bound simpler than \eqref{boundmaxq}, but sufficient for the present purpose, is obtained for a scalar field satisfying $\mu^2 \geq q^2\gamma^{-2}$, which corresponds to a BPS-like bound. In this case, the energy \eqref{RNenergy} is non-negative if $f>\gamma^2 A_{t_\ast}^2$. Since
\be
f-\gamma^2 A_{t_\ast}^2 = ( 1- \gamma^2 \phi^2 ) \,\frac{r^{d-3}-r_+^{d-3}}{r^{d-3}}\,,
\ee
the energy is non-negative if $\gamma^2 \phi^2 \leq 1$. The saturation of this bound corresponds to extremality, as seen in the $\ell \to \infty$ limit of \eqref{RNrMext}. Therefore, there is no instability of such a scalar field.

\section{Acknowledgements}

We are grateful to Shiraz Minwalla for useful discussions.
OJCD acknowledges financial support provided by the European
Community through the Intra-European Marie Curie contract
PIEF-GA-2008-220197. HSR is
a Royal Society University Research Fellow. This work was partially
funded by FCT-Portugal through projects PTDC/FIS/099293/2008,
CERN/FP/ 83508/2008 and CERN/FP/109306/2009.

\appendix

\section{\label{sec:KerrParameterSpace}Analysis of Kerr-AdS black holes}

The Kerr-AdS black hole has the peculiarity that, for small values
of the dimensionless horizon radius $r_+/\ell$, as the rotation
parameter $a$ or the ADM angular momentum $J$ increases, the angular
velocity $\Omega_H$ first increases and then decreases. Stated in
other words, it is possible to have two different values of $a$ with
the same value of the angular velocity $\Omega_H$. Recall that
$\Omega_H$ is given by \eqref{kerr:OmegaH} and it is the angular
velocity measured by an observer in a frame that does not rotate at
infinity. We do not find a detailed discussion of this property in the literature. We discuss explicitly the
Klein-Gordon equation for a massive scalar field in the Kerr-AdS
background whose study generates the results presented in Section~\ref{sec:CondensationKerrAdS}.

\begin{figure}[ht]
\centerline{\includegraphics[width=.45\textwidth]{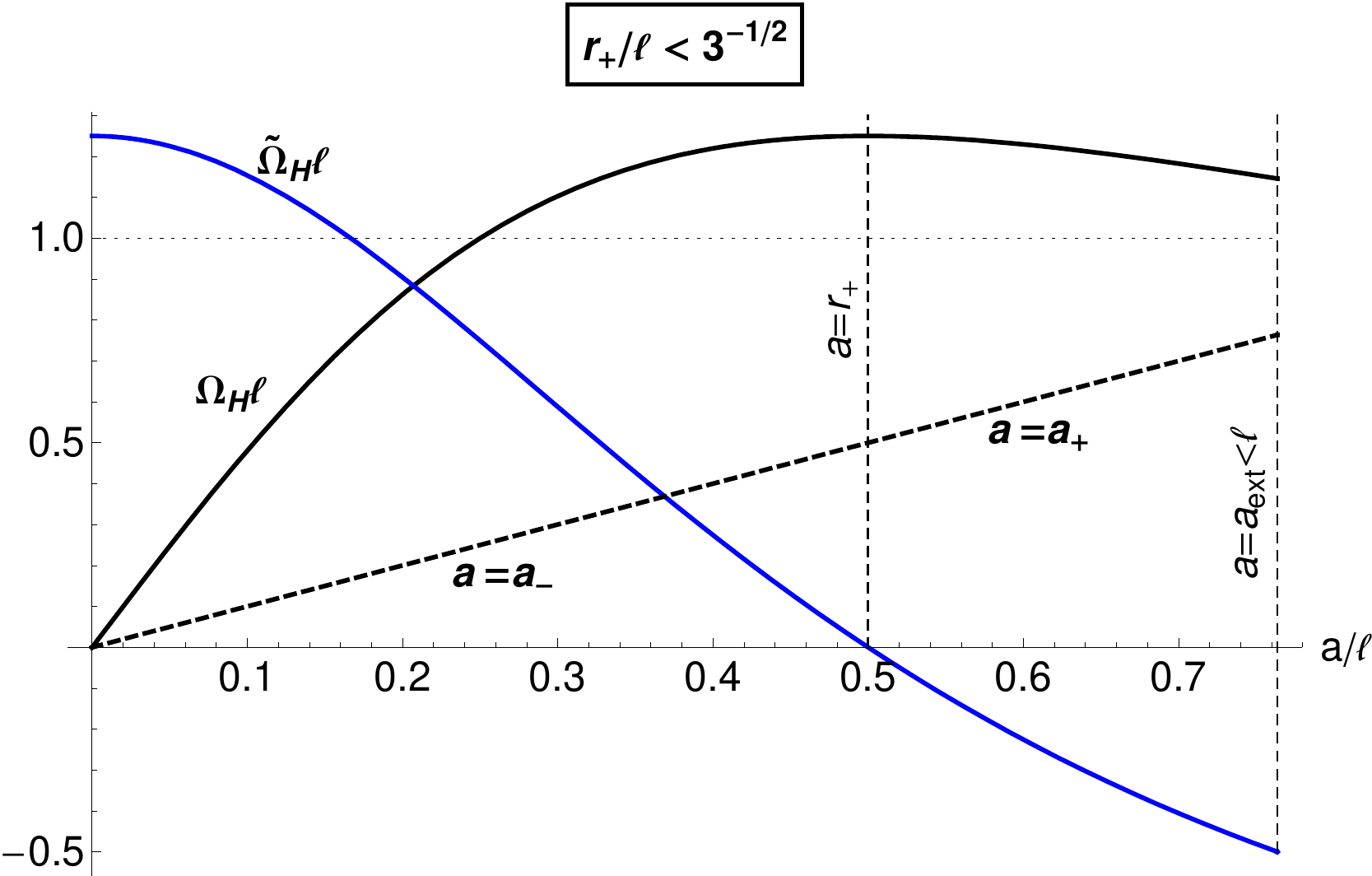}
\hspace{0.5cm}\includegraphics[width=.45\textwidth]{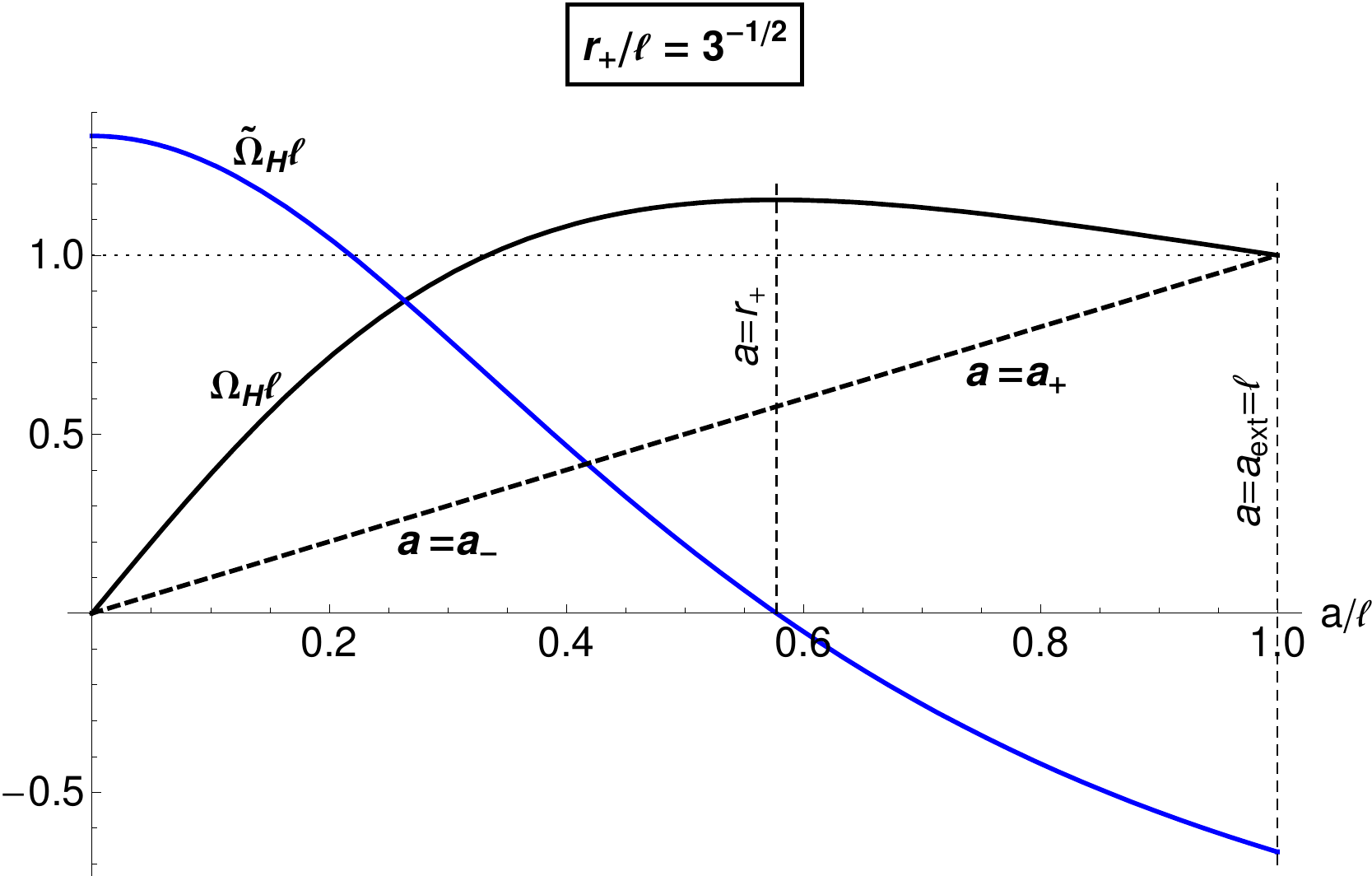}}
\caption{Dimensionless quantities
$\Omega_H\ell,\,\widetilde{\Omega}_H\ell,\,a_\pm/\ell$ as a
function of the dimensionless rotation parameter $a/\ell$ for
$r_+/\ell<3^{-1/2}$ (left) and $r_+/\ell=3^{-1/2}$ (right). The
transition between the curves $a=a_+$ and $a=a_-$ occurs at
$a=r_+$.} \label{Fig:kerrParameterSpaceA}
\end{figure}

\begin{figure}[ht]
\centerline{\includegraphics[width=.45\textwidth]{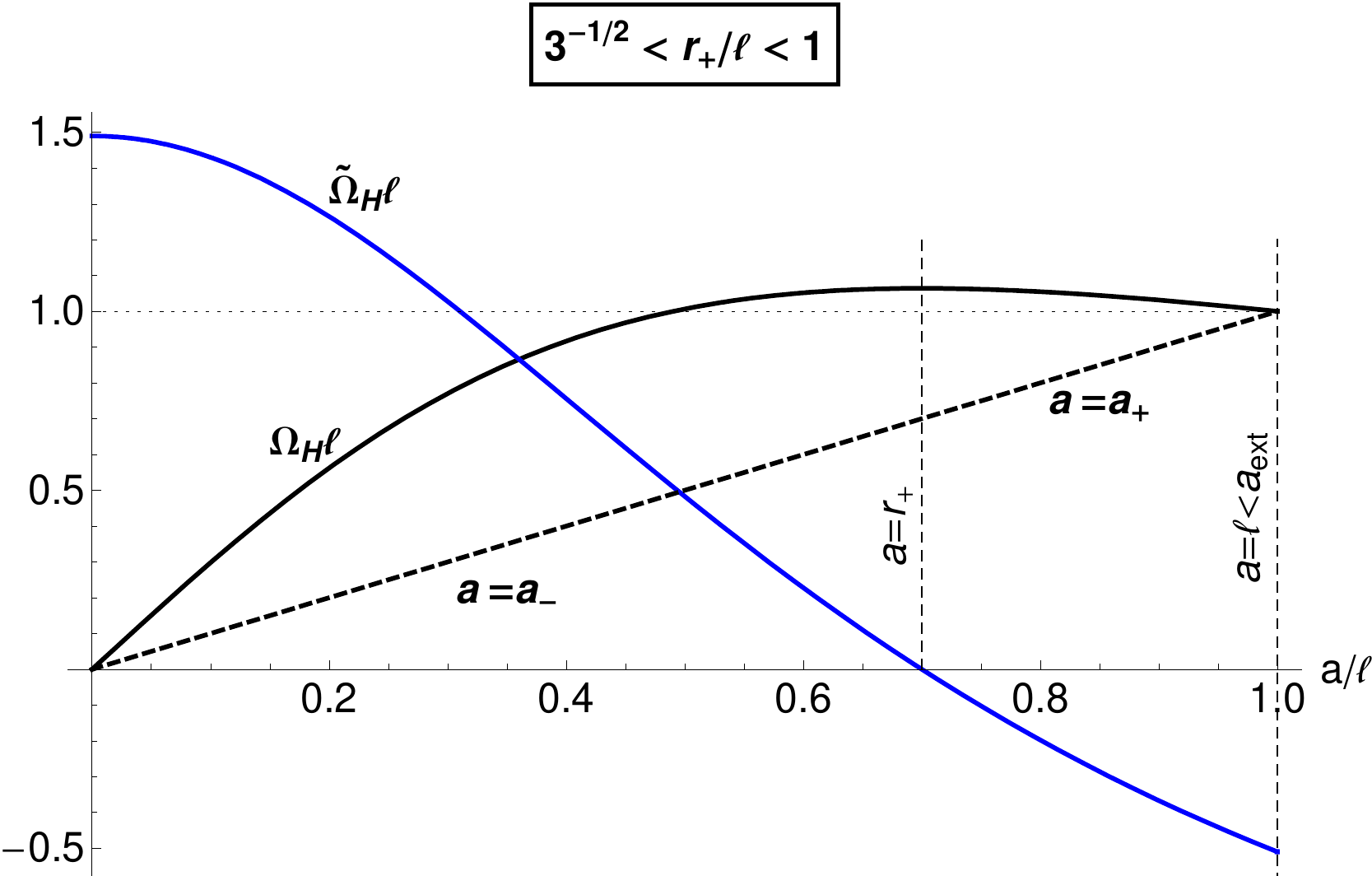}
\hspace{0.5cm}\includegraphics[width=.45\textwidth]{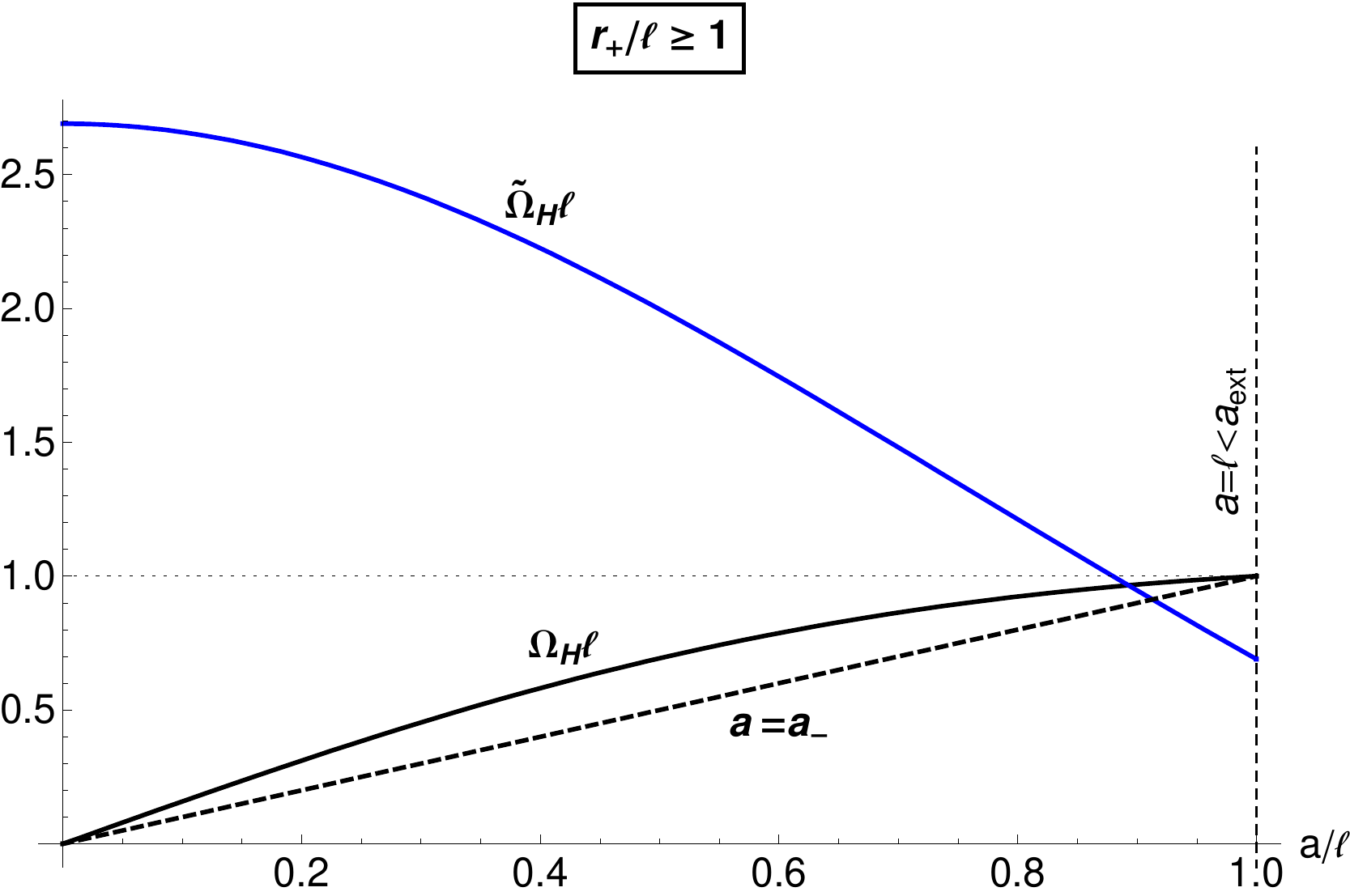}}
\caption{Similar to Figure \ref{Fig:kerrParameterSpaceA} but for
$3^{-1/2}<r_+/\ell<1$ (left) and $r_+/\ell\geq 1$ (right). Note that
for $r_+/\ell = 1$, $\widetilde{\Omega}_H=0$ at $a/\ell=1$.}
\label{Fig:kerrParameterSpaceB}
\end{figure}

For $r_+/\ell<3^{-1/2}$, as the rotation parameter $a$ increases, the
angular velocity $\Omega_H$ in \eqref{kerr:OmegaH} first increases,
then reaches a maximum at $a=r_+$ and then decreases monotonically
until $a=a_{\rm ext}$. Along this path the temperature always
decreases until it vanishes. For $r_+/\ell=3^{-1/2}$, one has
$\Omega_H^{\rm ext} \rightarrow 1/\ell$ when $a\rightarrow a_{\rm
ext}=\ell$, and this is a singular zero temperature configuration.
For $3^{-1/2}<r_+/\ell\leq 1$, $\Omega_H$ also has a maximum at
$a=r_+$ and then decreases to $\Omega_H\rightarrow 1/\ell$ as
$a\rightarrow\ell$. Along this path the temperature decreases but
never vanishes. Finally, for $r_+/\ell
>1$, the angular velocity always increases monotonically and the
temperature decreases without vanishing as $a$ approaches $\ell$.
This discussion is illustrated in Figures
\ref{Fig:kerrParameterSpaceA}-\ref{Fig:kerrParameterSpaceB}. We plot
the dimensionless function $\Omega_H\ell$ as a function of the
dimensionless rotation parameter $a/\ell$ for the five distinct
cases that can describe a black hole as $r_+/\ell$ runs from small
values to large ones. In all these plots, as $a/\ell$ grows the
angular momentum increases and the temperature \eqref{kerr:T}
decreases. Because of the constraint $a<\ell$, only the black holes
of Figure \ref{Fig:kerrParameterSpaceA} (left) can reach
extremality.

We can discuss these properties in a complementary way. Using
\eqref{kerr:OmegaH}, we express the rotation parameter $a$ in terms
of the angular velocity $\Omega_H$. There are two possible roots:
 \be
a_{\pm}=\frac{r_+^2+\ell^2\pm \sqrt{\left(r_+^2+\ell^2\right)^2-4
r_+^2 \ell^4 \Omega_H^2}}{2 \ell^2 \Omega_H}
 \ee
For $r_+/\ell \leq 1$, one has $a(\Omega_H)=a_-$ while $a\leq r_+$,
and $a(\Omega_H)=a_+$ for $r_+ \leq a \leq \hbox{min}\{ a_{\rm
ext},\ell\}$. For $r_+/\ell >1$, one always has $a(\Omega_H)=a_-$.
In Figures
\ref{Fig:kerrParameterSpaceA}-\ref{Fig:kerrParameterSpaceB} we also
identify which of the roots $a_\pm$ describes the black hole
in each situation.

For reasons that will be soon clear, we will find it useful to work
with a different quantity $\widetilde{\Omega}_H$
defined in terms of $\Omega_H$ by
 \be \label{kerr:OmegaEff}
\widetilde{\Omega}_H=
\frac{\varepsilon}{\ell^3}\sqrt{\left(r_+^2+\ell^2\right)^2-4r_+^2\ell^4
\Omega_H^2}\,,\qquad \hbox{with} \quad \varepsilon=\pm 1\quad
\hbox{if}\quad  a\lessgtr r_+\,,\quad \hbox{respectively.}
 \ee
This quantity decreases continuously and monotonically as $a$ grows
from zero to $\hbox{min}\{ a_{\rm ext},\ell\}$. It is positive if
$a<r_+$ and negative when $r_+ \leq a \leq \hbox{min}\{ a_{\rm
ext},\ell\}$.

\begin{figure}[ht]
\centerline{\includegraphics[width=.45\textwidth]{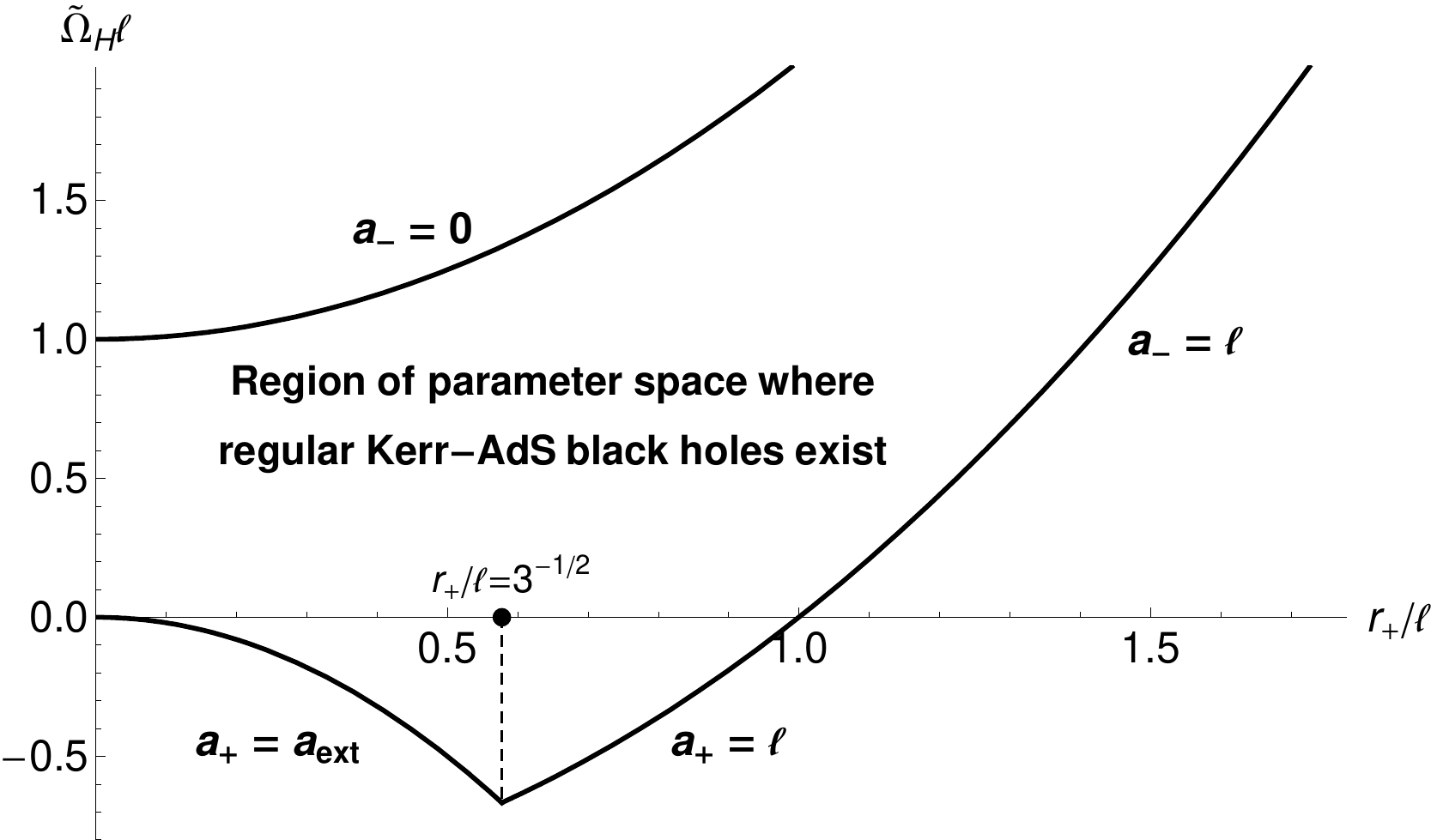}}
\caption{Region plot of the parameter space where Kerr-AdS black
holes exist (area in between the several curves), expressed in terms
of the eigenvalue variable $\widetilde{\Omega}_H \ell$ and the
dimensionless horizon radius $r_+/\ell$. Note that the $l=0$ and
$l=1$ zero-mode curves described in Figure \ref{Fig:Kerr3d} have as
endpoints the curves labeled as $a_+=a_{\rm ext}$ and $a_+=\ell$ in
the current figure.} \label{Fig:KerrRegionPlot}
\end{figure}

Figure \ref{Fig:KerrRegionPlot} on the other hand summarizes some of
the information of the previous plots: it gives a region plot of the
parameter space where Kerr-AdS black holes exist. We express this
information in terms of the variable $\widetilde{\Omega}_H$ that is
relevant for our eigenvalue problem below.


Consider the Klein-Gordon equation \eqref{KGeq} for a massive
scalar field in the Kerr-AdS background. We are interested only in
the stationary and axisymmetric zero-mode of the scalar condensation
instability, so we look for solutions
$\Phi=\Phi(r,\theta)$.\footnote{It is well-known that, under the
ansatz $\Phi(r,\theta)=R(r)S(\theta)$, the Klein-Gordon equation in
Kerr-AdS can be further separated into a radial equation for $R(r)$
and an angular equation for $S(\theta)$. The latter has the AdS
spin-0 spheroidal harmonics as solutions, $S(\theta)$.
Unfortunately, these harmonics cannot be constructed analytically,
neither can the associated angular eigenvalues of the separation
constant. It is possible to construct these approximately in a
series expansion for small values of the rotation and cosmological
constant, but we are interested precisely in the opposite regime of
large rotation. For this reason we do not find any advantage to
introduce this separation ansatz. We search directly for the
zero-modes of $\Phi(r,\theta)$.} Introducing the compact radial
coordinate, $y=1-r_+/r$, and the angular coordinate $x=\cos\theta$,
we find that the Klein-Gordon equation for $\Phi(y,x)$ can be
written as
 \be \label{kerr:KG}
 L \,\Phi(y,x) = \widetilde{\Omega}_H \ell\, \Lambda\, \Phi(y,x),
\ee where the second order differential operators $L$ and $\Lambda$
are given by ($y_+=r_+/\ell$)
\begin{eqnarray} \label{kerr:KGaux}
L \!\!&=&\!\! 4 (1-y) \left(1+y_+^2\right) {\biggl [} -(1-y) y
\left[(1-y)^2 y+(4-y (5-2 y)) y_+^2\right]
\partial_y^2 \nonumber \\
 && -2 \left[(1-y)^3 y+ y_+^2(2-y)[1-(1-y) y]\right] \partial_y
  -\left(1-x^2\right)
 (1-y) \left(1- y_+^2 x^2\right) \partial_x^2 \nonumber \\
 && +2 x (1-y) \left[1+y_+^2 \left(1-2 x^2\right)
\right] \partial_x{\biggr ]} + 4 \mu^2\ell^2  y_+^2 \left(1+y_+^2\right) \left[1+x^2 (1-y)^2\right],  \\
  \Lambda \!\!&=&\!\! 4 (1-y) {\biggl [} y [2-(3-y) y]
\left[(1-y)^2+y_+^2\right]
\partial_y^2 +2 \left[(1-y)^4+y_+^2\right] \partial_y\nonumber \\
 &&\hspace{-0.3cm} +\left(1-x^2\right) (1-y) \left(1+y_+^2 x^2\right)
\partial_x^2-2 x (1-y) \left[1-y_+^2\left(1-2
x^2\right)\right] \partial_x{\biggr ]} -4 \mu^2\ell^2  y_+^2
\left[1-x^2 (1-y)^2\right]. \nonumber
\end{eqnarray}
The ultimate reason why we introduced the quantity
$\widetilde{\Omega}_H$ instead of working with the black hole
angular velocity $\Omega_H$ is now clear. Indeed, in terms of the
former quantity, the Klein-Gordon equation \eqref{kerr:KG} is
explicitly an eigenvalue equation for $\widetilde{\Omega}_H$. The
reason to follow this strategy is motivated by the availability of
numerical techniques for solving eigenvalue equations of the form
\eqref{kerr:KG}, namely the spectral method already described
briefly in Section~\ref{sec:CondensationHHT}. We normalize all quantities in units of the
AdS length. The Klein-Gordon equation \eqref{kerr:KG} depends on the
three dimensionless parameters $r_+/\ell$,
$\widetilde{\Omega}_H\ell$ and $\Delta$ (or, equivalently,
$\mu\ell$). Running a spectral numerical code for several values of
$r_+/\ell$ and $\Delta$, we find the eigenvalues
$\widetilde{\Omega}_H\ell$ of \eqref{kerr:KG}, and thus the
associated physical angular velocity $\Omega_H \ell$ where the
zero-mode of the instability appears. The results are presented in
section \ref{sec:CondensationKerrAdS}.

\section{\label{sec:eomMPnonlinear} Coupled system of ODEs for the codimension-1 rotating hairy black hole}

In this Appendix we write explicitly the system of three coupled
$2{\rm nd}$ order ODEs for $\{ f,h,\Phi\}$ that the codimension-1
rotating hairy black hole must satisfy and that are symbolically
described in equation \eqref{HairyMP:eom123} of the main body of the
text. These are:
\begin{subequations} \label{HairyMP:eom3}
\begin{align}
 & 0=\left\{3 f \tilde{r}^2 \left[h^2 \tilde{r}^3 \left(2 (h-4) \ell^2+\tilde{r} \left(\mu^2\ell^2  \Phi^2-12\right)\right)+2 \ell^2 C_{\psi}^2\right]
  \right\}^{-1}\nonumber \\
 & \hspace{0.3cm}{\biggl\{}6 \ell ^2 C_{\psi }^2 \left[-3 f h+\tilde{r} \left(2 f h'-f' \left(3 h+\tilde{r} h'\right)+f \tilde{r} \left(2 h \Phi '^2
 +h''\right)\right)\right]\nonumber \\
 & \hspace{0.3cm} +\tilde{r}^3 {\biggl [} h \tilde{r} f' \left(3 h+\tilde{r} h'\right) {\bigl (}12 (h-1) h \ell ^2
  +\tilde{r} \left(6 (h-2) \ell ^2 +\tilde{r} \left(\mu ^2\ell ^2 \Phi^2-12\right)\right) h'{\bigr )}
  \nonumber \\
 & \hspace{0.3cm}+f {\bigl [}-4 h \tilde{r}^2 \left(6 (h-2) \ell ^2+\tilde{r} \left(\mu ^2\ell ^2  \Phi ^2
 -12\right)\right) h'^2
  +\tilde{r}^3 \left(-6 (h-2) \ell ^2-\tilde{r} \left(\mu ^2\ell ^2  \Phi ^2-12\right)\right) h'^3\nonumber \\
 & \hspace{0.3cm}
 -2 h^2 \tilde{r} h' {\bigl [}-6 (h-4) \ell ^2-3\tilde{r} \left(\mu ^2\ell ^2  \Phi ^2-12\right)
 +\tilde{r}^2 \left(6 (h-2) \ell ^2+\tilde{r} \left(\mu ^2\ell ^2  \Phi ^2-12\right)\right)
  \Phi '^2{\bigr ]}\nonumber \\
 & \hspace{0.3cm} +3 h^2 \left(-4 (h-1) h \ell ^2 \left(2 \tilde{r}^2 \Phi '^2-3\right)+\tilde{r}^2 \left(2 (h-4) \ell ^2
  +\tilde{r} \left(\mu ^2\ell ^2  \Phi ^2-12\right)\right) h''\right){\bigr ]}{\biggr ]}{\biggr\}}=0
 \,,  \label{HairyMP:eom3a} \\
 & 0= 3 f \tilde{r} \left[h^2 \tilde{r}^3 \left(2 (h-4)\ell^2+\tilde{r}\left(\mu^2\ell^2\Phi^2-12\right)\right)+2 \ell^2 C_{\psi}^2\right]f''\nonumber \\
 & \hspace{0.5cm}
 + \tilde{r} f'^2 \left[-6 \ell ^2 C_{\psi }^2+h \tilde{r}^3 \left(12 (h-1) h \ell ^2+\tilde{r} \left(6 (h-2) \ell ^2
 +\tilde{r} \left(\mu ^2\ell ^2  \Phi ^2-12\right)\right) h'\right)\right]\nonumber \\
 & \hspace{0.5cm}
 +f^2 \tilde{r}^2 \left(-12 (h-1) \ell ^2
 +\tilde{r} \left(\mu ^2\ell ^2  \Phi ^2-12\right)\right) \left(-3 h^2+\tilde{r} \left(h' \left(h+\tilde{r} h'\right)+2 h^2 \tilde{r} \Phi '^2\right)\right)\nonumber \\
 & \hspace{0.5cm}
 +f f' {\biggl\{} 6 \ell ^2 C_{\psi }^2+\tilde{r}^3 {\bigl[}2 h \tilde{r} \left(3 h \ell ^2-\tilde{r} \left(\mu ^2\ell ^2  \Phi ^2-12\right)\right) h'
 +\tilde{r}^2 \left(-6 (-2+h) \ell ^2-\tilde{r} \left(\mu ^2\ell ^2  \Phi ^2-12\right)\right) h'^2\nonumber \\
 & \hspace{0.5cm}+h^2 \left[12 (-8+5 h) \ell ^2
 +3 \tilde{r} \left(\mu ^2\ell ^2  \Phi ^2-12\right)-2 \tilde{r}^2 \left(6 (-2+h) \ell ^2+\tilde{r} \left(\mu ^2\ell ^2  \Phi ^2-12\right)\right) \Phi
 '^2\right] {\bigr ]}{\biggr\}} \,,
   \label{HairyMP:eom3b}\\
 & 0=  6 f \tilde{r} \left[h^2 \tilde{r}^3 \left(2 (-4+h) \ell ^2+\tilde{r}\text{  }\left(\mu ^2\ell ^2  \Phi ^2-12\right)\right)+2 \ell ^2 C_{\psi
 }^2\right]\Phi''   +   9 f h^2 \tilde{r}^3 \ell ^2 \mu ^2 \Phi  \nonumber \\
 & \hspace{0.5cm}
 +h \tilde{r}^4 f' \left(3 h+\tilde{r} h'\right) \left[3 \ell ^2 \mu ^2 \Phi
 +2 \left(6 (-2+h) \ell ^2+\tilde{r}\left(\mu ^2\ell ^2  \Phi ^2-12\right)\right) \Phi '\right]
 -f {\biggl\{} h \tilde{r}^4 h' {\biggl [}3 \ell ^2 \mu ^2 \Phi \nonumber \\
 & \hspace{0.6cm}
 +2 \left(6 (h-2) \ell ^2+\tilde{r}\left(\mu ^2\ell ^2  \Phi ^2-12\right)\right) \Phi '{\biggr ]}
 +\tilde{r}^5 h'^2 \left[3 \ell ^2 \mu ^2 \Phi +2 \left(6 (h-2) \ell ^2+\tilde{r}\left(\mu ^2\ell ^2  \Phi ^2-12\right)\right) \Phi '\right]\nonumber \\
 & \hspace{0.5cm}
 +2 \Phi ' {\biggl[}-6 \ell ^2 C_{\psi }^2
 +h^2 \tilde{r}^3 {\bigl[}12 (5-2 h) \ell ^2+\tilde{r} \left(72-6 \ell ^2 \mu ^2 \Phi ^2\right)\nonumber \\
 & \hspace{0.5cm}
 +\tilde{r}^2 \Phi ' \left(3 \ell ^2 \mu ^2 \Phi +2 \left(6 (h-2) \ell ^2
 +\tilde{r}\left(\mu ^2\ell ^2  \Phi ^2-12\right)\right) \Phi '\right){\bigr ]}{\biggr
 ]}{\biggr\}}\,.   \label{HairyMP:eom3c}
\end{align}
\end{subequations}



\begin{thebibliography}{99}

\bibitem{Gubser:2008px}
  S.~S.~Gubser,
  ``Breaking an Abelian gauge symmetry near a black hole horizon,''
  Phys.\ Rev.\  D {\bf 78} (2008) 065034
  [arXiv:0801.2977 [hep-th]].

\bibitem{Hartnoll:2008kx}
  S.~A.~Hartnoll, C.~P.~Herzog and G.~T.~Horowitz,
  ``Holographic Superconductors,''
  JHEP {\bf 0812} (2008) 015
  [arXiv:0810.1563 [hep-th]].

\bibitem{Breitenlohner:1982jf}
  P.~Breitenlohner and D.~Z.~Freedman,
  ``Stability In Gauged Extended Supergravity,''
  Annals Phys.\  {\bf 144} (1982) 249; \\
  P.~Breitenlohner and D.~Z.~Freedman,
  ``Positive Energy In Anti-De Sitter Backgrounds And Gauged Extended
  Supergravity,''
  Phys.\ Lett.\  B {\bf 115} (1982) 197.

\bibitem{Mezincescu:1984ev}
  L.~Mezincescu and P.~K.~Townsend,
  ``Stability At A Local Maximum In Higher Dimensional Anti-De Sitter Space And
  Applications To Supergravity,''
  Annals Phys.\  {\bf 160} (1985) 406.

\bibitem{Denef:2009tp}
  F.~Denef and S.~A.~Hartnoll,
  ``Landscape of superconducting membranes,''
  Phys.\ Rev.\  D {\bf 79} (2009) 126008
  [arXiv:0901.1160 [hep-th]].

\bibitem{Sonner:2009fk}
  J.~Sonner,
  ``A Rotating Holographic Superconductor,''
  Phys.\ Rev.\  D {\bf 80}, 084031 (2009)
  [arXiv:0903.0627 [hep-th]].

\bibitem{Hawking:1998kw}
  S.~W.~Hawking, C.~J.~Hunter and M.~Taylor,
  ``Rotation and the AdS/CFT correspondence,''
  Phys.\ Rev.\  D {\bf 59}, 064005 (1999)
  [arXiv:hep-th/9811056].

\bibitem{Horowitz:2009ij}
  G.~T.~Horowitz and M.~M.~Roberts,
  ``Zero Temperature Limit of Holographic Superconductors,''
  JHEP {\bf 0911} (2009) 015
  [arXiv:0908.3677 [hep-th]].

\bibitem{Basu:2010uz}
  P.~Basu, J.~Bhattacharya, S.~Bhattacharyya, R.~Loganayagam, S.~Minwalla and V.~Umesh,
  ``Small Hairy Black Holes in Global AdS Spacetime,''
  arXiv:1003.3232 [hep-th].

\bibitem{Bhattacharyya:2010yg}
  S.~Bhattacharyya, S.~Minwalla and K.~Papadodimas,
  ``Small Hairy Black Holes in $AdS_5 \times S^5$,''
  arXiv:1005.1287 [hep-th].

\bibitem{Holzegel:2009ye}
  G.~Holzegel,
  ``On the massive wave equation on slowly rotating Kerr-AdS spacetimes,''
  arXiv:0902.0973 [gr-qc].

\bibitem{Hertog:2006rr}
  T.~Hertog,
  ``Towards a Novel no-hair Theorem for Black Holes,''
  Phys.\ Rev.\  D {\bf 74}, 084008 (2006)
  [arXiv:gr-qc/0608075].

\bibitem{Kunduri:2006qa}
  H.~K.~Kunduri, J.~Lucietti and H.~S.~Reall,
  ``Gravitational perturbations of higher dimensional rotating black holes:
  Tensor Perturbations,''
  Phys.\ Rev.\  D {\bf 74} (2006) 084021.

\bibitem{Martinez:2004nb}
  C.~Martinez, R.~Troncoso and J.~Zanelli,
  ``Exact black hole solution with a minimally coupled scalar field,''
  Phys.\ Rev.\  D {\bf 70}, 084035 (2004)
  [arXiv:hep-th/0406111].

\bibitem{Klebanov:1999tb}
  I.~R.~Klebanov and E.~Witten,
  Nucl.\ Phys.\  B {\bf 556}, 89 (1999)
  [arXiv:hep-th/9905104].

\bibitem{Chan:1999sc}
  J.~S.~F.~Chan and R.~B.~Mann,
  ``Scalar wave falloff in topological black hole backgrounds,''
  Phys.\ Rev.\  D {\bf 59} (1999) 064025.

\bibitem{Aros:2002te}
  R.~Aros, C.~Martinez, R.~Troncoso and J.~Zanelli,
  ``Quasinormal modes for massless topological black holes,''
  Phys.\ Rev.\  D {\bf 67} (2003) 044014
  [arXiv:hep-th/0211024].

\bibitem{Nrecipes}
W. H. Press, S. A. Teukolsky, W. T. Vetterling, and B. P. Flannery,
``Numerical Recipes in C++: The Art of Scientific Computing,''
Cambridge University Press, UK, 2002.

\bibitem{Trefethen}
L.~N.~Trefethen, ``Spectral Methods in MATLAB,'' SIAM, Philadelphia,
2000.

\bibitem{Ashtekar:1999jx}
  A.~Ashtekar and S.~Das,
  ``Asymptotically anti-de Sitter space-times: Conserved quantities,''
  Class.\ Quant.\ Grav.\  {\bf 17} (2000) L17
  [arXiv:hep-th/9911230].

\bibitem{FernandezGracia:2009em}
  J.~Fernandez-Gracia and B.~Fiol,
  ``A no-hair theorem for extremal black branes,''
  JHEP {\bf 0911} (2009) 054
  [arXiv:0906.2353 [hep-th]].

\bibitem{Gibbons:2004uw}
  G.~W.~Gibbons, H.~Lu, D.~N.~Page and C.~N.~Pope,
  ``The general Kerr-de Sitter metrics in all dimensions,''
  J.\ Geom.\ Phys.\  {\bf 53} (2005) 49
  [arXiv:hep-th/0404008].

\bibitem{Gibbons:2004js}
  G.~W.~Gibbons, H.~Lu, D.~N.~Page and C.~N.~Pope,
  ``Rotating black holes in higher dimensions with a cosmological constant,''
  Phys.\ Rev.\ Lett.\  {\bf 93} (2004) 171102
  [arXiv:hep-th/0409155].

\bibitem{Gibbons:2004ai}
  G.~W.~Gibbons, M.~J.~Perry and C.~N.~Pope,
  ``The first law of thermodynamics for Kerr$-$anti-de Sitter black holes,''
  Class.\ Quant.\ Grav.\  {\bf 22} (2005) 1503
  [arXiv:hep-th/0408217].

\bibitem{Gutowski:2004ez}
  J.~B.~Gutowski and H.~S.~Reall,
  ``Supersymmetric AdS(5) black holes,''
  JHEP {\bf 0402}, 006 (2004)
  [arXiv:hep-th/0401042].


\bibitem{Monteiro:2009ke}
  R.~Monteiro, M.~J.~Perry and J.~E.~Santos,
  ``Semiclassical instabilities of Kerr-AdS black holes,''
  Phys.\ Rev.\  D {\bf 81} (2010) 024001
  [arXiv:0905.2334 [gr-qc]].

\bibitem{Dias:2009iu}
  O.~J.~C.~Dias, P.~Figueras, R.~Monteiro, J.~E.~Santos and R.~Emparan,
  ``Instability and new phases of higher-dimensional rotating black holes,''
  Phys.\ Rev.\  D {\bf 80} (2009) 111701
  [arXiv:0907.2248 [hep-th]].

\bibitem{Dias:2010eu}
 O.~J.~C.~Dias, P.~Figueras, R.~Monteiro, H.~S.~Reall and J.~E.~Santos,
  ``An instability of higher-dimensional rotating black holes,''
  JHEP {\bf 1005}, 076 (2010)
  [arXiv:1001.4527 [hep-th]].

\bibitem{Hawking:1999dp}
  S.~W.~Hawking and H.~S.~Reall,
  ``Charged and rotating AdS black holes and their CFT duals,''
  Phys.\ Rev.\  D {\bf 61} (2000) 024014
  [arXiv:hep-th/9908109].

\bibitem{Cardoso:2004nk}
  V.~Cardoso, O.~J.~C.~Dias, J.~P.~S.~Lemos and S.~Yoshida,
  ``The black hole bomb and superradiant instabilities,''
  Phys.\ Rev.\  D {\bf 70} (2004) 044039
  [Erratum-ibid.\  D {\bf 70} (2004) 049903]
  [arXiv:hep-th/0404096].

\bibitem{Cardoso:2004hs}
  V.~Cardoso and O.~J.~C.~Dias,
  ``Small Kerr-anti-de Sitter black holes are unstable,''
  Phys.\ Rev.\  D {\bf 70} (2004) 084011
  [arXiv:hep-th/0405006].

\bibitem{Cardoso:2006wa}
  V.~Cardoso, O.~J.~C.~Dias and S.~Yoshida,
  ``Classical instability of Kerr-AdS black holes and the issue of final
  state,''
  Phys.\ Rev.\  D {\bf 74} (2006) 044008
  [arXiv:hep-th/0607162].

\bibitem{Uchikata:2009zz}
  N.~Uchikata, S.~Yoshida and T.~Futamase,
  ``Scalar perturbations of Kerr-AdS black holes,''
  Phys.\ Rev.\  D {\bf 80} (2009) 084020.

\bibitem{Carter:1968ks}
  B.~Carter,
  ``Hamilton-Jacobi and Schrodinger separable solutions of Einstein's
  equations,''
  Commun.\ Math.\ Phys.\  {\bf 10} (1968) 280.

\bibitem{Maeda:2010hf}
  K.~Maeda, J.~i.~Koga and S.~Fujii,
  ``The final fate of instability of Reissner-Nordstr\'om-anti-de Sitter black
  holes by charged complex scalar fields,''
  arXiv:1003.2689 [gr-qc].

\end{thebibliography}

\end{document}